\documentclass[traditabstract]{aa} 

\usepackage{hhline}
\usepackage{caption}
\usepackage{booktabs}
\usepackage{bm}

\usepackage[table, x11names]{xcolor}
\usepackage{array, booktabs, boldline} %
\usepackage{cellspace}

\usepackage{soul} 
\usepackage{graphicx}
\usepackage{verbatimbox}
\usepackage{longtable,lscape}
\usepackage{rotating}
\usepackage{hyperref}
\hypersetup{colorlinks=true, linkcolor=blue, citecolor=black, filecolor=magenta, urlcolor=blue}
\usepackage{natbib}
\usepackage{amssymb}
\usepackage{txfonts}

\usepackage{fixltx2e}

\usepackage{adjustbox}
\usepackage{amsmath}
\usepackage{cleveref}
\usepackage{marginnote}
\usepackage{threeparttable}
\usepackage{comment}

\newcommand{\myemail}{antonello.calabro@cea.fr}
\newcommand{\SFRIR}{SFR$_{\text{IR}}$ }

\begin{document}


\title{Deciphering an evolutionary sequence of merger stages in infrared-luminous starburst galaxies at z $\sim$ 0.7}

\author{A. Calabr{\`o}\inst{1} 
\and E. Daddi\inst{1}
\and A. Puglisi\inst{1}
\and E. Oliva\inst{2}
\and R. Gobat\inst{3}
\and P. Cassata\inst{4}
\and R. Amor{\'i}n\inst{5,6}
\and N. Arimoto\inst{7}
\and M. Boquien\inst{8}
\and R. Carraro\inst{9}
\and I. Delvecchio\inst{1}
\and E. Ibar\inst{9}
\and S. Jin\inst{1,10}
\and S. Juneau\inst{11}
\and D. Liu\inst{12}
\and M. Onodera\inst{13,14}
\and F. Mannucci\inst{2}
\and H. M{\'e}ndez Hern{\'a}nez\inst{9}
\and G. Rodighiero\inst{4}
\and F. Valentino\inst{15}
\and A. Zanella\inst{16}}


\institute{CEA, IRFU, DAp, AIM, Universit{\'e} Paris-Saclay, Universit{\'e} Paris Diderot, Sorbonne Paris Cit{\'e}, CNRS, F-91191 Gif-sur-Yvette, France (\myemail)
\and INAF-Osservatorio Astrofisico di Arcetri, Largo Enrico Fermi 5, 50125 Firenze, Italy
\and Instituto de F{\'i}sica, Pontificia Universidad Cat{\'o}lica de Valpara{\'i}so, Casilla 4059, Valpara{\'i}so, Chile
\and Dipartimento di Fisica e Astronomia “G. Galilei”, Universit{\`a} di Padova, Vicolo dell'Osservatorio 3, 35122, Italy
\and Departamento de F\'isica y Astronom\'ia, Universidad de La Serena, Av. Juan Cisternas 1200 Norte, La Serena, Chile
\and Instituto de Investigaci\'on Multidisciplinar en Ciencia y Tecnolog\'ia, Universidad de La Serena, Ra\'ul Bitr\'an 1305, La Serena, Chile
\and Astronomy Program, Department of Physics and Astronomy, Seoul National University, 599 Gwanak-ro, Gwanak-gu, Seoul, 151-742, Korea
\and Centro de Astronom{\'i}a (CITEVA), Universidad de Antofagasta, Avenida Angamos 601, Antofagasta, Chile 
\and Instituto de Fisica y Astronom{\'i}a, Facultad de Ciencias, Universidad de Valpara{\'i}so, Gran Breta{\~n}a 1111, Playa Ancha, Valpara{\'i}so, Chile
\and Key Laboratory of Modern Astronomy and Astrophysics in Ministry of Education, School of Astronomy and Space Science, Nanjing University, Nanjing 210093, China
\and National Optical Astronomy Observatory, 950 N. Cherry Avenue, Tucson, AZ 85719, USA
\and Max Planck Institute for Astronomy, Konigstuhl 17, D-69117 Heidelberg, Germany
\and Subaru Telescope, National Astronomical Observatory of Japan, National Institutes of Natural Sciences (NINS), 650 North A'ohoku Place, Hilo, HI, 96720, USA
\and Department of Astronomical Science, SOKENDAI (The Graduate University for Advanced Studies), 650 North A'ohoku Place, Hilo, HI, 96720, USA 
\and Cosmic Dawn Center (DAWN), Niels Bohr Institute, University of Copenhagen, Juliane Maries Vej 30, DK-2100 Copenhagen \O; DTU-Space, Technical University of Denmark, Elektrovej 327, DK-2800 Kgs.\ Lyngby
\and European Southern Observatory, Karl Schwarzschild Stra\ss e 2, 85748 Garching, Germany
}
\date{Received 29 October 2018; Accepted}

\abstract
{Based on optical/near-IR Magellan FIRE spectra of 25 starburst galaxies at $0.5<$ z $<0.9$, \citet{calabro18} showed that their attenuation properties can be explained by a single-parameter sequence of total obscurations ranging from A$_V=2$ to A$_V=30$ towards the starburst core centers in a mixed stars and dust configuration. We investigate here the origin of this sequence for the same sample. We show that total attenuations anti-correlate with the starburst sizes in radio ($3$ GHz) with a significance larger than $5\sigma$ and a scatter of $0.26$ dex. More obscured and compact starbursts also show enhanced N2 (=[NII]/H$\alpha$) ratios and larger line velocity widths that we attribute to an increasing shock contribution toward later merger phases, driven by deeper gravitational potential wells at the coalescence. Additionally, the attenuation is also linked to the equivalent width (EW) of hydrogen recombination lines, which is sensitive to the luminosity weighted age of the relatively unobscured stellar populations. Overall, the correlations among A$_\text{V}$, radio size, line width, N2 and EW of Balmer/Paschen lines converge towards suggesting an evolutionary sequence of merger stages: all of these quantities are likely to be good time-tracers of the merger phenomenon, and their large spanned range appears to be characteristic of the different merger phases. Half of our sample at higher obscurations have radio sizes approximately $3$ times smaller than early type galaxies at the same redshift, suggesting that, in analogy with local Ultraluminous Infrared galaxies (ULIRGs), these cores cannot be directly forming elliptical galaxies. Finally, we detect mid-IR AGN torus for half of our sample and additional X-ray emission for 6 starbursts; intriguingly, the latter have systematically more compact sizes,  suggestive of emerging AGNs towards later merger stages, possibly precursors of a later QSO phase.} 

\keywords{ISM: extinction --- galaxies: evolution --- galaxies: formation --- galaxies: interaction --- galaxies: starburst --- galaxies: star formation --- galaxies: structure --- quasars: supermassive black holes}

\titlerunning{\footnotesize Towards deciphering an evolutionary sequence of merger stages}
\authorrunning{A.Calabr\`o et al.}
 \maketitle
\section{Introduction}\label{introduction}


Starburst galaxies (SBs) host the most powerful star-formation events in the Universe, with star-formation rates (SFR) that, following to the definition of \citet{rodighiero11}, are at least $4$ times higher compared to the average galaxy population at a given redshift, identified by the star-forming Main Sequence (MS) \citep{brinchmann04,noeske07,elbaz07,daddi07a}. 
In the Local Universe these systems are called Ultra Luminous Infrared galaxies (ULIRGs) and have a global infrared luminosity higher than $10^{12}$ L$_\odot$, due to their high stellar production. 
ULIRGs are characterized by peculiar and rather homogeneous properties: they show very compact starbursting cores of dense gas, with typical diameter sizes of $0.5$-$2$ kpc, and may contain double nuclei \citep{genzel98,diazsantos10}. As a consequence of the enhanced SFR density, these cores are also highly dust-rich and completely obscured in the optical/UV, with A$_\text{V}$ ranging $5$-$50$ mag in a foreground dust screen model \citep{genzel98,goldader02}. 

SBs represent a small fraction of the whole star-forming (SF) galaxy population ($2$-$3\%$), almost constant with redshift, according to \citet{sargent12,sargent14} and \citet{schreiber15}, but they constitute a key event in the life of a galaxy. Indeed, at $z\sim0$ the only viable option to explain these concentrations of gas, dust and SFR is through highly dissipative major merger events \citep{sanders96}, in which the gas of colliding galaxies loses angular momentum and energy during the interaction, falling into the coalescing center of the system. Here it serves as fuel for the starburst and for the growth of a supermassive black hole in the center. This scenario is sustained by studies of both local ULIRGs and higher redshift merging/starburst galaxies, showing the presence of an AGN with optical \citep{ellison11}, mid-IR \citep{daddi07b,weston17,satyapal17,goulding18}, X-ray \citep{brusa10,aird18} and radio \citep{best12,chiaberge15} diagnostics. 
The starburst activity and subsequent AGN feedback can cause gas consumption and removal through powerful winds \citep{sanders88,silk98}, leading to a passively evolving elliptical galaxy \citep{kormendy92,springel05b,hopkins08a}.

Hydrodynamical simulations indicate that the morphology of merger remnants is consistent with early type galaxies (ETGs), suggesting the scenario that gas-rich major mergers are primarily responsible for the formation of the massive ellipticals \citep{barnes91,mihos94,mihos96}. 
This evolutionary sequence is strengthened by observations of low surface brightness tidal tales and residual interacting features in local ETGs \citep{duc13}, which may come from major mergers. Other studies have shown that ULIRGs lie on or close to the fundamental plane (effective radius-velocity dispersion-surface brightness) of intermediate-mass ellipticals, S0 galaxies and bulges of local spirals, suggesting that these systems are the final step of gas-rich merger episodes \citep{genzel01}. 

The morphological transformation during the merger happens in different phases that are well represented by the so-called Toomre sequence \citep{toomre72}. Following the result of their simulations, several studies have tried to characterize this sequence by looking at different physical properties of the nuclear regions of merger galaxies, but found no correlations with evolutionary stages \citep[e.g.,][]{laine03}, apart from noticing that latest interaction times, during or after the coalescence, have among the highest infrared luminosities (L$_\text{IR}$) \citep{laine03,haan13}. 
The absence of clear correlations in these works can be interpreted either with the difficulty of identifying correctly the merger phase from the optical morphology (which becomes even more problematic beyond the Local Universe) and/or that other parameters are driving and tracing this transformation. 
For example, \citet{gao99} and \citet{leech10} studied a sample of local (U)LIRGs with double merging nuclei. They found lower molecular gas masses and higher star-formation efficiency and gas excitation (probed by the CO(3-2)/CO(1-0) line ratio) with decreasing separation between merger components, i.e., toward more advanced merger stages. However, since their sample only includes interacting pairs more distant than $\sim1$ kpc, this result is limited to relatively early stage mergers and does not necessarily apply also to coalesced systems. Furthermore, some spatially resolved studies on local (U)LIRGs have found that there is a higher contribution of shocks accompanied by an increased velocity dispersion of the gas toward later merger stages  \citep{monrealibero06,monrealibero10,rich14,rich15}. 

While in the Local Universe the relation between starbursts and mergers is well settled observationally \citep[e.g.][]{murphy96,luo14}, at higher redshift the interpretation becomes less clear. Due to the enhanced gas fractions and disk instabilities of high redshift star-forming galaxies, mergers as a result might increase SFRs less dramatically \citep{fensch17}, and starburst galaxies may be triggered by anomalous gas accretion events \citep{scoville16}. However, other studies have shown that the most extreme starbursts are still merger-driven, displaying disturbed morphologies \citep{elbaz03} and increased star-formation efficiencies compared to MS galaxies \citep{sargent14,silverman18a,silverman18b,silv15}. 
Similarly, the connection between mergers and AGNs is still debated. Even though it may still hold for the most luminous cases \citep{combes03}, some studies (focused especially on optical wavelengths) do not find systematic differences in merger fraction and galaxy distortions between active and non-active systems \citep{cisternas11,kocevski12}, and not all AGNs are triggered by mergers, according to, e.g., \citet{draper12} and \citet{ellison15}. 

The solution to these long-standing problems is further complicated due to the intrinsic faintness of interacting features (e.g., tidal tales, bridges) and their elevated obscurations, which hamper their identification and physical characterization. \citet{puglisi17} have selected a sample of starbursts ($\times 4$ above the MS) at z $\sim1.6$, showing that optical lines, including H$\alpha$ and H$\beta$, only probe a small nearly-unattenuated component of the galaxies, accounting for $\sim 10\%$ of the total SFR$_\text{IR}$. Therefore, a possible solution to study the properties of the dusty starburst population at high redshift is to observe at longer wavelengths, targeting near-infrared rest-frame lines, such as the Paschen line series of hydrogen. Current ground-based spectrographs can observe Pa$\beta$ (the second brightest Paschen recombination line) in K band up to a redshift of $\sim 0.9$. Motivated by this idea, we followed-up of a sample of $25$ SBs at $0.5<z<0.9$ with FIRE (Folded Port InfraRed Echellette), a near-IR spectrograph mounted at the Magellan telescope, in order to detect optical/near-IR lines ranging $0.6<\lambda<1.3$ \AA\ rest-frame. This sample was presented in \citet{calabro18}, which hereafter is referred as Paper I.  

By comparing H$\alpha$, Pa$\beta$ fluxes and bolometric IR luminosities, we found that the attenuation properties of intermediate redshift SBs are not consistent with the predictions of dust-foreground extinction models, but rather with those of a geometrical model in which dust and stars are homogeneously mixed (Paper I). We also found that they are highly obscured on average, with median A$_{\text{V,tot}}$= $9$ mag in a mixed model, while independently derived dust-column densities suggest A$_{\text{V,tot}}$ even higher for a large fraction of them, up to $75$ magnitudes. This means that they have extremely obscured cores and that optical-near-IR lines only probe an external skin containing a fraction ($\sim10$-$30 \%$) of the total SFR. Even more, we argued that the presence of optically thick cores are themselves striking evidence of the merger origin of intermediate-z SBs (like in local ULIRGs), as no other mechanisms are known to produce such large obscuring column densities of gas and dust. This can be used to identify other mergers in the high-redshift Universe \citep{calabro18}, where other methods based on morphology become unfeasible. 

Despite their higher average obscurations compared to MS galaxies, we found that our sample, while showing a nearly constant Pa$\beta$/H$\alpha$ ratios, spans a large range of A$_\text{V,tot}$ between $2$ and $30$ (see Figure~4 in Paper I), thus forming a sequence of increasing attenuations.
This indicates that a substantial intrinsic diversity should exist among them, possibly related to different phases of the merger, to the gas properties (i.e., total gas mass, gas fraction) or to the morphology of pre-existing colliding galaxies. In alternative, the sequence may be driven also by the orbital geometry of the merger. For example, \citet{dimatteo08} have shown that the inclination of the two encounters affects both the duration and intensity of the star formation, regulating the amount of gas that is funneled toward the coalescing center. 

In this paper we investigate the origin of the obscuration sequence that was found in Paper I, analyzing the attenuations A$_\text{V,tot}$ of our starbursts in relation with other physical properties derived from our Magellan spectra and from available multiwavelength images. In Section~2 we describe how the starbursts are selected from the parent photometric catalog along with our Magellan-FIRE observations, spectra reduction (which includes also the analysis of public optical spectra available for a fraction galaxies) and subsequent emission line fluxes (and equivalent width) measurements. This will be followed by a description of the morphological properties, radio size measurements and AGN identification within our sample. In Section \ref{results} we will show the main results of this work, which are discussed extensively in Section~4. A summary with conclusions will be presented in the last Section~5. We adopt the \citet{chabrier03} Initial Mass Function, AB magnitudes, and standard cosmology ($H_{0}=70$ $\rm km s^{-1}Mpc^{-1}$, $\Omega_{\rm m} = 0.3$, $\Omega_\Lambda = 0.7$). We also assume by convention a positive equivalent width (EW) for emission lines and a negative EW for lines in absorption.

\section{Methodology}\label{methodology}

In this section we outline the general starburst selection procedure, and then describe the observations, the spectral reduction and the emission line measurements that were performed for a representative subset of $25$ of them. We will then present their morphological classification, radio size measurements and AGN identification procedure. 

As described in Paper I, we have already calculated for this subset the absolute dust attenuation in V band towards the center ($=$ A$_{\text{V,tot}}$) in two steps. First, we considered the ratio (SFR$_\text{IR}$/SFR$_{Pa\beta,\text{obs}}$), where SFR$_\text{IR}$ is the intrinsic SFR from the infrared and $\text{SFR}_{Pa\beta ,\text{obs}}$ is derived from the observed Pa$\beta$ luminosity, adopting an intrinsic ratio Pa$\beta$/H$\alpha=0.057$ and a standard \citet{kennicutt94} calibration, valid for case B recombination and T$_e=10^4K$. A$_{Pa\beta,\text{IRX}}$ can be eventually inferred from that quantity as $=2.5 \times log_{10}(1+\text{SFR}_{\text{IR}}/\text{SFR}_{Pa\beta,\text{obs}}$).  
In the final step, the attenuation in V band is obtained by solving for A$_{\text{V,tot}}$ the following equation valid for a mixed geometry model of dust and ionizing stars:
\begin{equation}\label{eqAV}
\frac{SFR_{Pa\beta,\text{obs}}}{SFR_{IR}} = \frac{log_{10}(e)}{0.4} \times \left(\frac{1-10^{-0.8 A_{\text{tot}}(Pa\beta)}}{2 A_{\text{tot}}(Pa\beta)}\right)
\end{equation}
where $A_{\text{tot}}(Pa\beta)$ is the total absolute attenuation at Pa$\beta$ towards the center defined as k(Pa$\beta$) $\times$ A$_\text{V,tot}$/R$_\text{V}$. In the last expression, k(Pa$\beta$) corresponds to the local extinction at Pa$\beta$, assuming a \citet{cardelli89} law (R$_\text{V}=3.1$). More details on the derivation and the implications of Eq.\ref{eqAV} can be found in Paper I. 

\subsection{Sample selection and Magellan FIRE observations}\label{sample_selection}

The starburst candidates were selected from the IR+(sub)mm catalog of \citet{jin18} by first requiring the spectroscopic redshift (coming from optical surveys, Salvato et al. in preparation) to be in the range between $0.5$ and $0.9$. This guarantees that Pa$\beta$ and H$\alpha$ fall, respectively, in the K and YJ bands, thus within the wavelength coverage of FIRE ($0.82$ $\mu$m - $2.4$ $\mu$m). For two galaxies, we found that their previous spectroscopic redshift measurements were not correct. Even though they are outside of our selection range, we did not remove them since they satisfy the other selection criteria. 

Secondly, we required SFR$_{\text{norm}}$ to be more than a factor of $4$ higher than SFR$_{\text{MS,z=0.73}}$, as in \citet{rodighiero11}\footnote{Two starbursts that we have previously selected and observed turned out to have slightly lower SFRs than the $\times4$ limit, because of subsequent updates of the photometric catalog.}, where SFR$_{\text{norm}}$ is the total SFR (SFR$_\text{tot}$=SFR$_\text{IR}$+SFR$_{\text{UV,unobscured}}$), normalized to the median redshift of the sample ($0.73$) using the evolution of \citet{sargent14}: 
\begin{equation}
SFR_{\text{norm}}=(SFR_\text{IR}+SFR_{\text{UV,unobs}}) \times \left(\frac{1+0.73}{1+z}\right)^{2.8},
\end{equation}\label{SFRmed}
while SFR$_{\text{MS,z=0.73}}$ is the SFR of the Main Sequence at redshift $0.73$ from \citet{schreiber15}. This latter was found in Paper I to agree well with the MS derived independently for our parent sample through a running median on $10$ stellar mass bins from $10^{10}$ to $10^{12}$ M$_\odot$. We chose this selection procedure in order to compare all the galaxies with a single MS (cf. Fig.~1 of \citet{calabro18}), instead of considering different relations for each galaxy redshift. However, the two procedures yield the same number of starbursts among the whole sample. For the targets studied in this paper, the mean distance from the MS ($\text{dist}_\text{MS}=$ log$_{10}$(SFR$_\text{tot}$/SFR$_\text{MS}$)) would change by $1.8\%$, thus it would not affect in any case the results.

The \SFRIR was derived following the methodology of \citet{jin18} and \citet{liu18}, which is based on fitting the photometry from IRAC to radio $1.4$ GHz with four components: a stellar \citet{bruzual03} SED (with age $200$ Myr, constant star-formation history, solar metallicity Z$_\odot$, Chabrier IMF and Calzetti attenuation law), a warm+cold dust emission template from \citet{draine07} and \citet{magdis12}, and a mid-infrared AGN SED from \citet{mullaney11} to separate the IR contributions of the AGN and SF. 
The unobscured UV-based SFRs (SFR$_{\text{UV,unobscured}}$) were calculated instead from the u-band magnitudes \citep{laigle16}, which probes the UV rest-frame at our redshift, following \citet{heinis14}. Overall, the contributions of the AGN and of the UV-unobscured SFR to the total SFR are on average small (3$\%$ and 1$\%$, respectively).

Finally, as a last requirement, we asked for M$_\ast$ to be greater than $10^{10}$ M$_\odot$, where M$_\ast$ comes from \citet{laigle16} and is computed at the photometric redshift. This condition ensures that, within the mass and redshift constraints adopted here, all SBs would be IR-detected with a S/N$_{\text{IR}}>5$ (cf. Fig.~13 in \citet{jin18}), thus we have a SFR complete sample of SBs. 
Then, for the whole sample, we considered the spectroscopic redshifts (when available) instead of the photometric values, however this does not significantly affect the stellar masses: the two estimates for the parent sample are in agreement within a 1$\sigma$ scatter of $0.11$ dex, compatible with the uncertainties reported by \citet{laigle16}. 
Only for one SB analyzed in this work (ID $685067$, z$_\text{spec}=0.37$), the new stellar mass was remarkably lower ($-0.56$ dex), due to the large difference with previous photometric redshift (z$_\text{phot}=0.71$). Therefore, $\text{dist}_\text{MS}$ was even higher, strengthening its starburst selection.

This criteria yielded 152 SBs, 25 of which were observed during 4 nights at the Magellan 6.5 $m$ Baade Telescope ($17$-$18$/$03$/$2017$ and $22$-$23$/$03$/$2018$). The observed $25$ galaxies were chosen from the pre-selected SB sample according to a priority list. 
We preferentially observed sources close to bright stars (J < $19$-$20$ mag), so as to facilitate target acquisition, although we eventually avoided blind offsets, since our galaxies are already sufficiently bright (peak magnitudes $<19$ mag) to be detected in $\sim 20$-$60$ s in the good seeing conditions of those observations. In addition, we targeted intrinsically brighter sources first, maximizing SFR/D$_L^2$(z) ratio (D$_L$ is the luminosity distance), and assuming no prior knowledge about the dust attenuation of the system, which was set to 0 in all cases. This introduces a small bias in our selection toward the more massive objects. However, our galaxies span the full range of stellar masses above $10^{10} M_\odot$. We refer to Fig.~1 in Paper I, where we presented the redshift, the SFR and the stellar mass distribution of our observed starbursts and our parent galaxy sample.

Our targets were observed with the single-slit echelle spectrograph FIRE \citep{simcoe13}, which has a wavelength range of $0.82$-$2.4 \mu$m. We refer to \citet{simcoe13} for the full technical description of the instrument.
We chose a slit width of $1''$, (yielding a spectral resolution of R$=3000$) to minimize slit losses (the average intrinsic FWHM angular size in Ks-band for our sample is $\sim 0.6''$) and reduce the impact of OH sky emission. In all the cases, the slits were oriented along the semi-major axis of each galaxy, as determined from HST i-band images. Additionally, we benefited from good seeing conditions over all the four nights, with an average of $0.7''$ and a minimum of $0.45''$. 
The majority of our starbursts were observed in AB sequence, with single exposure times of $15$-$20$ minutes \footnote{We chose single frame integrations of $20$ minutes during the first run and $15$ minutes in the second run, which significantly reduces saturation of OH lines in K band, thus helping spectral reduction in that band.}. We decided to double the integration times (completing the ABBA sequence)\footnote{In practice, doing an AB sequence is irrelevant for the spectral reduction, as the pipeline reduces each frame separately (See Section \ref{spectroscopic_reduction}), though it allows us to easily derive 2D emission line maps with the standard IRAF tasks \textsc{imarith} and \textsc{imcombine}.} for galaxies with a lower S/N of the H$\alpha$+[\ion{N}{II}] complex (based on real-time reduction), to improve the detection of fainter lines. 

\subsection{Spectroscopic reduction}\label{spectroscopic_reduction}

The spectra were fully reduced using the publicly available IDL-based FIREHOSE pipeline \citep{gagne15}. For each exposure, we used internal quartz lamps (one for each observing session) to trace the $21$ orders of the echelle spectra and to apply the flat field correction. Then, the wavelength calibration was performed by fitting a low (1-5) order polynomial (depending on the spectral order) to ThAr lines of lamp exposures (taken close in time to the corresponding science frames). We checked that the residuals of the fitted lines to the best-fit wavelength solution are less than $1$ pixel in all the cases, and is $< 0.1$ pixel for the majority of the orders. This translates into an average wavelength accuracy of $\Delta \lambda / \lambda$ $\simeq 5 \times 10^{-5}$, nearly constant across the entire spectral range. Finally, the sky subtraction was applied independently for each single frame following the method of \citet{kelson03}. In this step, the OH lines in the spectra are used to refine the wavelength calibration.

The 1-D spectra are extracted from the 2-D frames using an optimal extraction method \citep{horne86}. However, this procedure cannot be applied when there is a rapidly varying spatial profile of the object flux \citep{horne86}, as in the presence of spatially extended and tilted emission lines. We used in these cases a boxcar extraction procedure, with a sufficiently large extraction aperture (always $> 1.3''$) in order to include all the line emission from the 2-D exposure. We adopted the boxcar extraction for $3$ galaxies in our sample, which are the IDs $245158$, $493881$ and $470239$. Given the agreement within the uncertainties between the fluxes measured with the two approaches for the remaining galaxies, the use of the boxcar procedure does not appear to introduce a systematic flux bias. 

After the spectral extraction, we applied the flux calibration to each 1-D extracted spectrum, using telluric spectra derived from the observations of A0V stars. Before dividing the object and telluric spectra in the pipeline, we could interactively refine the wavelength matching between the two by minimizing the $rms$ of the product. However, at infrared wavelengths slightly different times and/or airmasses between science and standard star observations can produce non-negligible telluric line residuals, affecting the subsequent analysis. We found that this problem was more relevant in K-band, where strong telluric features are present in the observed wavelength ranges $19950$-$20250$ \AA\ and $20450$-$20800$ \AA. The residuals in these regions can produce artificial variations of the real flux up to a factor of $2$, while it is less significant at shorter wavelengths (Y to H). In order to remove these artifacts, we followed the procedure described in \citet{mannucci01}: we first considered a standard star at an airmass of $\sim1.5$ and calibrated it with two different stars observed at significantly lower and higher airmasses (e.g., $1.2$ and $1.9$). Then the two obtained spectra are divided, yielding a global correction function (which is different from $1$ only in the regions of strong telluric features defined above) that applies to all the single-exposure spectra, each of them with a different multiplicative factor until the telluric line residuals disappear 
\footnote{We fitted a linear function in nearby regions free of telluric regions and emission lines, and then determined the correction function through minimizing the rms of the difference between the corrected spectrum and the afore-mentioned continuum fit}. 
Finally, we combined (with a weighted-average) all the 1-D calibrated spectra of the same object. 

The error on the flux density f$_\lambda$ obtained from the FIRE pipeline was checked over all the spectral range, analyzing the continuum of each galaxy in spectral windows of $200$ \AA\ and steps of $100$ \AA, masking emission lines. In each window, we rescaled the rms noise so as to have the $\chi_{\text{reduced}}^2 =$ 1 when fitting the continuum with a low-order ($\lesssim 1$) polynomial\footnote{A spline of order $1$ spanning the whole wavelength range was used as a correction function}. This criterion, equivalently, ensures that the noise level matches the $1$-$\sigma$ dispersion of the object spectrum in each window. Typical corrections are within a factor of $2$, variable across the spectral bands. 

Due to slit losses, variable seeing conditions and the spatial extension of our objects, which are typically larger than the slit width ($1''$), part of the total flux of the galaxies is lost. In order to recover the total absolute flux, we matched the whole spectrum to the photometric SED. This was done by applying a $5\sigma$ clipping and error-weighted average to the Magellan spectrum inside z++, Y, J, H and K$_s$ photometric bands, and comparing the obtained mean $f_\lambda$ in each filter to the corresponding broad-band photometry \citep{laigle16}. Since the SED shapes derived from the spectra are generally in agreement with the photometric SED shapes, we rescaled our spectra with a constant factor, determined through a least squares minimization procedure. The aperture correction factors for our sample range between $0.8$ and $3$, with a median of $1.4$. They are subject to multiple contributions, i.e., slit position with respect to the object, seeing conditions during the target and the standard stars observations. The few cases in which the aperture correction was lower than $1$ could be due indeed to a much better seeing of the standard star compared to the target observation. 
We remind that this procedure assumes that lines and continuum are equally extended, which is clearly an approximation. Spatially resolved near-IR line maps (e.g., with SINFONI) would be required to test possible different gradients of the two emission components, and to derive better total flux corrections.

\subsection{Complementary optical spectra}\label{complementary_optical_spectra}

\begin{figure*}[t!]
  \centering
  \raggedright{\textbf{HST F814W} \quad \qquad \textbf{UltraVISTA H} \qquad \textbf{VLA 3 GHz}}
  \includegraphics[angle=0,width=17.5cm,trim={0cm 0.cm 0cm 127cm},clip]{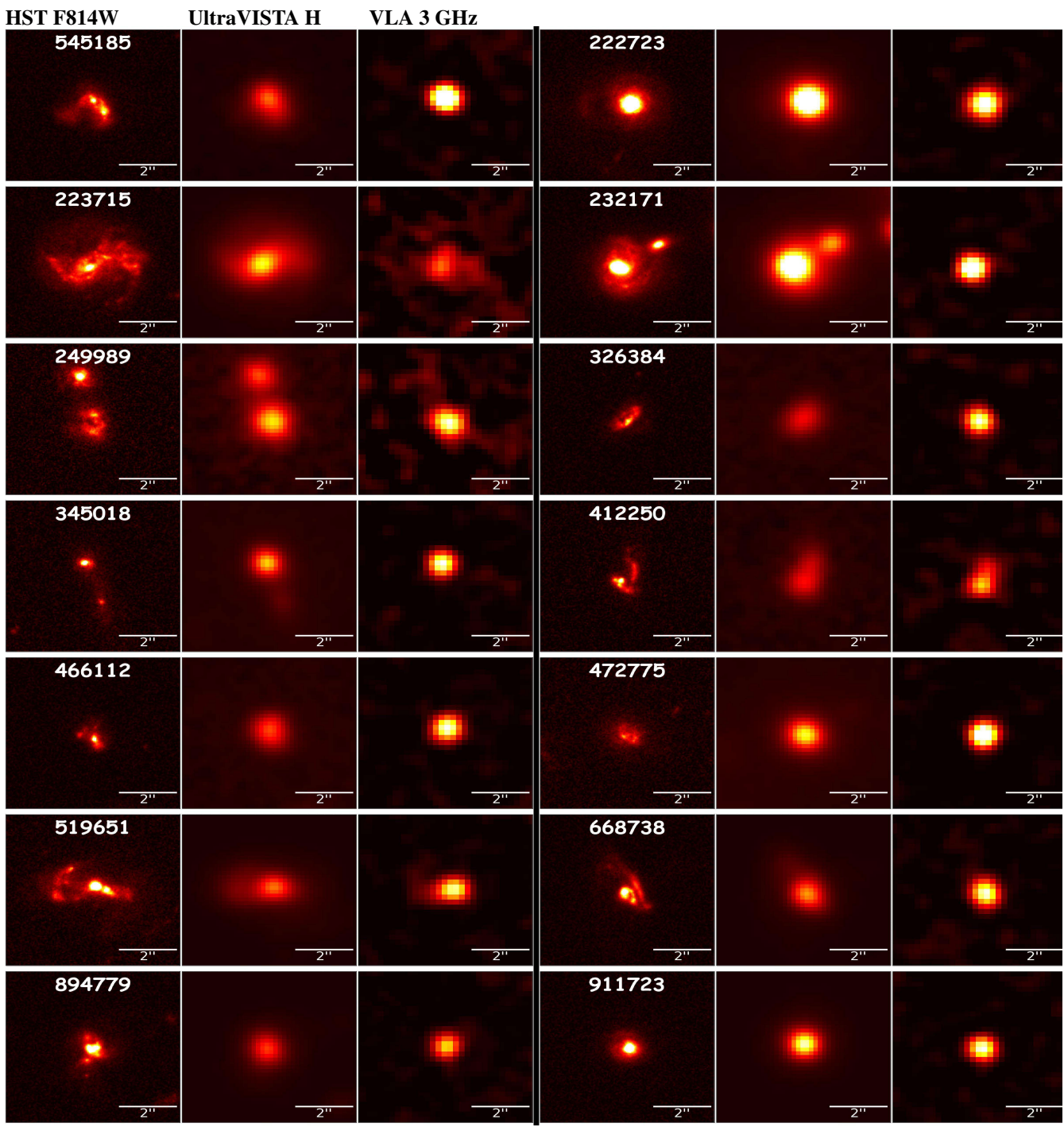} 
  \caption{\small For each of the galaxies observed in our second observing run at Magellan we show (from left to right): HST-ACS F814W images (FWHM$_{\text{res}}=0.095''$), H-band UltraVISTA cutouts (same f.o.v. and FWHM$_{res} \sim 0.75''$) and 3 GHz radio images from VLA-COSMOS 3GHz Large Project \citep{smolcic17} (FWHM$\sim 0.75''$). 
  Cutout images for the sample of the first observing run were presented in Paper I.}\label{stamps}
\end{figure*}

A subset of our Magellan sample also has publicly available optical spectra: $10$ starbursts in our sample have been observed with the VIMOS spectrograph \citep{lefevre03} by the zCOSMOS survey \citep{lilly07}, and their optical spectra are publicly available through the LAM website (\url{cesam.lam.fr/zCOSMOS}). They span the range $5550<\lambda<9650$ \AA, which includes [\ion{O}{II}]$\lambda 3727$ \AA, H$\gamma$, H$\beta$ and [\ion{O}{III}]$\lambda 5007$ \AA\ lines for our galaxies. The spectral resolution is on average $R=600$, constant across the whole range, while the noise level increases from the blue to the red part of the spectrum, due to the increased OH sky emission at longer wavelengths. Due to the absence of the noise spectrum in the public zCOSMOS release, we used instead a sky spectrum at the same resolution of VIMOS, rescaled with a spline to match the flux standard deviation in spectral regions free of emission lines, similarly to what has been done before on the Magellan spectra. 
Even though it is a first approximation, this procedure allows to reproduce the increasing noise in correspondence of OH lines, i.e., where strong sky-subtraction residuals are expected, and take into account the higher average noise level of the red part of the spectrum. 
For one galaxy in our sample (ID $493881$), we took its optical spectrum from SDSS-III DR9 \citep{ahn12}, which spans a wider wavelength range $3600<\lambda<10400$ \AA\ at higher resolution (R$\sim 2000$), thus allowing a better sky subtraction and a higher SNR. 
In both cases, the spectra were already fully reduced, wavelength and flux calibrated, as described in the respective papers. We apply only an aperture correction by matching the observed spectrum to the photometric SED \citep{laigle16} with the same methodology adopted for the Magellan spectra. However, we warn that there could be some mis-matches compared to our Magellan observations in the slit 
centering and orientation, as also in the seeing conditions, thus the spectra may not be exactly representative of the same regions.

\subsection{Line measurements}\label{line_measurements}

We measured emission line fluxes, line widths and uncertainties on fully calibrated and aperture corrected spectra using the python anaconda distribution (Mark Rivers, 2002\footnote{GitHub Repository: \href{stsci.tools/lib/stsci/tools/nmpfit.py}{stsci.tools/lib/stsci/tools/nmpfit.py}}) of the IDL routine MPFIT (Markwardt 2009). 
Given the FWHM resolution of FIRE for $1''$ slit width ($=100$ km/s), all our emission lines are resolved, owing to intrinsically higher velocity widths. 

We fit the lines with a single Gaussian on top of an underlying order-1 polynomial continuum. In each case, we require a statistical significance of the fit of $95\%$, as determined from the residual $\chi^2$. When a single Gaussian does not satisfy the above condition, we use a double Gaussian (with varying flux ratio and same FWHM, in km/s, for the two components), which instead provides a better fit, placing its $\chi^2$ within the asked confidence level. The flux uncertainties were derived by MPFIT itself, and they were always well behaved, with best-fit $\chi_\text{reduced}^2 \simeq 1$. For non detected lines (i.e., SNR $<2$ in our case), we adopt a $2\sigma$ upper limit \footnote{We remark that we are guided by the wavelength position, line width and flux ratio (for double line components) of H$\alpha$, which is always detected at $>4\sigma$. The Gaussian amplitude remains thus the only variable to constrain the fit for the other lines.}. However, we highlight that our detected emission lines have always high S/N ratios: H$\alpha$, [\ion{N}{II}]$\lambda$6584$\AA$\ and Pa$\beta$ are identified on average at 9.3, 8.4 and 7.4 $\sigma$, respectively (lowest SNRs are 4, 5.3 and 3.3 for the same lines). 

We fitted a double Gaussian for $12$ galaxies in our sample. 
As we will see later in Section \ref{pre-coalescence} by combining the informations of their 1D and 2D spectra, in $6$ of them we interpret the two Gaussians as coming from different merger components. For the remaining galaxies, in $2$ cases the lines are consistent with global rotation, while for the last $4$ we were not able to derive firm conclusions, even though we favour the contribution of multiple system parts to their emission. In Appendix \ref{emissionlinefits}, we show the 1D emission line fits for all our $25$ starbursts, and we discuss in more detail the origin of double Gaussians line components. 

We applied the stellar absorption correction on Balmer and Paschen emission lines, rescaling upwards their fluxes. In order to determine the level of absorption for these lines, we adopted \citet{bruzual03} synthetic spectra with solar metallicity and constant star-formation history for $200$-$300$ Myr, which are the typical merger-triggered starburst timescales \citep{dimatteo08}. 
The current starburst activity imposed by our selection suggests that the final coalescence of the major merger occurred relatively recently, certainly within the last 200 Myr. 
Numerical simulations of major mergers with different masses and dynamical times indicate indeed that star formation stops within 100-200 Myr after the coalescence, even without AGN quenching \citep{springel05a,bournaud11,powell13,moreno15}. 
Averaging the results over this interval, we applied an EW$_{abs}$ of $5$, $2.5$, $2.5$ and $2$ \AA\ for H$\beta$, H$\alpha$, Pa$\gamma$ and Pa$\beta$, respectively\footnote{The EW$_{abs}$ of H$\beta$ and H$\alpha$ are consistent with those adopted in previous works \citep[e.g.][]{valentino15}, while it was not possible to make comparisons for Pa$\gamma$ and Pa$\beta$.}. In the same order, we estimated for these lines an average absorption correction of $35\%$, $7\%$, $26\%$, $13\%$ of the total flux. If we allow an uncertainty of $\pm 1$ \AA\ on the EW$^{abs}$ correction of either Pa$\gamma$ and Pa$\beta$, this will produce variations on the final fluxes that are $6\%$ on median average, and thus it will not significantly affect our results. 

All the lines in the Magellan spectra, either in emission or in absorption, were analyzed based on the following steps. Firstly, H$\alpha$ and [\ion{N}{II}]$\lambda\lambda$ $6548$,$6583$, which are the lines with the highest S/N, were fitted together assuming a common linear continuum and a fixed ratio of the [\ion{N}{II}] doublet of $3.05$ \citep{storey88}. From this fit we derived the redshift of the galaxy, the intrinsic FWHM of H$\alpha$ (in terms of velocity), and the flux ratio of the two H$\alpha$ components for double gaussian fits. The instrinsic total line widths were obtained by subtracting in quadrature the FIRE resolution width ($100$ km/s) from the best-fit observed FWHM. For double Gaussians, the total FWHM was calculated adding the single FWHM and the separation between the two component peaks, as this quantity is more representative of the whole system. 

Then, the three parameters defined above were fixed and used to fit all other emission lines, including those in the optical zCOSMOS and SDSS spectra, after rescaling the FWHM to account for the different spectral resolutions. For the galaxies with the highest S/N of Pa$\beta$ ($>8\sigma$), we verified that even without fixing its FWHM a-priori, the fit yields a line velocity width consistent within the errors with the value found from H$\alpha$, indicating that our assumption is generally valid.
Given the \textit{rms} wavelength calibration accuracy (see Section \ref{spectroscopic_reduction}), we allowed the line central wavelength to vary in the fit by $3\sigma$, corresponding to $1.5$ \AA\ at $10000$ \AA, and $3$ \AA\ at $20000$ \AA.
For each measured flux, we also added in quadrature an error due to the uncertainty of the absolute flux calibration. This additional uncertainty ranges between $5\%$ and $10\%$, and is determined as the maximum residual difference between the average fluxes estimated from the photometry and from the aperture corrected spectra, among all the bands ranging from z++ to K$_S$. 
Finally, the equivalent widths of the lines were derived following its definition ($=\int (F(\lambda)-F_{cont}(\lambda))$/$F_{cont}(\lambda)$), where $F(\lambda)$ is the best-fit gaussian flux distribution and $F_{cont}$ is the fitted underlying continuum.  
Since the fluxes of H$\alpha$ and Pa$\beta$ were presented in Paper I, here we show in Table \ref{table2} the FWHM of the lines, the EWs of H$\alpha$, H$\delta$ and Pa$\beta$, the fluxes of [\ion{O}{III}]$5007$ and H$\beta$ that have been used in the BPT diagram.

\subsection{Ancillary data}\label{ancillary_data}

Almost all of our starbursts (24) were observed by HST-ACS in the F814W filter \citep{koekemoer07} at an angular resolution of $0.095''$ ($\sim0.7$ kpc at z$=0.7$). The UltraVISTA survey \citep{mccracken12} observed our galaxies in YJHK bands at a spatial FWHM resolution of $\sim0.75''$, comparable to the average seeing during our Magellan observations. Two galaxies in the subset have higher resolution ($0.19''$) F160W HST images from the DASH program \citep{momcheva16}. 
Finally, all our galaxies are well detected in radio $3$ GHz VLA images \citep{smolcic17} (owing a similar a spatial resolution of $\simeq 0.75''$), with an average S/N of $18$. 
In \citet{calabro18} we showed the HST F814W, H-band UltraVISTA and radio $3$ GHz VLA cutout images for only $10$ galaxies in our whole sample, for reason of space. Therefore, we include here in Fig.\ref{stamps} the same types of images for the remaining sample of $14$ starbursts, all of which have been observed during the second Magellan run (we remind that galaxy ID 578239 was not observed by HST, so we did not show it).

\subsection{Morphological classification}\label{morphological_classification}

Even though the light emission of dusty starbursts at optical/NIR wavelengths might be still severely affected by dust, the high resolutions offered by HST F814W images allows us to investigate the global structure of these systems. In Paper I we showed that the morphology of $18$ galaxies has been already classified by \citet{kartaltepe10} (K10), revealing a merger origin for the majority of them. Adopting the same criteria of K10, we have classified visually the remaining $6$ galaxies (one has no HST coverage), but the results do not change significantly: $61\%$ of our total sample are major mergers, as revealed by their highly disturbed morphology, tidal tales and bridges, $23\%$ are classified as minor mergers from the presence of only slightly perturbed structures (e.g., warped disks, asymmetric spiral arms, etc.) without large companions, $11\%$ are classified as spheroidal/S0 galaxies and the remaining $5\%$ as spirals. The major merger subset is additionally divided in five smaller classes according to their merger state (I: First approach, II: First contact, III: pre-merger, IV: Merger, V: Old-merger/merger remnant), following K10. 

However, we remind that the merger recognition and, even more, the merger stage classification from the optical morphology is very uncertain and more difficult at higher redshifts, due to lower resolution and to surface brightness dimming, which hampers the detection of faint tidal tales/interacting features, especially after the coalescence. The galaxy ID $245158$ represents a show-case example of this uncertainty: it has been classified as spiral/minor merger from its global morphology, but it clearly shows a double nucleus in the central region, further confirmed by a double component H$\alpha$ in the 2-D and 1-D spectrum, indicating rather an ongoing merger system.

\subsection{Radio size measurements}\label{radio_size_measurements}

\begin{figure}[t!]
    \centering
    \includegraphics[angle=0,width=\linewidth,trim={0.cm 0.cm 0.cm 0cm},clip]{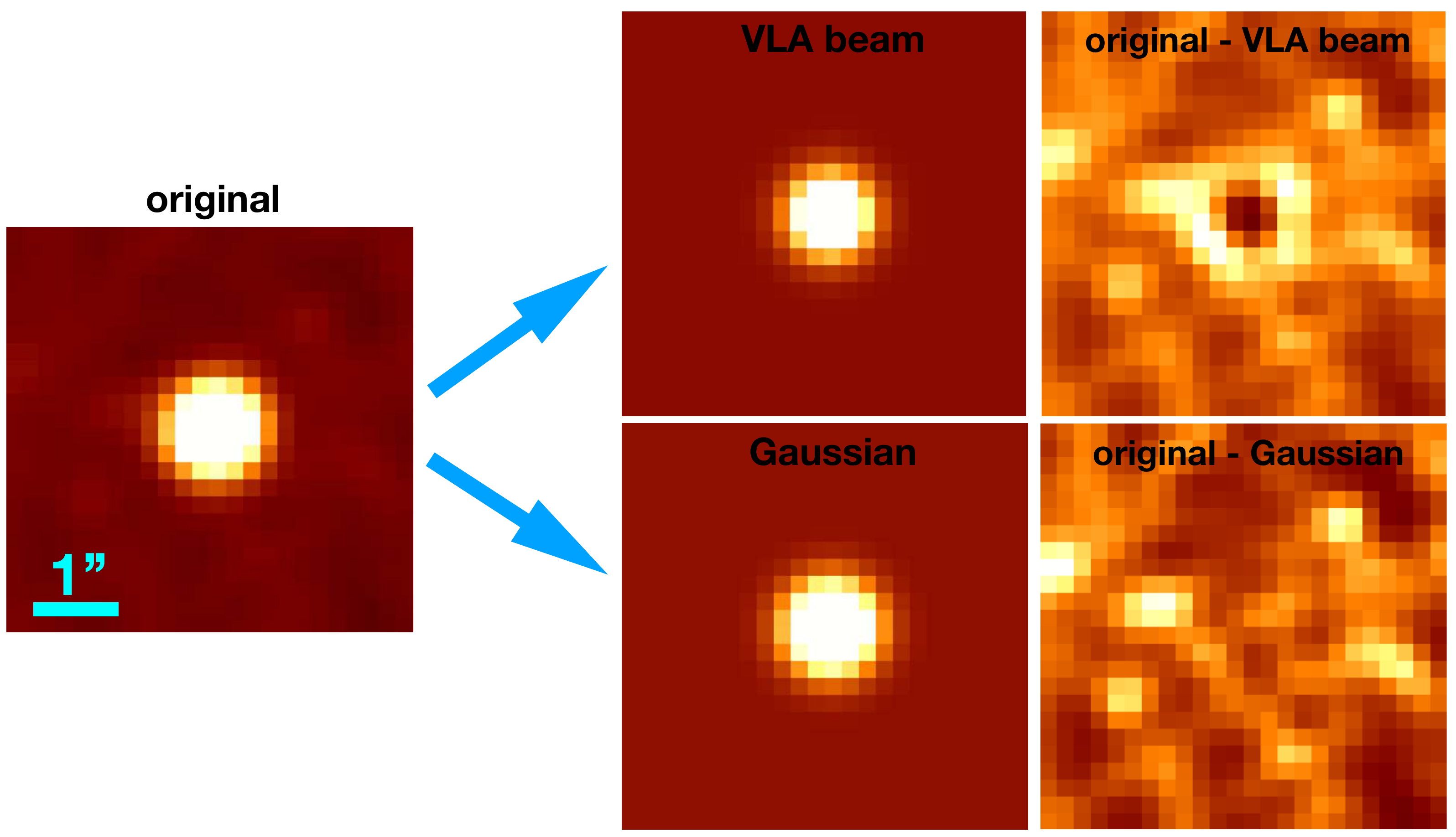}
    \caption{\small GALFIT fitting of the radio VLA image (3 GHz) of the galaxy ID 685067 ($z = 0.37$) with the VLA synthesised beam in the \textit{upper} row and with a Gaussian profile (convolved with the beam) in the \textit{bottom} row. In horizontal sequence are shown, from left to right, the original image, the fitted model and the residual (original-model). For this galaxy, we derive an angular size of $0.20 \pm 0.04''$ (the pixel scale is $0.2''$/pixel), which corresponds to a physical size of $1.06 \pm 0.19$ kpc. We notice that here GALFIT converges when fixing the position angle (PA) and axis ratio (q) parameters, to 0 and 1, respectively (Table \ref{table2}). This example illustrates the possibility to reliably measure the radio sizes of our objects even when they are smaller than the FWHM resolution of VLA ($0.75''$). In this case, the difference between the two models is recognized by looking at the residual images (i.e., original-model).}\label{galfit_residuals}
\end{figure}

The radio continuum emission has been used as a dust-free tracer of the SFR in galaxies in the absence of contamination from an AGN \citep[e.g.,][]{condon92}. Since all our galaxies do not show either radio jets or radio flux excess compared to that expected from the SFR only (as we will show in Section \ref{AGN_identification}), we used $3$ GHz VLA images for measuring the FWHM size of their starburst cores, where the bulk of star-formation is taking place. For each SB, we used GALFIT \citep{peng10} to fit a 2D function, convolved with the VLA synthesised beam, to their radio emission. We tested several 2D profiles, which include a Gaussian, a S\'ersic function and the VLA beam itself, requiring a significance of the fit (probed by the $\chi^2$) of at least $95\%$ confidence level, as for the emission line measurements. In addition to the $\chi^2$ analysis, all the GALFIT residuals of the fit (original-model) were checked by eye inspection, and the excluded fittings have always worst residuals. 
The best-fit profiles obtained for our sample are summarized as follows:
\begin{itemize}
    \item A single 2D Gaussian with varying FWHM, axis ratio and position angle provides the best fit for $13$ starbursts. In one case (ID 470239), the required conditions are obtained only by fitting a single S\'ersic profile with varying parameters, but its half light diameter (calculated as $2 \times r_e$, with $r_e$ the effective radius) is only $3\%$ different from the total FWHM of a single gaussian fit, thus we assume the latter as the final value. \footnote{We also tried to fit a single and double S\'ersic profile for all the other sources. However, given the larger number of parameters of this profile and the limited VLA resolution, we do not obtain convergence for the majority of them, or the resulting $\chi_\text{reduced}^2$ are too high.}
    \item A double 2D Gaussian is required by $3$ galaxies (ID 245158, 412250 and 519651), allowing to resolve them and measure single components FWHM and their relative separation.
    \item Fitting the VLA synthesised beam yields the best solution for $6$ galaxies, which are then unresolved with current resolution ($0.75''$)
    \item A single 2D Gaussian with fixed axis ratio and position angle (to $1$ and $0$, respectively) is used for $2$ sources (ID 578239 and 685067). We remark that, in case the $95\%$ significance level of the fit is satisfied with either this or the previous approach (as in the case of some very compact sources), we adopt the Gaussian solution only if the associated $\chi^2$ probability level is at least double compared to the fit with the VLA beam.
\end{itemize}

As shown later, for a few galaxies we have measured angular sizes that are much smaller than the synthesised beam FWHM ($\sim 0.75''$), down to $\sim0.2''$ and to a physical scale of $1$ kpc. To demonstrate that it is possible to reliably determine the sizes even for these extreme, compact sources, we show in Fig. \ref{galfit_residuals} a comparison between the GALFIT residuals obtained when fitting the image with a Gaussian (convolved to the VLA beam) (upper row) and with the radio synthesised beam itself (bottom row). It is evident that a Gaussian provides a better fit of the original source profile and a cleaner residual compared to the VLA beam alone. 

The uncertainties on the sizes were recalculated for all the starbursts from their radio SNRs, using the fact that better detected radio sources also have the smallest radio size uncertainties, as shown, e.g., in \citet{coogan18}. We used then the same formulation as:   
\begin{equation}
    FWHM_{err}\simeq 1 \times \frac{FWHM_{beam}}{SNR},
\end{equation}\label{errorsize} 
where FWHM$_{beam}$ is the circularized FWHM of the VLA synthetized beam, and the multiplying coefficient was determined from simulations, following \citet{coogan18}. All the size measurements with relative uncertainties and the method used for their determination are included in Table \ref{table2}. 

Since our galaxies are well detected in radio band (average SNR of $18$) and the VLA synthetized beam is well known, we always obtain a good fit for the resolved sources. Among them, we were able to fit a double Gaussian for $3$ objects. In these cases, their total FWHM (adopted throughout the paper) were determined as the sum of the average single FWHM sizes and the separation between the two components. However, we will also consider the single sizes in some cases, e.g., in Fig. \ref{Mass_Size}. This finding suggests that also some other galaxies may represent double nuclei that are blended in $0.75''$ resolution VLA images. 
For the unresolved galaxies instead (i.e. those fitted with the VLA beam, as explained above), we adopted a $3\sigma$ upper limit on their FWHM. Within the most compact starbursts, some of them may be affected by pointlike emission from an AGN, which decreases artificially the observed size. However, we tend to discard this possibility since, as we will see later, none of our AGN candidates show a radio-excess compared to the radio emission due to their SFR.  

\subsection{AGN identification}\label{AGN_identification}
 
We started searching for AGN components in the mid-IR. Through the multi-component SED fitting of IR+(sub)mm photometry (described in Section \ref{sample_selection}) we detected at $>3\sigma$ the dusty torus emission component for a subset of $12$ SBs. The significance of the detection was derived from the ratio between the total best-fit dusty torus luminosity ($=L_\text{AGN,IR}$) and its $1\sigma$ uncertainty, inferred as the luminosity range (symmetrized) yielding a variation of the $\chi_{red}^2$ $\leq1$ with respect to the minimum value of best-fit \citep{avni76}. More details about the torus estimation method are described in \citet{liu18} and \citet{jin18}. Among the $12$ mid-IR AGNs, we detected the dusty torus emission at high significance level ($> 5\sigma$) for $6$ starbursts (ID 777034, 519651, 222723, 232171, 466112, 894779), while for the remaining objects we obtained a lower significance ranging $3\sigma<$ $L_{AGN,IR}$ $<5\sigma$ (see Table \ref{table2}). The SED fitting of all the galaxies can be found in the Appendix \ref{additional_plots}.  

Within the sample of IR-detected AGNs, $6$ galaxies (ID 777034, 222723, 232171, 635862, 578239, 911723) are also detected in X-rays at more than $3\sigma$ by XMM-Newton, Chandra or NuStar \citep{cappelluti09,marchesi16,civano15}. Throughout the paper, we will consider the X-ray luminosities L$_X$ measured by \citet{lanzuisi17}, integrated over the energy range $2$-$10$ keV. 
To estimate the contribution of star-formation to the total intrinsic L$_\text{X}$, we used the relation between SFR and L$_\text{X,SFR}$ of \citet{mineo14}, rescaled to a Chabrier IMF and applying a correction factor of $0.6761$ to convert the X-ray luminosity from the $0.5$-$8$ keV to the $2$-$10$ keV band.

We remind that the column densities N$_H$ inferred from their hardness ratios \citep{lanzuisi17} are consistent with those derived from the dust attenuations (toward the centers) assuming a mixed model (Paper I), suggesting that also the X-ray emission is coming from the nucleus, where the AGN is expected to be located. 
Furthermore, all our starbursts do not show radio jets in VLA images, and do not have significant radio excess than expected from their SFR, assuming a typical IR-radio correlation with q$_\text{IR} = 2.4$, as in \citet{ibar08}, \citet{ivison10} and \citet{liu18}. 


\section{Results}\label{results}

\begin{figure*}[h!]
    \centering
    \includegraphics[angle=0,width=0.48\linewidth,trim={0.cm 3cm 0.cm 0cm},clip]{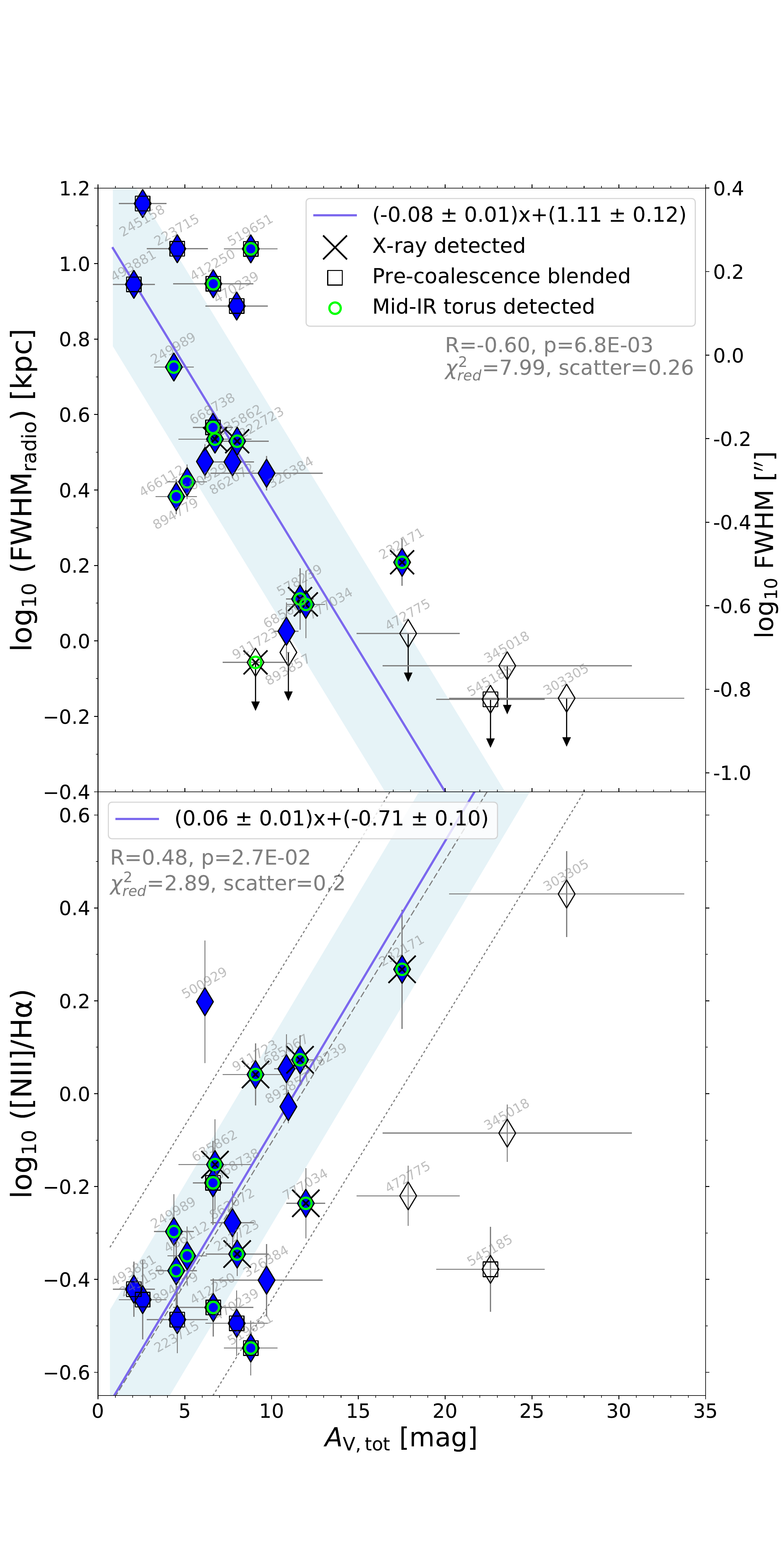}
    \includegraphics[angle=0,width=0.48\linewidth,trim={0.cm 3cm 0.cm 0cm},clip]{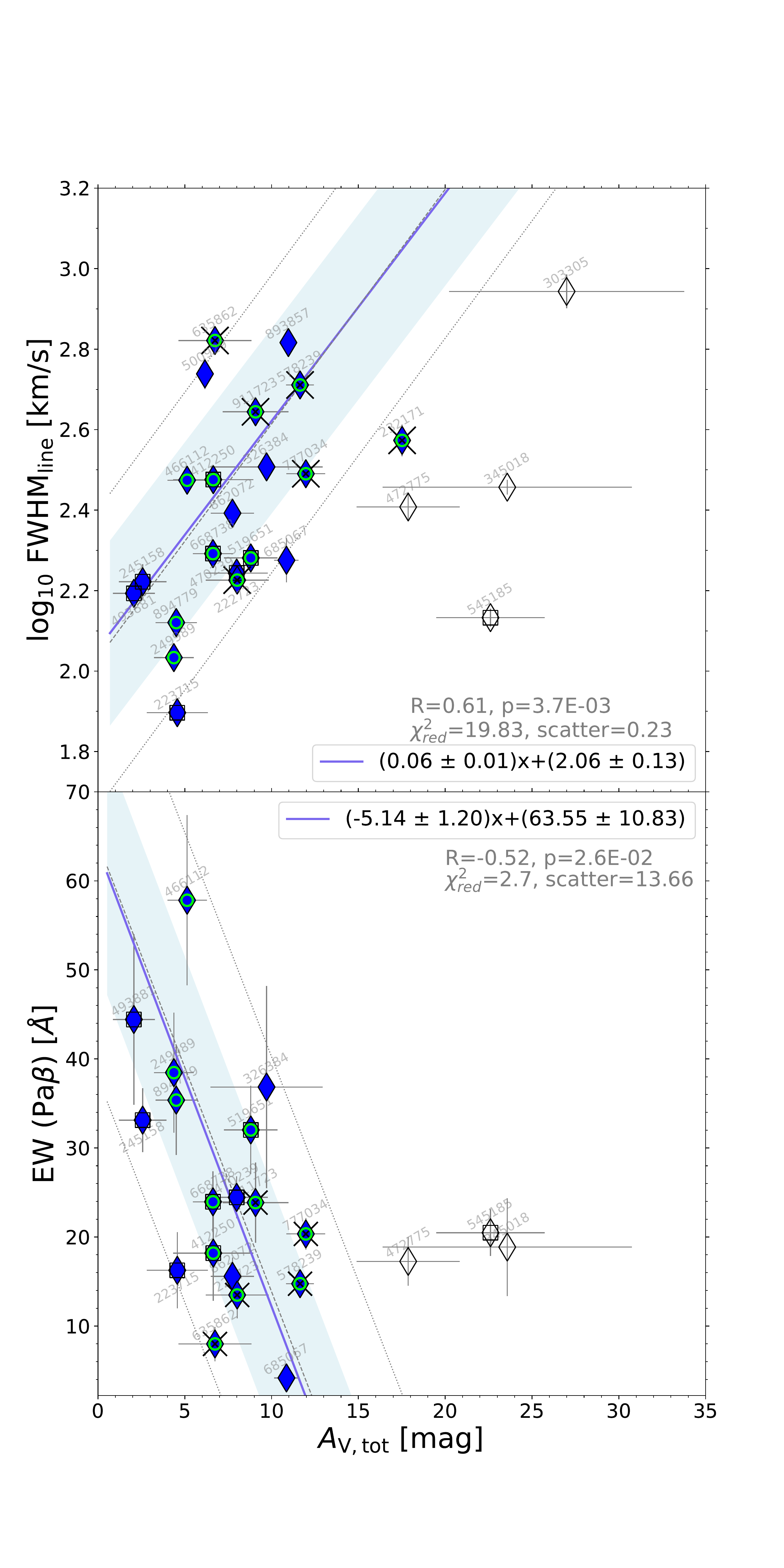}
    \caption{\small \textit{Left:} Correlations of A$_\text{V,tot}$ with the radio size (top), and with the N2 index (bottom). We show with filled blue diamonds all the Magellan SBs that were used to derive the best-fit linear relation (blue continuous line) and the $\pm$1$\sigma$ dispersion (blue shaded area), while empty diamonds represent galaxies excluded from those calculations. The latter comprise the $4$ outliers discussed in the text (ID 303305, 345018, 472775, 545185) and all the upper limits in the A$_\text{V,tot}$ vs FWHM$_\text{radio}$ plot.
    In the corners, we show in black the equation of the linear fit (which includes $1\sigma$ error of the two best-fit parameters), and in gray the Spearman correlation coefficient (R) with the corresponding p-value (p), the reduced chi-square of the fit ($\chi_\text{red}^2$), and the $1\sigma$ scatter of our SBs around the best-fit line, all of which do not take into account upper limits and the $4$ outliers mentioned above.
    For comparison, the linear fit and $1\sigma$ dispersion including the $4$ outlier galaxies are highlighted with a gray continuous dashed line and two dotted lines of the same color. \textit{Right:} Correlations of A$_\text{V,tot}$ with the line velocity width (top), and with the EW(Pa$\beta$) (bottom). In the last diagram, $4$ galaxies without EW(Pa$\beta$) measurements are not considered.}\label{correlations}
\end{figure*}

The spectra that we have obtained at the Magellan telescope, along with longer wavelength radio images, provide us key information 
to understand both the attenuation sequence and the variety of morphological classes of our starbursts. First of all, since dissipative mergers are able to funnel the gas from the large scales of Milky-way-like disks ($\sim10$ kpc) to sizes that are more than one order of magnitude smaller \citep{dimatteo05}, it is useful to analyze the characteristic star-forming sizes of our starbursts. 
Besides this, from the galaxy integrated Magellan spectra we can study together the excitation and kinematic state of the gas, and the aging of the stellar population in the outer starburst cores, traced respectively by the [\ion{N}{II}]/H$\alpha$ ratios, the intrinsic (resolution corrected) line velocity widths of single 1-D Gaussian components ($=$FWHM$_{\text{line}}$, which is a proxy for the velocity dispersion in the system) and the Balmer/Paschen line equivalent widths ($=$EW$_{H\alpha,Pa\beta}$). 

In Fig. \ref{correlations} we present the main result of this analysis, showing that the FWHM$_{\text{radio}}$, the N2 parameter, the FWHM$_{\text{line}}$ and the EW$_{Pa\beta}$ are all correlated to the total dust attenuation A$_\text{V,tot}$, which is used here as the reference quantity for comparison. 
This suggests that our starbursts can be described as, at first order, a one-parameter sequence: similar correlations at different significance levels are indeed found also when comparing on a single basis each pair of the above parameters.

We tested these correlations using three different approaches with all the available data, excluding from the calculations only the upper limits and missing EW(Pa$\beta$) measurements.
Firstly, we calculated the Spearman rank correlation coefficient R (the higher R, the stronger the correlation) and the corresponding p-value, which represents the probability of obtaining an equal (or stronger) R if no correlation exists. We defined a threshold probability of $5\%$ to accept the correlation. Overall, we found that the p-values are nearly always lower than $0.05$, meaning that the correlations are significant according to our criteria. In only one case (EW$_{Pa\beta}$ vs FWHM$_{\text{radio}}$) we determined a slightly higher p-value of $0.1$ (thus a higher probability of no correlation), which could be partly affected by the lowest number of data (i.e. lowest statistics) available here compared to the other diagrams. However, the other methods indicate instead a stronger physical connection between the two quantities.

In the second approach, we fitted the galaxies in each diagram with a linear relation (in log-log space, except for the last diagram where the y-axis is in linear scale), by using an Orthogonal Distance Regression procedure (ODR), which allows to take into account measurement uncertainties in both axis (we discuss later possible outliers or different fitting functions). We determined the SNR of the angular coefficient (i.e., how much it differs from 0), finding significant correlations at more than $3\sigma$ in 8 cases, while they are less strong ($2<$ SNR $<3$) for the remaining two diagrams. In the four correlations shown in Fig.\ref{correlations}, we obtained a significance of $5.8$, $5$, $4.3$ and $3.65\sigma$ for A$_\text{V,tot}$ vs N2, FWHM$_{\text{line}}$ and EW$_{Pa\beta}$, respectively. With this method, we also determined the 1$\sigma$ dispersion of our data with respect to the best-fit linear relation. 

\begin{table*}[!htb]
    \centering
    \begin{tabular}{||l||l||l||l||l||}
    \hlineB{2}
    & \bfseries FWHM$_{\text{radio}}$ & \bfseries N2 & \bfseries FWHM$_{\text{line}}$ & \bfseries EW$_{\text{Pa}\bm{\beta}}$ \\
    \hlineB{2}
    \bfseries A$_{\text{V,tot}}$ & -0.6 (0.007) & 0.48 (0.027) & 0.61 (0.0037)  & -0.52 (0.026) \\
    \rowcolor{SeaGreen3!30!} & 5.8$\sigma$  & 6$\sigma$  & 4.3$\sigma$ & 4.3$\sigma$ \\
    \rowcolor{Tan3!30!} & 0.028\% & $<0.001\%$  & $2.9\%$ & $0.12\%$ \\
    \hlineB{1.5}
    \bfseries FWHM$_{\text{radio}}$  &  & -0.71 (0.0006)  & -0.46 (0.049) & 0.45 (0.1) \\
    \rowcolor{SeaGreen3!30!} &  & 5.74$\sigma$  & 2.93$\sigma$ & 4.9$\sigma$ \\
    \rowcolor{Tan3!30!} &  & $<0.001\%$  & $0.033\%$  & $3\%$ \\
    \hlineB{1.5}
    \bfseries N2 &  &  & 0.67 (0.0003) & -0.43 (0.05)  \\
    \rowcolor{SeaGreen3!30!}  &  &  &  5.45$\sigma$ & 3.85$\sigma$ \\
    \rowcolor{Tan3!30!}  &  &  &  $<0.001\%$ & $0.15\%$ \\
    \hlineB{1.5}
    \bfseries FWHM$_{\text{line}}$  &  &  &  & -0.46 (0.05) \\
    \rowcolor{SeaGreen3!30!}  &  &  &  &  2.5$\sigma$  \\
    \rowcolor{Tan3!30!}  &  &  &  &  $12.9\%$  \\    
    \hlineB{2}
    \end{tabular}
\caption{\small Correlation coefficients among the total attenuation towards the center in a mixed model (A$_\text{V,tot}$), the 3GHz radio FWHM size (FWHM$_{\text{radio}}$), the line velocity width (FWHM$_{\text{line}}$) and the equivalent width of Pa$\beta$ (which tightly correlates also with the EW of H$\alpha$, H$\beta$ and H$\delta$). In each case we show in three colored lines: \textbf{(white)} the Spearman correlation coefficient and corresponding p-value; \textbf{(green)} the significance of the correlation derived from the ratio of the linear best-fit angular coefficient and its uncertainty; \textbf{(orange)} the probability of having a significance lower than 2$\sigma$ if a random $20\%$ of the sample is removed. For the calculations we excluded the upper limits, missing EW(Pa$\beta$) measurements, and the $4$ outlier starbursts (ID 303305, 345018, 472775, 545185) in the three diagrams relating A$_\text{V,tot}$ to N2, FWHM$_{\text{line}}$ and EW(Pa$\beta$).}\label{table1}
\end{table*}

Finally, we also performed Monte Carlo simulations: for each relation, we run 100k simulations, removing each time at random $20\%$ of the points, recalculating the significance of the correlation using our second approach. We then estimated the rate ($\sim$ probability) at which such correlations completely disappear with a significance falling below $2\sigma$. This analysis allows to test the systematics and scatter of the correlations, ensuring they are robust and not driven by a few outliers. Overall, we find low probabilities (less than $5\%$) to obtain a less than $2\sigma$ significance when removing a random $20\%$ of the galaxies, indicating that our correlations do not cancel out and are not found by chance. In the four diagrams of Fig.\ref{correlations}, we obtained probabilities of $0.028\%$, $0.001\%$, $4.7\%$ and $0.7\%$, in the same order as above. 

We remind that A$_\text{V,tot}$ are determined from the Pa$\beta$ observed fluxes and the bolometric L$_{IR}$ (assuming a mixed model geometry) \footnote{As explained in Paper I, for $4$ galaxies in our sample where Pa$\beta$ resides in opaque atmospheric regions or out of the FIRE coverage, we estimated the attenuation (ID 245158) or its upper limit (ID 303305, 500929, 893857) through the Pa$\gamma$ line, adopting a flux ratio Pa$\beta$/Pa$\gamma=2.2$. This is the average expected observed ratio for all the attenuation values in our range, assuming either a mixed model or a foreground dust-screen geometry, and it is verified by $9$ starbursts with simultaneous detection of Pa$\gamma$ and Pa$\beta$.}. However, for $4$ galaxies in the sample (ID 303305, 500929, 893857 and 232171) we do not detect either Pa$\beta$ or Pa$\gamma$, thus in these cases we derived A$_\text{V,tot}$ in a similar way from their H$\alpha$ fluxes (so to avoid upper limits), adding a representative error of $0.1$ dex determined from the remaining sample as the scatter of the correlation between Pa$\beta$ and H$\alpha$ based A$_\text{V,tot}$.
We also verified that including the upper limits in the calculations does not significantly alter the fitted trends.  
Hereafter, we discuss in detail on a single basis the most important findings. 

In the first (top-left) panel of Fig. \ref{correlations}, the FWHM radio sizes, while spanning a wide range from less than $600$ pc to $\sim$12 kpc, are tightly anti-correlated to the dust obscuration level A$_\text{V,tot}$ (R=-0.6, p-value=0.007, and a scatter of $0.26$ dex). Towards the smaller sizes and higher obscurations (A$_\text{V,tot}>20$ mag), three galaxies are unresolved with VLA, thus they may be actually closer to the best-fit linear relation derived from the remaining sample. In this diagram, X-ray detected AGNs are found both at small and large radii, and have a similar distribution compared to the other galaxies, suggesting that radio size measurements and hence the result in Fig. \ref{correlations} are not contaminated by AGNs. 

In the last three panels of Fig. \ref{correlations}, the [\ion{N}{II}]/H$\alpha$ ratio, the line velocity width (FWHM$_\text{line}$) and the EW of Pa$\beta$\footnote{We use this line for comparison since, being at longer wavelength, it is more representative of the whole system, allowing to probe a larger fraction of starburst cores if a mixed geometry holds. However, in the Appendix \ref{additional_plots} we show that EW(Pa$\beta$) is tightly correlated to the EW of H$\alpha$, H$\beta$ and H$\delta$, all of them being strongly sensitive to the age of the stellar population (at fixed SFH), thus similar results are obtained also if choosing a different line for the EW.} are also correlated to the total attenuation at more than 3$\sigma$ significance level (R coefficients and p-values are $0.51$(0.009), $0.48$(0.015) and $-0.46$(0.034),  respectively). 
However, we notice that $4$ galaxies (ID$=$ 303305, 472775, 345018 and 545185) are outside the 1$\sigma$ dispersion of the best-fit relations in all the three diagrams (gray dashed and dotted thin lines). They show lower N2, FWHM$_\text{line}$, and higher EW than expected from their dust obscuration level. Alternatively, they have a larger A$_\text{V,tot}$ for their N2, FWHM$_\text{line}$ and EW values. 

In order to understand the nature of these galaxies, we simulated $100$k different realizations of the last three diagrams of Fig.\ref{correlations}, with N2, FWHM$_\text{line}$, EW(Pa$\beta$) and A$_\text{V,tot}$ of $25$ galaxies distributed according to the best-fit relations and the corresponding $1\sigma$ dispersions.
Then we computed the probability of having at least $4$ galaxies ($3$ for the last plot) with an orthogonal distance from the best-fit relation (gray dashed line) equal or greater than the $4$ (or $3$) outliers described above. 
We found, in the same order presented above, a probability of $0.2\%$, $0.025\%$ and $0.005\%$, indicating that those $4$ galaxies are real outliers and cannot be simply explained by the $1\sigma$ scatter of the best-fit lines. 

Given their deviant behavior, we excluded these outliers and derived again the best-fit relations, which are shown in Fig.\ref{correlations} with a blue continuous line. We found on average a reduction of the $1\sigma$ dispersion (shown with a light blue shaded area) by $\sim0.1$ dex and a slight improvement of the correlation significance compared to the previous calculations. However, the best-fit linear equations are not significantly different, thus we give only the analytic expressions of this second fit where the outliers are not considered. The new results for the three diagrams, and all the diagnostics for the remaining $7$ correlations are presented in Table \ref{table1}. We notice that the $4$ divergent starbursts have an upper limit on their radio size, and are not outliers in other diagrams that do not involve A$_\text{V,tot}$, thus the latter are not affected by this analysis. 
A possible physical explanation of the diverging behavior of these $4$ galaxies will be discussed in Section \ref{outliers}. 

Finally, if we look at all the correlations in Table \ref{table1}, we can notice the presence of a subset of quantities that correlate better than others. Apart from the previously discussed A$_\text{V,tot}$ vs. FWHM$_{\text{size}}$, the line width, N2 and radio size are tightly and robustly correlated with each other. Indeed, from bootstrapping analysis,
the probability that there is no correlation is less than $0.033\%$. As we will see in Section \ref{velocity_enhancement}, this result hides a deeper physical link among them. 

\section{Discussion}

The results presented in the previous section show that the wide range of attenuations measured in Paper I translate into a wide range of other physical properties, i.e., radio sizes, N2, velocity width, Balmer/Paschen EW, and even more, all these quantities appear to be connected to each other, defining a one-parameter sequence. In this Section we propose a physical interpretation of this sequence, and show that the correlating properties considered before are consistent with being primarily reflecting different evolutionary merger stages. Then we discuss the role played by each parameter into this sequence.

\subsection{Identification of early-phase, pre-coalescence mergers}\label{pre-coalescence}

\begin{figure}[t!]
    \centering
    \includegraphics[angle=0,width=\linewidth,trim={0.cm 7.4cm 1.8cm 1cm},clip]{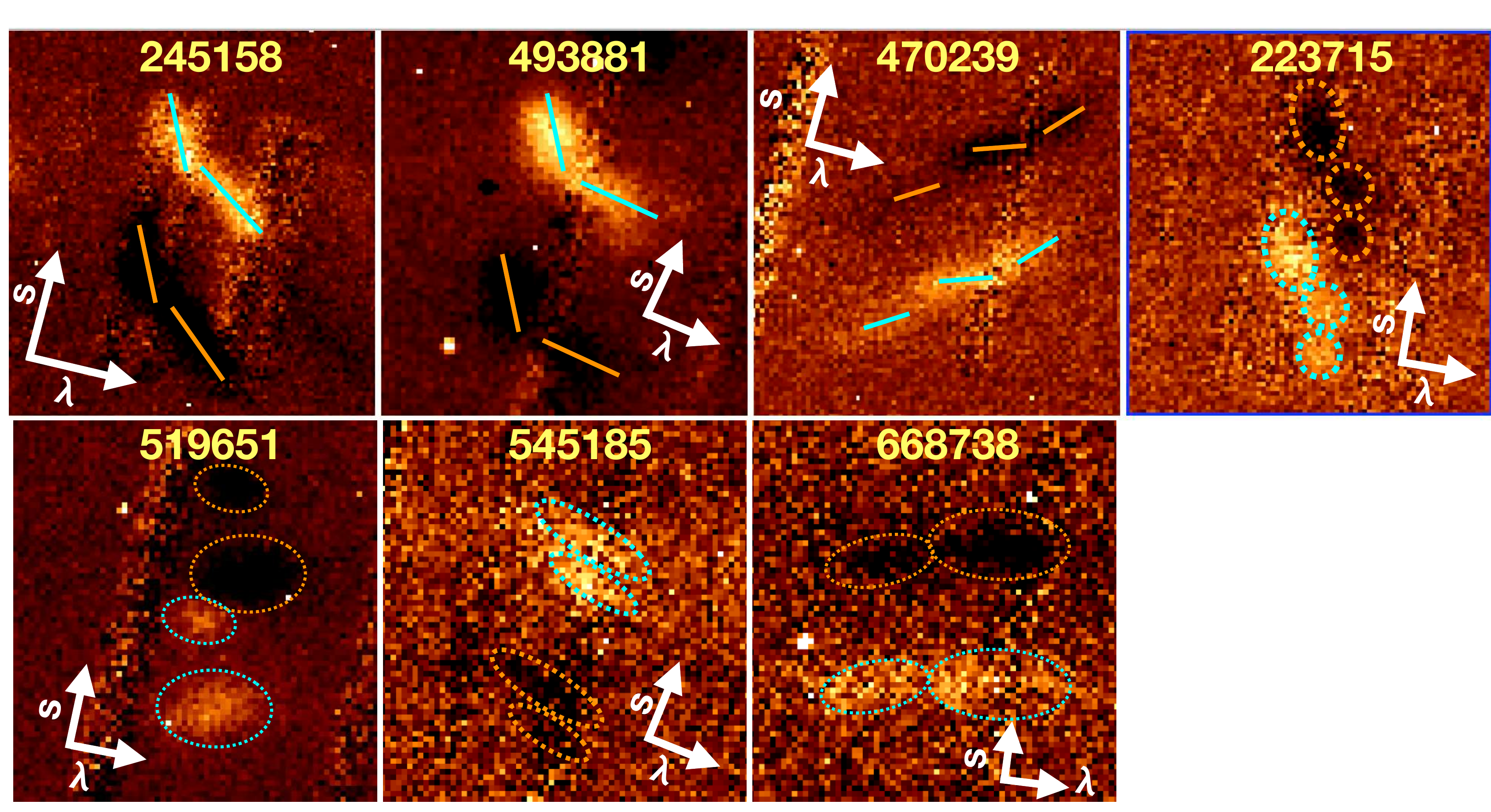}
    \caption{\small Close-up view of the H$\alpha$ emission line profiles for the galaxies satisfying one of the two pre-coalescence criteria defined in Section \ref{pre-coalescence}. 
    In each panel, as a consequence of the sky-subtraction procedure applied to the 2D spectra, the lines appear twice in different slit positions, in the first with a positive flux (in yellow) and in the second with a negative flux (in black). For each cutout, we show with two white arrows the slit position (s) and the dispersion direction ($\lambda$), which are slightly rotated due to the curved Magellan spectra.
    In order to clarify the classification criteria adopted in this work for finding pre-coalescence SBs, we highlight: (1) with continuous lines the different tilting angles of H$\alpha$ line profiles (first $3$ SBs); (2) with dotted ellipses the spatially separated H$\alpha$ lines coming from different merger components (last $4$ SBs). 
    }\label{2Dspectra}
\end{figure}

A first guess for a physical understanding of what is guiding the large spread of properties comes from the morphology. Indeed we have already seen that our sample comprises mergers at different stages of evolution (MI to MV), though this classification is very uncertain and sometimes misleading, as shown in Section \ref{morphological_classification}: the faintness of tidal tails and residual interacting features make systems at the coalescence difficult to recognize, while multiple optical components and double nuclei in HST images may just reflect the dust attenuation pattern rather than the true distribution of SFR and M$_\ast$.  
As shown in Section \ref{radio_size_measurements}, a double gaussian component fit on radio images allowed to resolve three sources, suggesting that they may be composed of two interacting nuclei. However, the limited resolution of VLA ($0.75''$ of FWHM) does not allow us to derive solid conclusions on the remaining sample, which might contain more close pairs. New maps and ALMA follow-ups would increase the resolution and hopefully resolve these blended pre-coalescence systems. 

A complementary way to find close interacting pairs in early merger stages comes from the analysis of their 2D spectra. 
With that aim, we performed a crude sky-subtraction procedure: we subtracted the 2D spectra taken for the same object but at different positions along the slit (A and B, separated by $2.2''$), in order to remove the sky lines and allow a visual inspection of the emission line profiles. 
For construction, the lines appear twice in each sky-subtracted frame and exactly with the same shape: one time with a positive flux (when the object is in position A), and the other with a negative flux (when the galaxy is in position B). 
By looking at these line profiles, we identified interacting pairs by requiring one of the following conditions:
\begin{enumerate}
    \item detached H$\alpha$ line components along the spatial direction, coming from separated merger components located at different slit positions (e.g., ID 223715, 519651, 545185, 668738 in Fig. \ref{2Dspectra}).  
    \item tilted H$\alpha$ line with two different inclination angles (based on visual inspection), indicating the presence of two emitting regions with independent kinematic properties, inconsistent with a single rotating disk (e.g., ID 245158, 493881, 470239 in Fig. \ref{2Dspectra}).
\end{enumerate}

In our sample, we identified from the two above conditions $7$ close-pair pre-coalescence starbursts, which are shown in Fig.\ref{2Dspectra}. For an additional source with a double radio emission component (ID $412250$), one of the two nuclei was not falling inside the slit, thus it was not observable with FIRE. However, this SB should be considered a merging pair at the same level of the others. 

As we can see in Fig. \ref{correlations}, the selected pre-coalescence starbursts are preferentially found at larger half-light radii, and all the systems with FWHM$_\text{radio}>6$ kpc belong to this category. This result has two main implications. 
Firstly, the sizes measured in radio are not necessarily those of single merger components, but they should be interpreted primarily as separation between the two pre-coalescence starburst units 
(e.g., for all the three systems resolved in radio (ID 245158, 412250, 519651), their separation is larger than the size of single nuclei). Secondly, the early evolutionary phases are also characterized by lower dust obscurations, suggesting that the merger induced gas compaction (i.e., the increase of hydrogen column density in the center) has not yet completed.

This pre-coalescence subset identification provides an immediate physical interpretation for $6$ galaxies of those that were simultaneously fitted with a double Gaussian in the 1D spectrum (Section \ref{line_measurements}), explaining this profile as coming from different merger components.
However, we warn that these diagnostics are not identical and the connection between the line profiles in the 1D and in the 2D spectrum is not straightforward. Starbursts with multiple spatial emission lines do not necessarily display double Gaussians in the 1D spectrum, because this is subject to projection effects and depends on the distribution in wavelength of each spatial component. Indeed, the lines of one of the galaxies shown in Fig. \ref{2Dspectra} (ID 519651) were still fitted with a single Gaussian in the 1D.

Furthermore, our subset of $6$ pre-coalescence starbursts identified from the 2D spectra is not necessarily complete, as many galaxies (e.g., ID 635862, 777034, 472775, 685067) have sky-subtracted 2D spectra with low S/N, not allowing to apply the visual criteria 1) and 2) presented above in this Section. We would have required longer integration times or spatially resolved observations to build a complete sample of starbursts before the coalescence. Similarly, if the two merger nuclei are too close, it would be impossible to detect them even in the 2D spectra, and would need a significant improvement of spatial resolution to identify the pair.

\subsection{Velocity enhancement and shocks toward the coalescence}\label{velocity_enhancement}

\subsubsection{BPT diagram and shocks}

\begin{figure}[h!]
    \centering
    \includegraphics[angle=0,width=8.4cm,trim={0.1cm 3.3cm 2.6cm 5.5cm},clip]{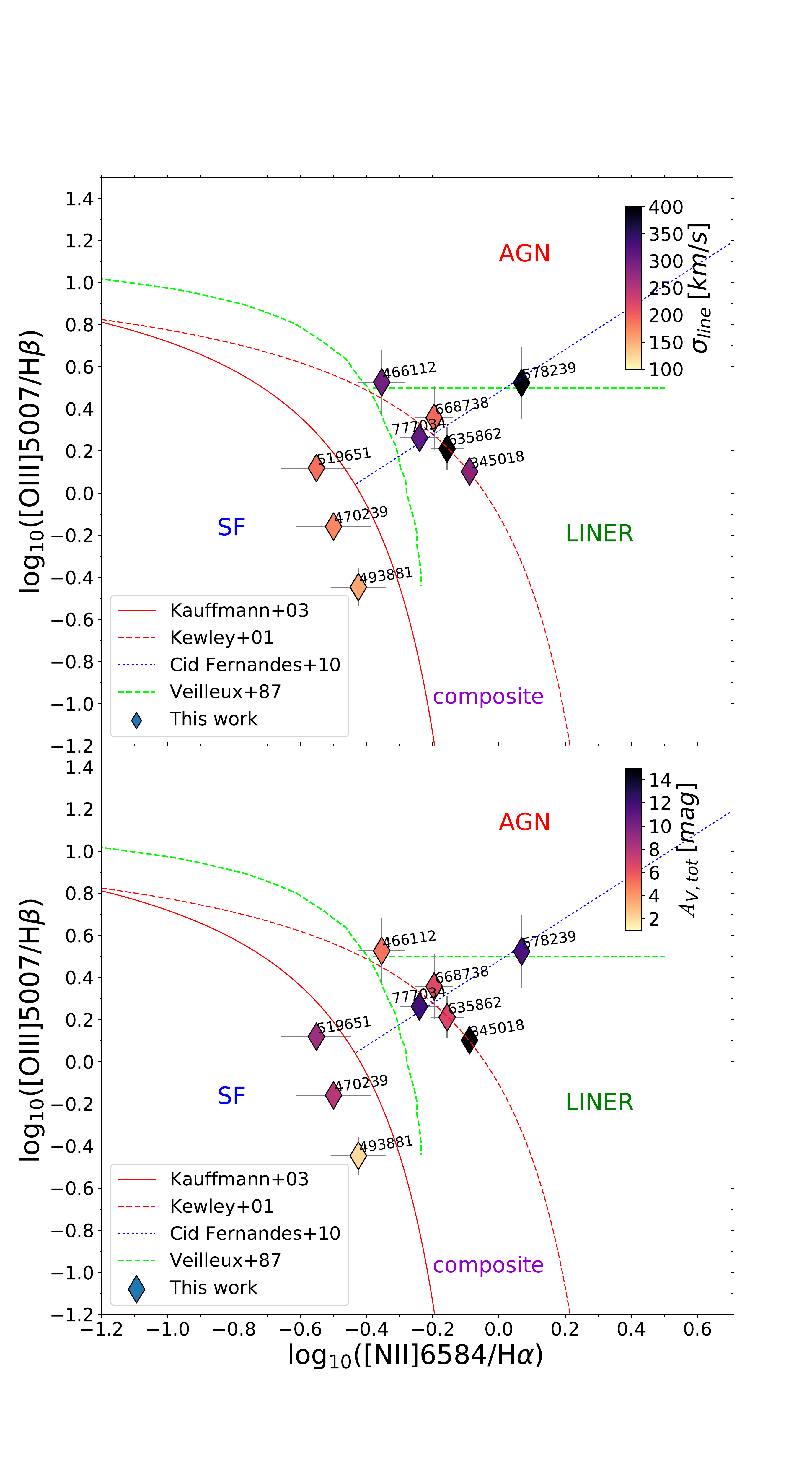}
    \caption{\small \textit{Top:} BPT diagram of $9$ starbursts in our sample with optical spectra available (for one galaxy included in zCOSMOS, we did not detect both [\ion{O}{III}]5007 and H$\beta$). While $3$ sources lie in the SF excitation region, the remaining galaxies are not consistent with SF, and their spectra show a mixture of composite, AGN and LINER properties. The color coding indicates that galaxies with higher N2 which are closer to the AGN/LINER regions also have increasingly higher line velocity widths ($\sigma_{\text{line}}$). \textit{Bottom:} Same diagram as above, but here the galaxies are color coded according to their total dust attenuation A$_\text{V,tot}$. More obscured starbursts preferentially display AGN/LINER properties.}\label{BPT1}
\end{figure}

The second (bottom-left) panel of Fig.\ref{correlations} shows that more obscured starbursts tend to have higher N2 relative to H$\alpha$, reaching [\ion{N}{II}]/H$\alpha$ ratios higher than 1, which are more typical of AGN and LINERs. Indeed, the classical BPT diagnostic diagram in Figure \ref{BPT1} \footnote{Two variants of the BPT using the [SII]6717+6731/H$\alpha$ or the [OIII]$\lambda$5007/[OII]$\lambda$3727+3729 ratios (S2BPT or O2BPT, respectively) are shown in Fig. \ref{BPTdiagram} in the Appendix. We remind that, due to an enhanced ionization parameter and lower metallicity (at fixed mass) in the ISM at higher redshifts, the average star-forming galaxies population at z$=0.7$ occupies a region in the BPT diagram which is shifted rightwards by $\lesssim +0.1$ dex compared to z$=0.1$ \citep{faisst18,masters16}. However, there are currently no studies addressing how this will affect the separation lines among SB, AGN and LINERs. If we suppose that at z$\sim0.7$ the same shift applies also to these lines, galaxies at intermediate obscurations and line widths would still fall in the composite region with dominant LINER/AGN-like properties. Also, this would not affect our subsequent conclusions based on the comparison with shock models.}, 
performed on $9$ galaxies with [\ion{O}{III}] and H$\beta$ available measurements, confirms that SBs with higher obscuration and line velocity width are found in the composite, AGN or LINER classification regions, according to empirical separation lines derived in the local Universe \citep{kauffmann03,kewley01,cidfernandes10,veilleux87}. 

Notably, the location of this subset of galaxies (which are shifted to the right compared to the purely SF region) is consistent with the predictions of shock models, with varying shock contribution and velocity (compare with Fig. 10 and Fig. 2 of \citet{rich11,rich14}, respectively). Additionally, \citet{lutz99} argue that LINER-like spectra in infrared selected galaxies are due to shocks, possibly related to galactic superwinds. 
The presence of increasing widespread shocks provides the most immediate interpretation for the spectra in our sample with enhanced [\ion{N}{II}]/H$\alpha$, given that AGN emission would be highly suppressed (Paper I). 

However, we cannot exclude some residual influence by an AGN. Hydrodynamical simulations performed by \citet{roos15} show that even in the case of high obscuration 
an AGN can ionize the gas very far from the nucleus, reaching kpc scales and the circum-galactic medium. Furthermore, the accreting black hole might not be in the center, but that sounds unplausible: the attenuations towards the center derived independently from the X-ray detected AGNs are consistent with those derived from the mixed model (see Paper I) and, even further, \citet{rujopakarn18} show that the AGN position correlates with that of active star forming regions. Finally, we also notice that two galaxies (which simultaneously have X-rays and mid-IR dusty torus detection) were fitted with broad H$\alpha$ components (line width of $\sim 1000$ km/s). Such large velocity widths have been observed in both shock-dominated regions (possibly supernova driven, \citet{ghavamian17}) and AGNs \citep{peterson97,gaskell09,netzer15}. IFU data would be needed to disentangle shock or AGN emission, as we expect the latter to be much more concentrated in the central part of the system.

\begin{figure}[ht!]
    \centering
    \includegraphics[angle=0,width=\linewidth,trim={0.1cm 0.2cm 1.7cm 2.3cm},clip]{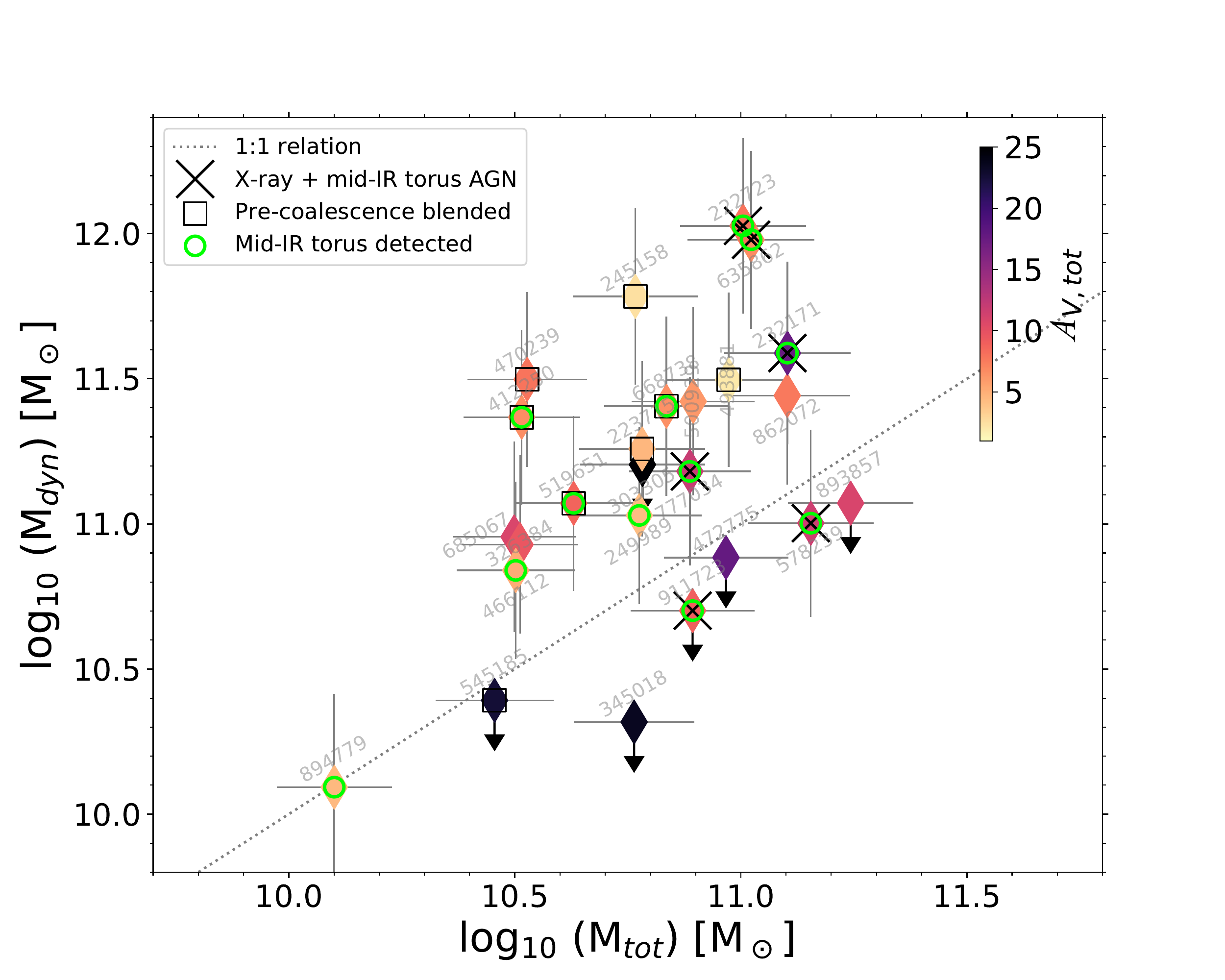}
    \includegraphics[angle=0,width=\linewidth,trim={0.1cm 0.2cm 1.7cm 2.3cm},clip]{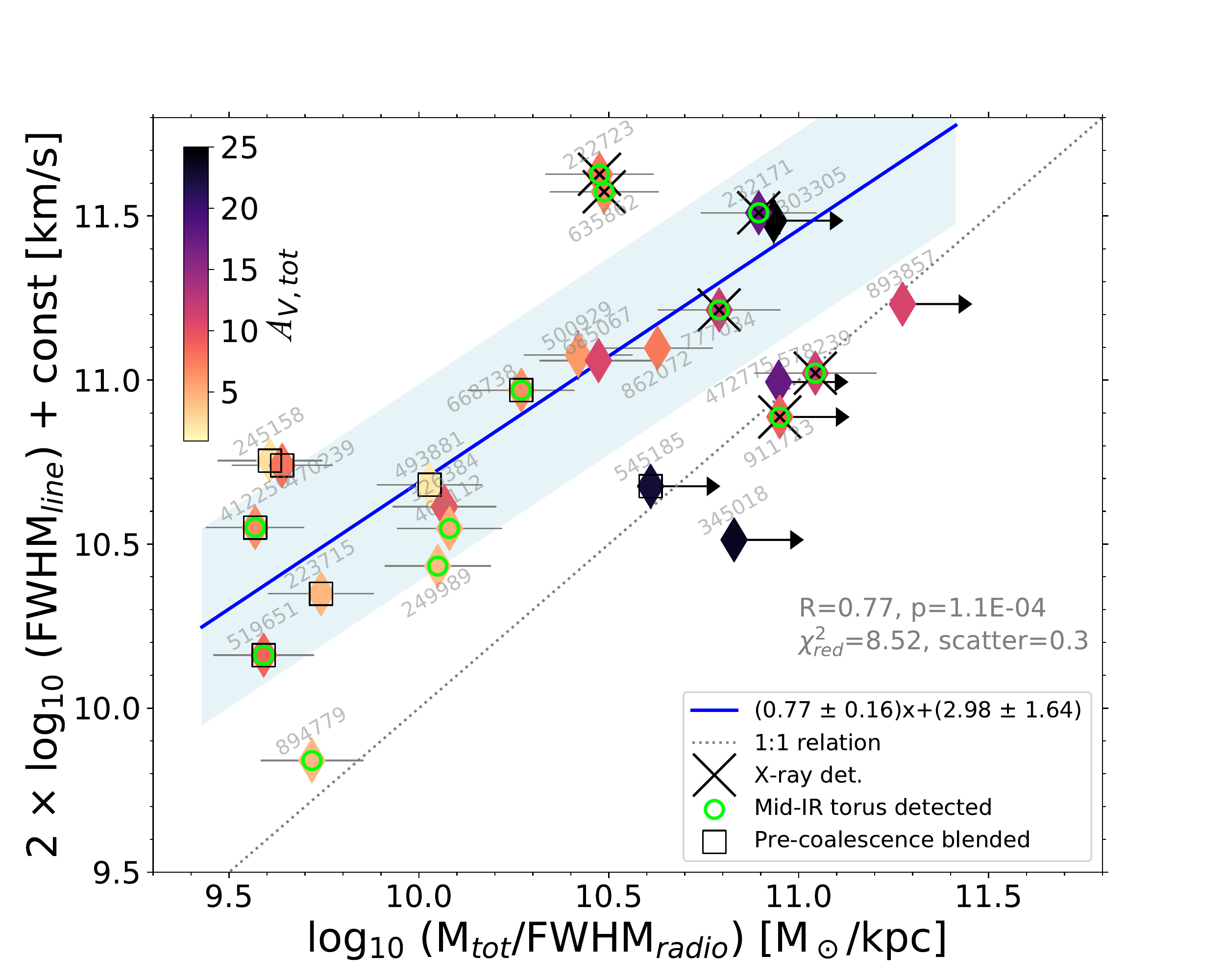}
    \caption{\small \textit{Top:} Comparison between the dynamical mass M$_{\text{dyn}}$ and the total mass content M$_{\text{tot}}$ ($=$ M$_{gas}$+M$_{\ast}$+M$_{\text{dark matter}}$) for our SBs sample, color coded by their total attenuation A$_\text{V,tot}$. \textit{Bottom:} Diagram showing the square of the total FWHM velocity width as a function of M$_{\text{tot}}$/FWHM$_{\text{radio}}$, using the same color coding based on A$_\text{V,tot}$. On the y-axis, const$=1.3 \times G$/($4$<sin$^2$(i)>) groups the coefficients in Equation~3 so as to facilitate comparison with the virialized case (1:1 relation, shown as a grey dotted line). The blue continuous line represents a linear fit to our sample, excluding galaxies with an upper limit on their radio size, while the blue shaded area shows the $\pm1\sigma$ limits of this best-fit relation. Both panels of the figure suggest that our galaxies may be approaching virialization, and more obscured starbursts are closer to the equilibrium.}\label{virial}
\end{figure}

\begin{figure}[t!]
    \centering
    \includegraphics[angle=0,width=8.4cm,trim={0.1cm 0.cm 1.3cm 2.cm},clip]{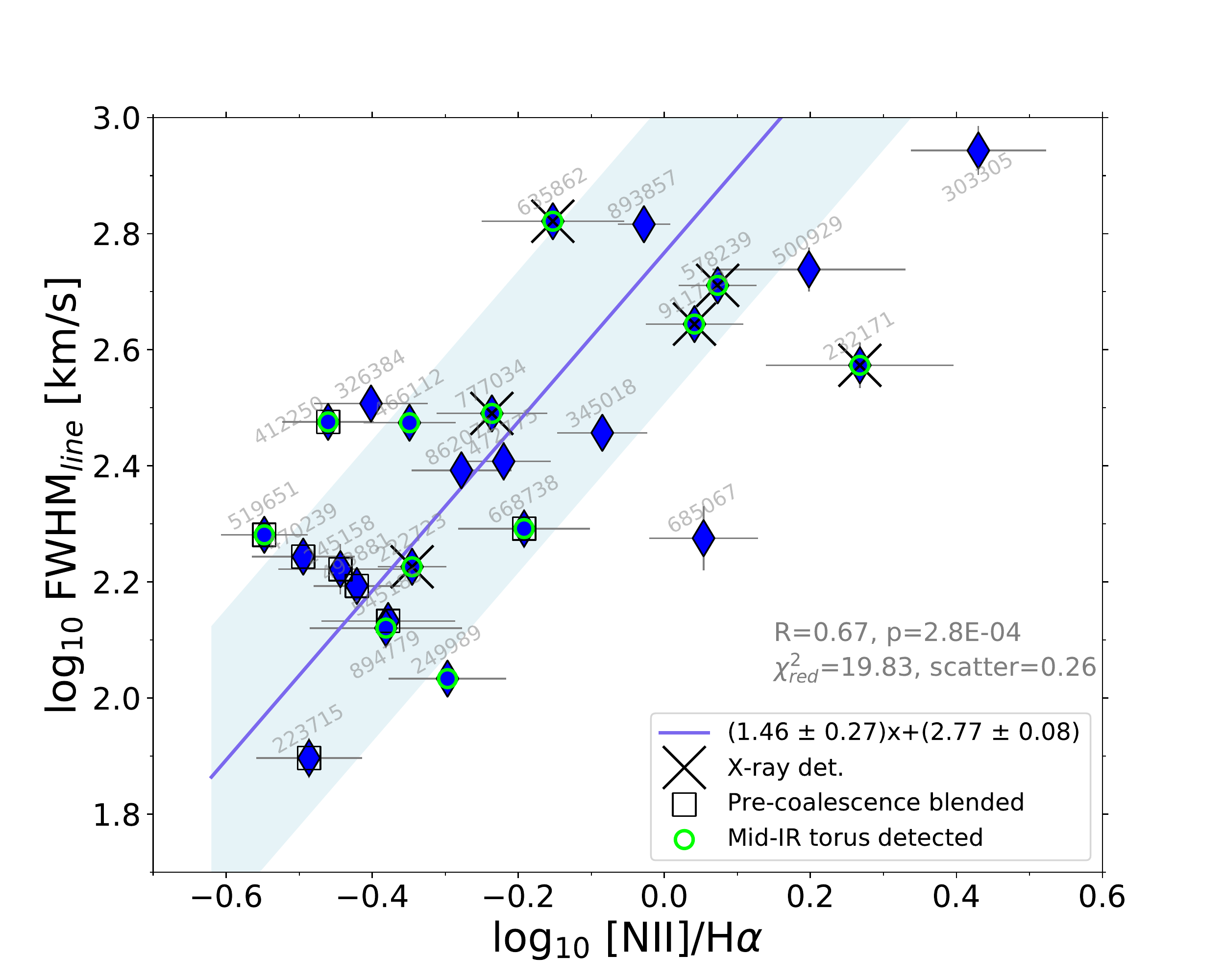}
    \caption{\small Correlation between the line velocity width of single Gaussian components (FWHM$_{\text{line}}$) and the N2 index (both measured from our Magellan-FIRE spectra), indicating that the two quantities are tightly (scatter $=$ 0.26 dex) physically related.}\label{sigmaline}
\end{figure}

\subsubsection{The dynamical masses of our sample}\label{dynamical_masses}

In order to better understand the dynamical status of our starbursts and how far they are from relaxation, we compared their total masses M$_{\text{tot}}$ to the dynamical masses M$_{\text{dyn}}$ estimated from the line velocity widths and radio sizes. For the latter, we used the formulation of \citet{daddi10} as :
\begin{equation}
    M_{\text{dyn}}=1.3\times\frac{ \text{FWHM}_\text{radio} \times (\text{FWHM}_\text{line,total}/2)^2}{\text{G sin}^2(i)}
\end{equation}
where FWHM$_\text{line,total}$ is the one-dimensional total line velocity width (accounting for both rotation and dispersion), while sin$^2$(i) is the correction for inclination that we take as the average value for randomly oriented galaxies ($57^\circ$). In order to determine the total uncertainty on M$_{\text{dyn}}$, we considered an additional error on the inclination factor of $0.3$ dex, as in \citet{coogan18}. This represents the main contribution to the error ($\sim 90\%$ in median), since the line width and radio sizes are always well measured with high S/N. 

Then we compared this quantity to the total mass content (baryonic + dark matter) of the systems, estimated as :
\begin{equation}
    \text{M}_{\text{tot}}=\text{M}_\ast+\text{M}_{\text{gas}}+\text{M}_\text{dark matter}
\end{equation}
in which M$_\text{gas}$ was determined, as described in Paper I, as M$_\text{gas} = 8.05+0.81 \times$ log(SFR$_\text{IR}$) \citep{sargent14}, valid for a starburst regime, and we assumed M$_\text{dark matter} = 10\% \pm 10 \%$ of M$_\ast$. Since this contribution is highly uncertain, it was set nearly uncontrained. However, this range is consistent with studies of high-$z$ ($>0.5$) massive star-forming galaxies, which found a modest to negligible dark matter fraction inside the half-light radius \citep[e.g.,][]{daddi10,genzel17}. In any case, given the small contribution, its exact value does not affect the results of this paper. 
For the error determination, we considered the above uncertainty on M$_\text{dark matter}$, a $0.1$ dex error on M$_\ast$ \citep{laigle16}, and $20\%$ incertitude on the gas mass (even though its contribution is negligible given that M$_{\text{gas}} \simeq 0.1$ M$_{\ast} $ on average for our sample).

The comparison between M$_\text{dyn}$ and M$_\text{tot}$ in Fig. \ref{virial} shows that, on average, our galaxies are not completely virialized: while $\sim$ half of the sample is consistent within $2\sigma$ with the 1:1 relation, the remaining part is located above at higher M$_\text{dyn}$. The largest departures from virialization are observed for the pre-coalescence and less obscured systems, i.e. supposedly earlier stage mergers. On the contrary, the systems with better agreement may be fully coalesced starburst cores with higher A$_\text{V,tot}$. 

The tight connection between velocity and gravitational potential is clarified in the bottom panel of Fig. \ref{virial}, as the FWHM$_\text{line}^2$ and M$_\text{tot}/$FWHM$_\text{radio}$ correlate at 5$\sigma$ significance (with R=0.74 and p-value=0.0002). Also here, while pre-coalescence mergers have larger displacements from the 1:1 relation, they are confined in a region at lower velocity widths and shallower potential wells. This suggests that also other starbursts (ID 249989, 466112, 326384) in this region may be pre-coalescence mergers that we were not able to securely identify, due to their lower S/N 2D spectra, and indeed their optical morphology strengthens this suspicion. In the upper-right part of Fig. \ref{virial}-\textit{bottom}, separated from the previous sample, are clustered the more obscured starbursts, i.e., supposedly coalesced mergers. We notice also that all X-ray detected AGNs are localized in this region of the diagram, indicating a possible link between evolutionary phase and AGN properties, that we will further investigate in the following Section.

Overall, the above results suggest a time-evolutionary scenario, in which more advanced, already coalesced mergers are close to virialization, and the increased central potential wells (due to the contribution of both merger components) are responsible for the enhancement of both the kinetic energy content and shocks towards the later stages of the interaction. The tight relation between the line velocity width of single Gaussian components (a proxy for the velocity dispersion in the system) and shock production (traced by the N2 index) is further indicated by the color coding of the BPT diagram in Fig. \ref{BPT1}-\textit{top}, and by the correlation between FWHM$_{\text{line}}$ and \ion{N}{II}/H$\alpha$ in Fig. \ref{sigmaline}, which has a significance higher than $5\sigma$ (R=0.67, p-value=$3\times10^{-4}$) and a dispersion of $0.26$ dex.  


\subsection{Lower line equivalent widths toward late merger stages}\label{lower_line_equivalent_widths}

\begin{figure*}[t!]
    \centering
    \includegraphics[angle=0,width=17.5cm,trim={4.8cm 0.1cm 3.cm 0.3cm},clip]{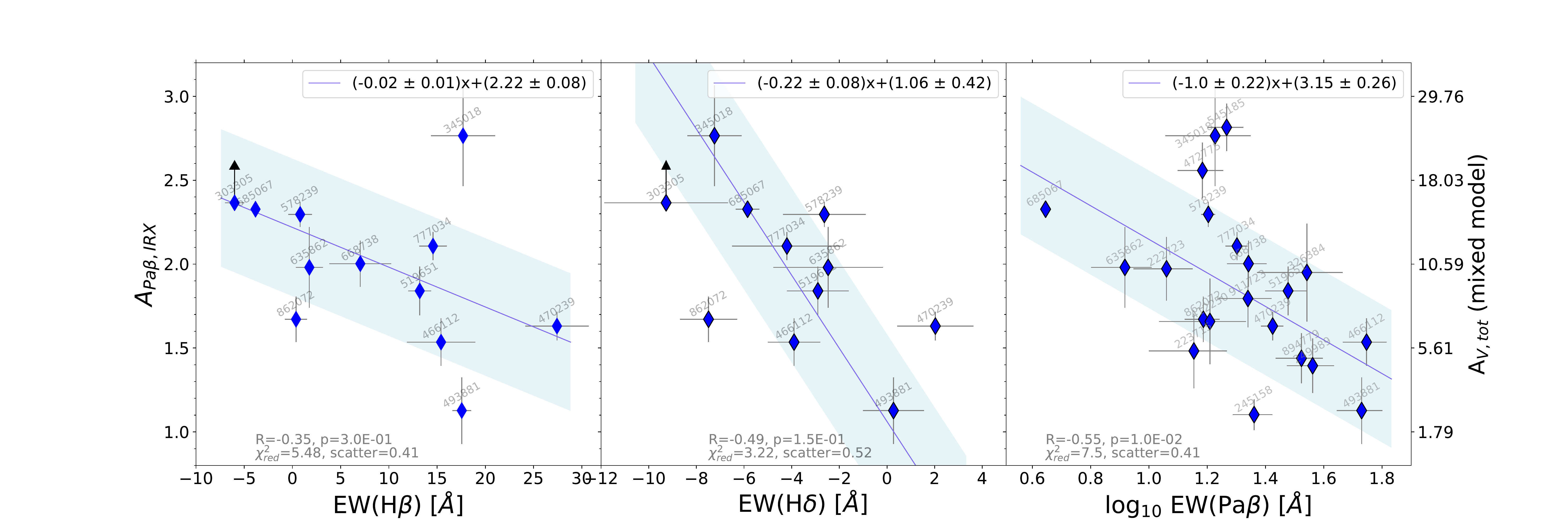}
    \vspace{-0.25cm}
    \includegraphics[angle=0,width=17.5cm,trim={4.8cm 0.cm 3.cm 1.4cm},clip]{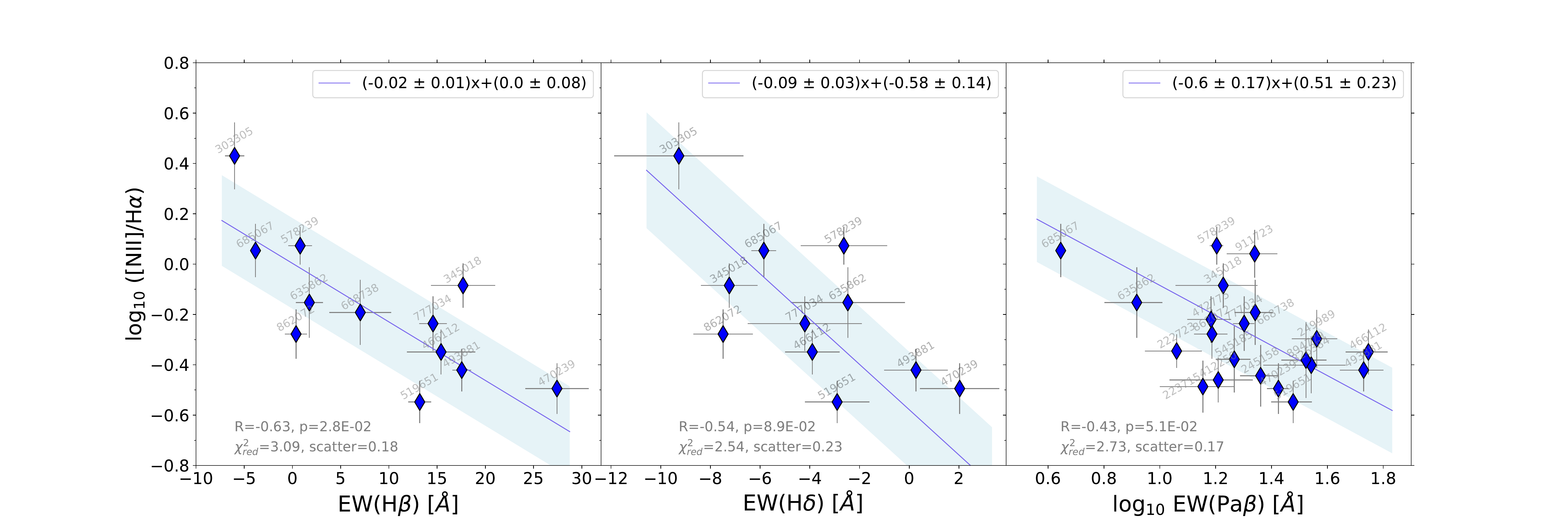}
    \caption{\small \textit{Top:} Comparison between the EW of H$\delta$, H$\beta$ and Pa$\beta$ lines to the dust attenuation parameter A$_{Pa\beta,\text{IRX}}$, defined as $2.5 \times log_{10}(1+\text{SFR}_{\text{IR}}/\text{SFR}_{Pa\beta,\text{obs}}$ (Paper I). We remark that the last panel is equivalent to the bottom-right plot in Fig. \ref{correlations}, even though a different scale has been used (the relation between A$_{Pa\beta,\text{IRX}}$ and A$_\text{V,tot}$ is given in Eq. \ref{eqAV}). \textit{Bottom:} Correlations between the EW of H$\delta$, H$\beta$, Pa$\beta$ lines and the N2 index ($=log_{10}([NII]/H\alpha)$). The blue continuous lines are the best-fit linear relations, determined as explained in Section \ref{results}, while the blue shaded area show the $\pm1\sigma$ scatter of our data around the best-fit relations.}\label{EW_ALL}
\end{figure*}

The equivalent widths (EW) of hydrogen recombination lines give a relatively dust-unbiased picture (assuming that stars and emission lines are equally extincted) of the contribution of the SFR to the stellar mass content 
, and they are sensitive to the luminosity weighted age of the stellar populations, so that they could provide useful information about the evolutionary stage of the merger. However, these EWs would only probe what is happening in the outer parts of the system, since the core is completely obscured in optical/near-IR.

We show in the last diagram in Fig. \ref{correlations} and the upper part of Fig. \ref{EW_ALL} that, when the starbursts become more obscured, the EWs of Pa$\beta$, H$\delta$ and H$\beta$ decrease, indicating a gradual SFR decline in the outer skin of more obscured and compact starbursts. Additional correlations are found also independently between those EWs and the other quantities, such as the N2 index (Fig. \ref{EW_ALL}-\textit{bottom}). We additionally remark that the different Paschen and Balmer lines correlate each other (see Fig. \ref{EW_EW} in the Appendix), for which reason our results, derived adopting the Pa$\beta$ line as a reference (because it is the least attenuated), are also valid when considering the H$\delta$, H$\beta$ and H$\alpha$ lines. 

In our sample, we also found that the Balmer EWs, while having a large dynamical range, can reach very low values: in five galaxies (ID 303305, 685067, 777034, 862072 and 345018) we measure an EW(H$\delta$) < $-4\AA$ (i.e. in absorption), which are typically found in E+A dusty galaxies \citep{poggianti00}.  
Low EW hydrogen recombination lines (in strong absorption) are clear signatures of the prevalence of A-type stars, indicating that a recent ($<$ 1 Gyr ago) massive star-formation episode has taken place during the past $10^8$-$10^9$ yr, while the youngest stellar  populations (mainly OB stars) are nearly all obscured by dust in the inner starburst core. 
Our dusty starburst systems should also not be confused with post starburst (PSB) galaxies, which have similar absorption EWs (e.g., EW(H$\delta$) lower than $\sim -5 \AA$ as in \citet{goto07} and \citet{maltby16}), but are thought to be nearly (or already) quenched systems, with much lower SFR levels and lower dust content compared to our sample (see \citet{pawlik18} for a full discussion of the different types of PSB galaxies). We caution that the quenched PSB selection from only the Balmer EW can actually return real starbursts and not post starburst systems. 

Putting all together, the time-evolutionary scenario that we have suggested has the advantage of explaining in a simple way these new results. If we follow the merger evolution towards the coalescence, the outer starburst skin becomes increasingly dominated by A-type stars, recognizable through the deep absorption lines in the optical/near-IR and which were formed at earlier times when the separation between the merging nuclei was larger. At the same time, the star-formation in the skin is being suppressed, possibly driven by supernova feedback. 


\subsection{Outliers}\label{outliers}

We found in Section \ref{results} (Fig. \ref{correlations}) that $4$ galaxies are outside the $1\sigma$ dispersion of the best-fit relations between the dust attenuation A$_\text{V,tot}$ and, simultaneously, the [NII]/H$\alpha$ ratio (N2), the line velocity width (FWHM$_\text{line}$) and the EW(Pa$\beta$). 
In particular, they have lower N2, FWHM$_\text{line}$ and EW(Pa$\beta$) than expected from their A$_\text{V,tot}$, suggesting that, compared to other highly obscured galaxies, there is a minor impact from shocks or a dominant contribution of star-formation to the emission lines.

Within our SB sample, we recognize that these $4$ outliers have the largest dust obscurations A$_\text{V,tot}\geq 18$ mag, and are among the most compact, with radio FWHM sizes below $1$ kpc. These extreme and peculiar features suggest they may represent the very end stages of the merger evolution, and that the correlations with A$_\text{V,tot}$ may saturate toward these late phases. We also notice that the same objects are not systematically outliers when we consider their N2, FWHM$_\text{line}$ and EW(Pa$\beta$) values, confirming the close physical connection among these quantities, as shown in the two previous Sections \ref{velocity_enhancement} and \ref{lower_line_equivalent_widths}.

\subsection{The complete sequence of merger stages at intermediate redshift}\label{cartoon_section}

\begin{figure*}[t!]
  \centering
  \includegraphics[angle=0,width=0.9\linewidth,trim={0cm 0.0cm 0.cm 0cm},clip]{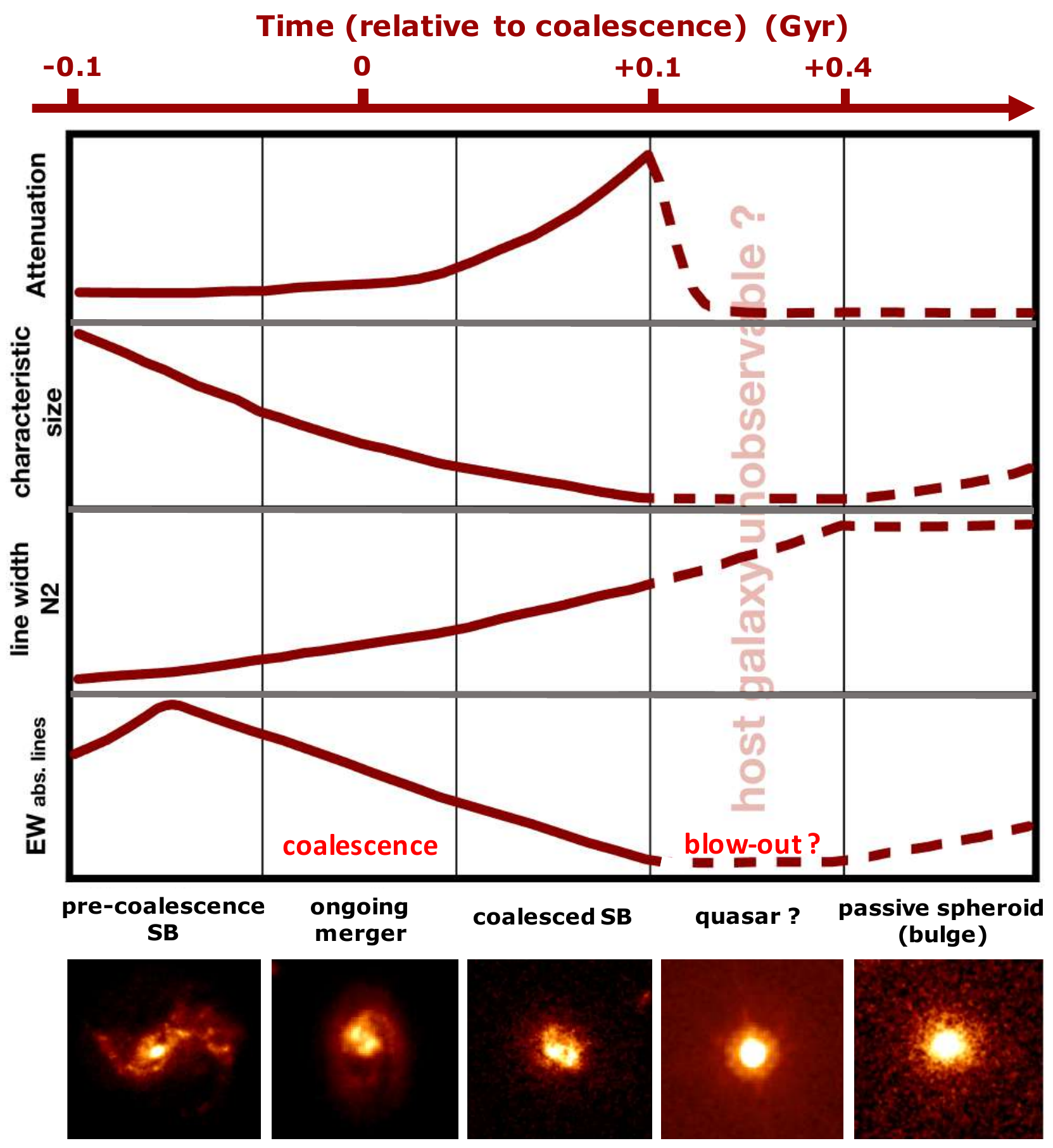} 
  \caption{\small Schematic illustration of the time-evolutionary behavior of the physical parameters studied in the text: dust attenuation, characteristic size of the system, line velocity width (or, equivalently, the N2 index) and the EW of Balmer and Paschen lines. The time sequence is divided into 5 fundamental merger stages, with the QSO and passive spheroidal system representing the final stages according to the classical merger paradigm \citep{sanders96,hopkins08a,hopkins08c}. Solid lines are qualitative trends during the SB phase inferred from our results, while dashed lines are predictions about the future evolution of the $4$ parameters shown on the y-axis (line width and N2 index behave similarly). In the upper part of the figure, we show a qualitative merger timescale following Fig.~1 of \citet{hopkins08c}, assuming for the merger a total starburst duration of $200$ Myr. For each phase, we show in the bottom part the ACS-F814W cutout of a representative case. The first three images are SB galaxies from our sample: ID 223715, ID 777034 and ID 472775. They were chosen as having increasing dust attenuations and radio compactness, suggestive of more advanced merger phases: the first was identified as a pre-coalescence merger in Section \ref{pre-coalescence}, while the latter is unresolved in radio and is highly obscured (A$_\text{V,tot}=18$ mag). The last $2$ cutouts show a quasar at $z=0.73$ and an ETG at $z=0.66$, selected in COSMOS field from the catalogs of \citet{prescott16} and \citet{tasca09}, respectively. 
  }\label{cartoon}
\end{figure*}

Our observations and results, presented in previous sections, suggest we are starting to see an evolutionary sequence in high-redshift mergers. This can be traced through a variety of physical measurable quantities of our galaxies, including the total attenuation towards the center, the characteristic size of the starburst region, the EW of hydrogen absorption lines, and finally the [\ion{N}{II}]/H$\alpha$ ratios and line velocity widths, which behave similarly.
In Fig. \ref{cartoon} we schematize with a cartoon all the results that we have found so far, showing with a red continuous line the qualitative trend of the different physical quantities as a function of time. We divided the time axis into five merger evolutionary stages, which are arranged in relation to the two most crucial transformation events during the merger: the coalescence and the blow-out/QSO appearance.

We notice that the first phase may not necessarily represent the beginning of the interaction, i.e., when the two galaxies approach for the first time. Even though the whole merger episode may last 1-1.5 Gyr in total, from the first encounter to the formation of a passive spheroidal system, the starburst activity is typically shorter, ranging 200-300 Myr \citep{dimatteo08} and may be triggered intermittently at various stages of the evolution. Furthermore, whether or not a strong burst is already activated at the first approach depends on many factors, including the impact geometry, the morphology, the stellar mass ratio and the gas content of the colliding galaxies \citep{dimatteo08}. 


Besides the observable starburst phases studied in this work, can we also make some predictions on the future evolution of these systems ? In general, it is very hard to demonstrate visually a connection between mergers and their descendants. Indeed, not all merger-induced starbursts exhibit morphological disturbances \citep{lotz08}, and when merger residual signatures are present, they fade rapidly, becoming almost invisible beyond the local Universe even in the deepest optical images \citep[e.g.][]{hibbard97}. We can in principle rely on hydrodynamical simulations, which allow to trace the full time-sequence of mergers, even though they also present limitations due to the many assumptions, initial conditions and physical complexity involved in such events. 

In the classical theoretical merger paradigm, the infalling gas triggers obscured AGN accretion \citep{bennert08a}, whose peak of activity typically occurs $\sim 250$ Myr after the onset of the starburst \citep{wild10}, and $\sim 100$ Myr after the peak of SFR \citep{davies07,hopkins12}. It is during these later starburst phases that the AGN feedback can blow out with strong feedback winds the surrounding dust and gas cocoon, eventually revealing itself as a bright QSO \citep{hopkins08c}. This phase is generally very short, lasting for $\lesssim 100$ Myr \citep{hopkins10c}, and has been claimed since a long time: \citet{lipari03} suggested that QSOs could be indeed young IR active galaxies at the end phase of a strong starburst. 


Since the QSO dominates the luminosity of the system at all wavelengths, it would be extremely hard to analyze the physical properties of the host galaxies during this phase. Indeed, \citet{zakamska16} show that even in radio-quiet QSOs both the infrared and the radio emission are dominated by the quasar activity, not by the host galaxy. 
An alternative possibility is to look far from the central bright source. Recent works are revealing Ly-$\alpha$ nebulae surrounding high-redshift quasars, with extension that can reach tens of kpc ($\lesssim$ 50 kpc) from the center \citep{arrigonibattaia18}.
On the other hand, one may focus on local samples, increasing simultaneously the images resolution. For example, \citet{lipari03} and \citet{bennert08b} discovered with HST the presence of outflows, arcs, bubbles and tidal tales in optical band in a sample of local QSOs, possibly formed through strong galactic winds or merger processes. Again in nearby (z $<0.3$) QSOs, near-IR H band adaptive optics observations \citep{guyon06} revealed that $\sim 30\%$ of their hosts show signs of disturbances, and the most luminous QSOs are harbored exclusively in ellipticals or in mergers (which may become ellipticals soon). Furthermore, while the SFRs of the hosts are similar to those of normal star-forming galaxies, their mid- and far-IR colors resemble those of warm ULIRGs, strengthening a connection between these two objects. 

In the following two Sections \ref{mass-size-section} and \ref{QSO_in_formation}, we discuss separately the two ending stages of the merger sequence, and investigate how our work can provide some clues to understand what are the physical properties of the systems into which our starbursts will evolve. In the cartoon of Fig. \ref{cartoon}, the predicted evolution for all the quantities studied in this paper (see Section \ref{results}) is shown with a dashed line. These qualitative trends are motivated mainly from simulations, and are not confirmed observationally.

\subsection{Mass-size relation and comparison with higher and lower-z starbursts}\label{mass-size-section}

\begin{figure*}[t!]
  \centering
  \includegraphics[angle=0,width=\linewidth,trim={0.1cm 0.2cm 2cm 1.3cm},clip]{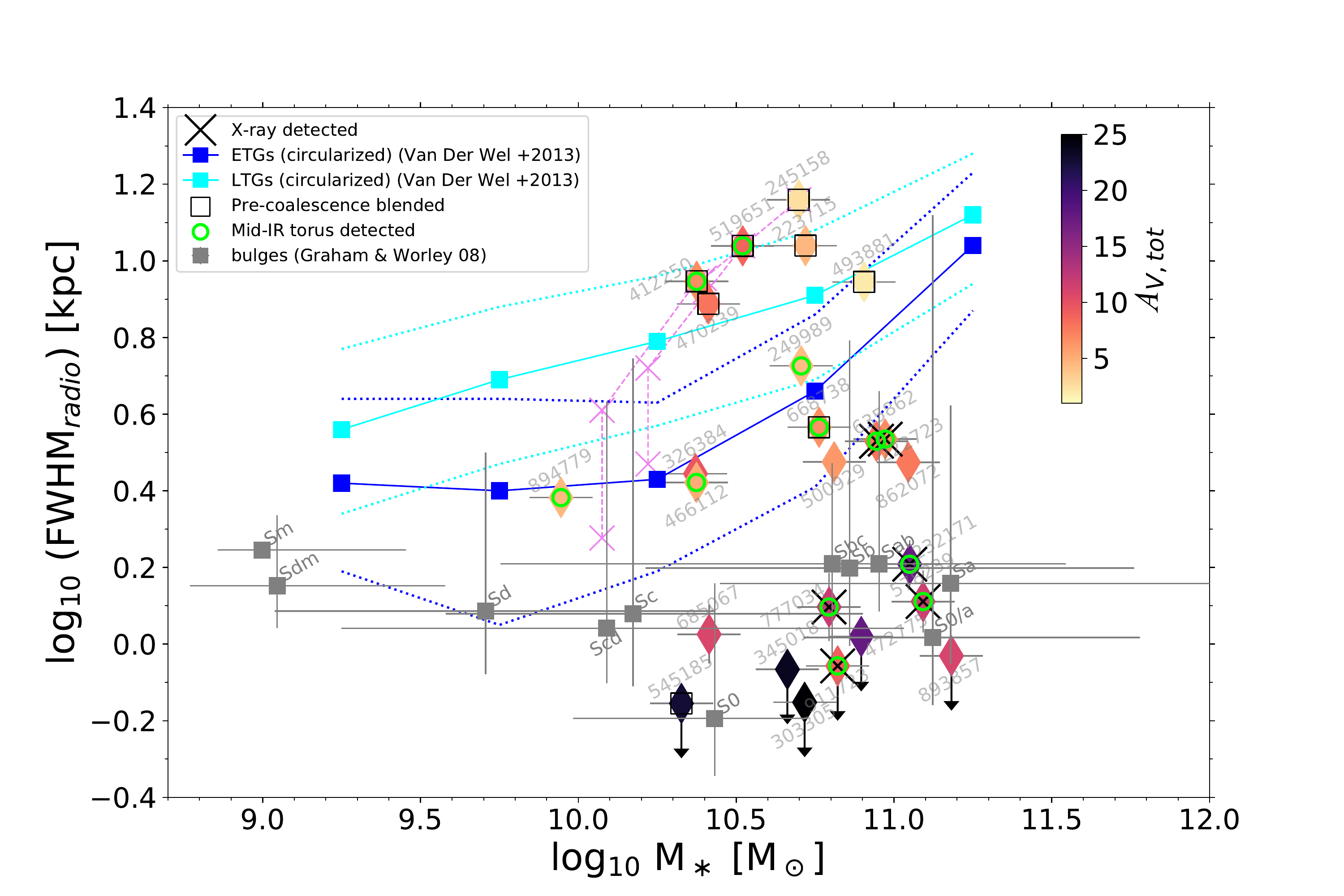} 
  \caption{\small Diagram showing the radio size vs stellar mass for our sample. We compare our results to the stellar mass - stellar size relation of LTGs (cyan line with 1$\sigma$ dispersion) and ETGs (blue line) at z$=$0.7 \citep{vanderwel14}, and with the bulge properties of low redshift (z$\sim$0.1) spiral galaxies from \citet{graham08} (grey squares). The eleven points shown here for the bulges represent the median ($\pm1\sigma$) of their distributions of stellar masses and stellar sizes (in K band) as a function of galaxy type, from S0 to Sm spirals. For $3$ galaxies in our sample fitted with a double Gaussian, we also represent the `deblended' radio sizes of each single component with violet crosses, connecting them with a dashed violet line. In these cases, we assigned to each component half of the total stellar mass of the system, even though a precise estimation requires a separate fit on deblended photometric data.}\label{Mass_Size}
\end{figure*}

The merger-induced starbursts are supposed to end up in a passive system, but we do not know the exact physical properties (e.g., size, stellar mass, morphology) of these merger remnants. Sub-millimeter galaxies (SMGs) at high redshift ($>2$), which are commonly viewed as higher luminous counterparts of lower redshift ULIRGs, have been suggested to be direct progenitors of massive ETGs \citep{tacconi08,toft14}. 
We can investigate this connection by comparing in Fig. \ref{Mass_Size} the stellar masses and the characteristic sizes of our starbursts with those of disk galaxies and spheroids at z $\sim0.7$ \citep{vanderwel14}. To be conservative, we are adopting here the M$_\ast$-size relations for circularized radii. If we consider instead the non-circularized cases, the same relations would slightly shift upwards by $\sim 0.1$ dex.
In addition, we remark that we are comparing our radio (starburst) extensions to optical rest-frame sizes tracing the stellar mass distribution of disks and elliptical galaxies. Indeed, we implicitly assume that, after the gas in our starburst cores is converted into stars, the extensions of these cores will represent also the stellar component sizes of their passive remnants. On the other hand, they may still represent the dense star-forming gas components of post-starburst systems if some residual is left after the merger, as they may remain compact for at least 1 Gyr \citep{davis18}. 

In the diagram of Fig. \ref{Mass_Size}, $6$ SBs are consistent with the late-type galaxy (LTG) relation at z $\sim0.7$. However all of them are pre-coalescence SBs and, as we have seen before, they should not be considered disk galaxies as their size is primarily reflecting the separation between the merging components. For two of the three galaxies resolved in radio, the single values return below on the early-type galaxy (ETG) relation. The characteristic sizes of this subset (FWHM$_{\text{size}}$ ranging 3-15 kpc in diameter, with median FWHM$_{\text{size}}$ of 8 kpc) are similar to those typical observed in SMGs \citep{casey11,tacconi08,biggs08}, which suggests that SMGs at high-redshift (or at least a fraction of them) could be indeed intermediate-phase mergers composed of unresolved double nuclei, as argued by \citet{iono09} and \citet{arribas12}. 

In the bottom part instead, we can immediately notice that a major fraction of our sample (13 galaxies, i.e., $52\%$ of the total sample) is not consistent with the ETG relation (taking $1$-$\sigma$ dispersion), and is located well below it by $\sim0.5$ dex, with an average size of $\leq 1.2$ kpc, indicating that they are much more compact than their stellar envelopes and than typical ellipticals at z $\sim0.7$. We underline that such difference would be even higher if we compare this subset to the M$_\ast$-size relation at redshifts lower than $0.7$, as the ETG sizes at z $=0.25$ are a factor of 1.5 higher than those at z $=0.75$, at our median stellar mass \citep{vanderwel14}.
This sample of very compact starbursts has typical extensions that are similar to those of dense star-forming regions in local ULIRGs \citep{genzel98,piqueraslopez16}, including Arp 220 \citep{sakamoto17} and M82 \citep{barker08}, suggesting they are driven by the same merger mechanisms (as also argued in Paper I). 

If we take for each galaxy its distance from the LTG relation (dist$_{\text{LTG}}=\log_{10}$(FWHM$_{\text{size}}$/FWHM$_{\text{LTG}}$)), we can also use this quantity in place of the radius to trace the same sequence found in Section \ref{results}, taking into account the mild dependence on stellar mass. 
As the merger proceeds, the system moves from the LTG to the ETG relation and then even below at significantly smaller sizes (by $\sim 0.5$ dex at least), meaning that the compact starburst cores that form at the coalescence cannot produce directly the ellipticals seen at redshift $0.7$ and below. 

The sizes of our starbursts instead resemble those of typical bulges in lower redshift spirals and lenticular galaxies \citep{graham08,laurikainen10}, indicating a possible evolutionary link with mergers, as suggested by other works \citep[e.g.][]{sanders96,lilly99,elichemoral06,querejeta15}. 
This idea is consistent with the typical observed gas fractions of our starbursts (derived as M$_{gas}$/(M$_\ast$+M$_{gas}$), with M$_{gas}$ calculated in Section \ref{dynamical_masses}), which range between $0.02$ and $0.25$ ($\sim 0.1$ in median). Assuming that all the remaining gas is consumed before the passivization and that the same amount of gas has been already converted into stars (which depends on the merger phase and dynamics), it means that the current starburst cores can produce approximately $20\%$, and up to $50\%$, of the final stellar mass of the galaxies. Higher resolution radio images targeting specific emission lines can further constrain the kinematic properties of the starbursting cores, by looking for rotation, or their luminosity profile, e.g., measuring their Sersic index.
How this old stellar component is affected by the merger can depend on many conditions difficult to model in detail, including the geometry of the interaction, the gas content and the mass ratio of the colliding galaxies.   


\begin{figure*}[]
  \centering
  \includegraphics[angle=0,width=0.338\linewidth,trim={0cm 0.cm 0cm 0cm},clip]{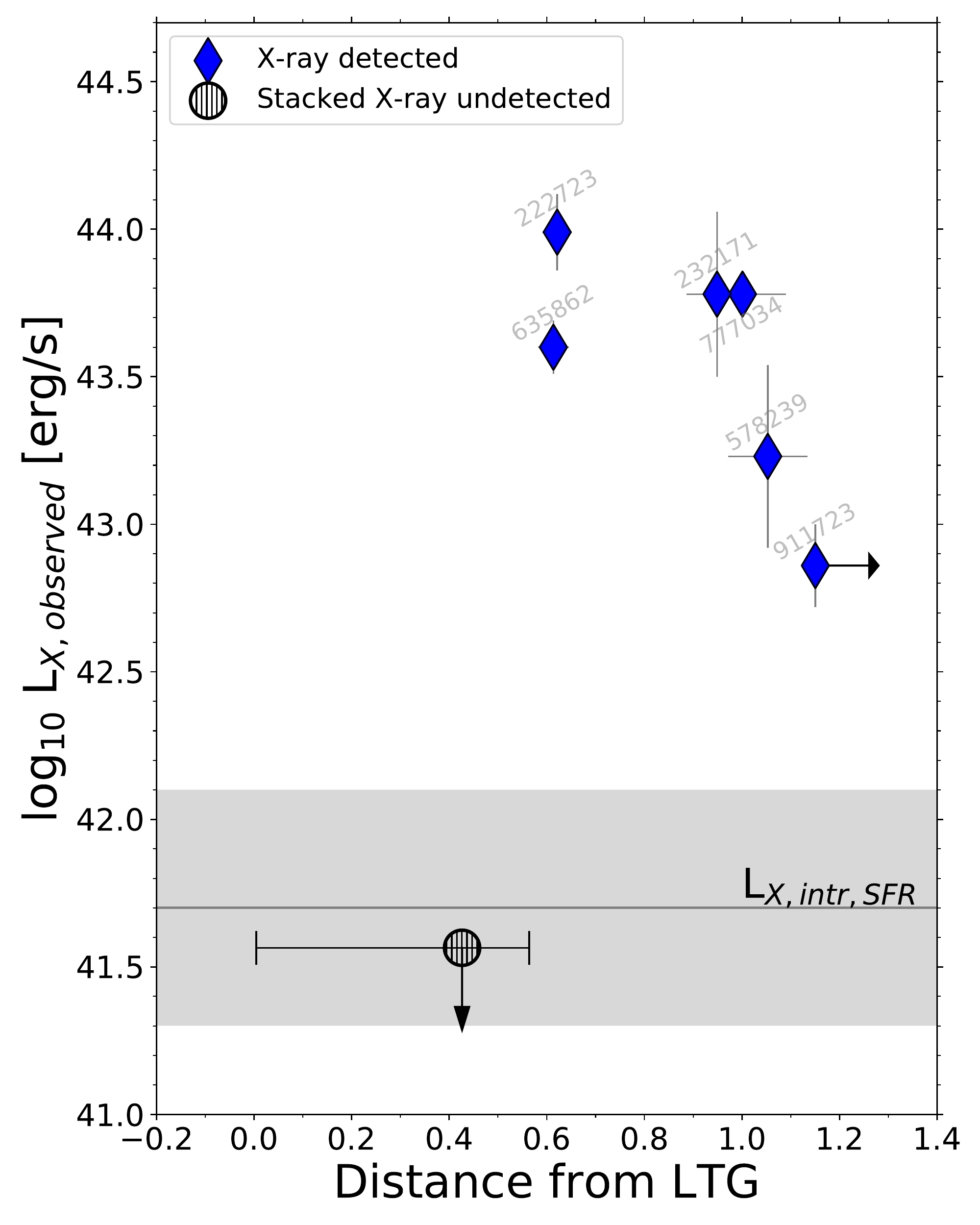}
  \includegraphics[angle=0,width=0.65\linewidth,trim={1.2cm 0.7cm 3.cm 2cm},clip]{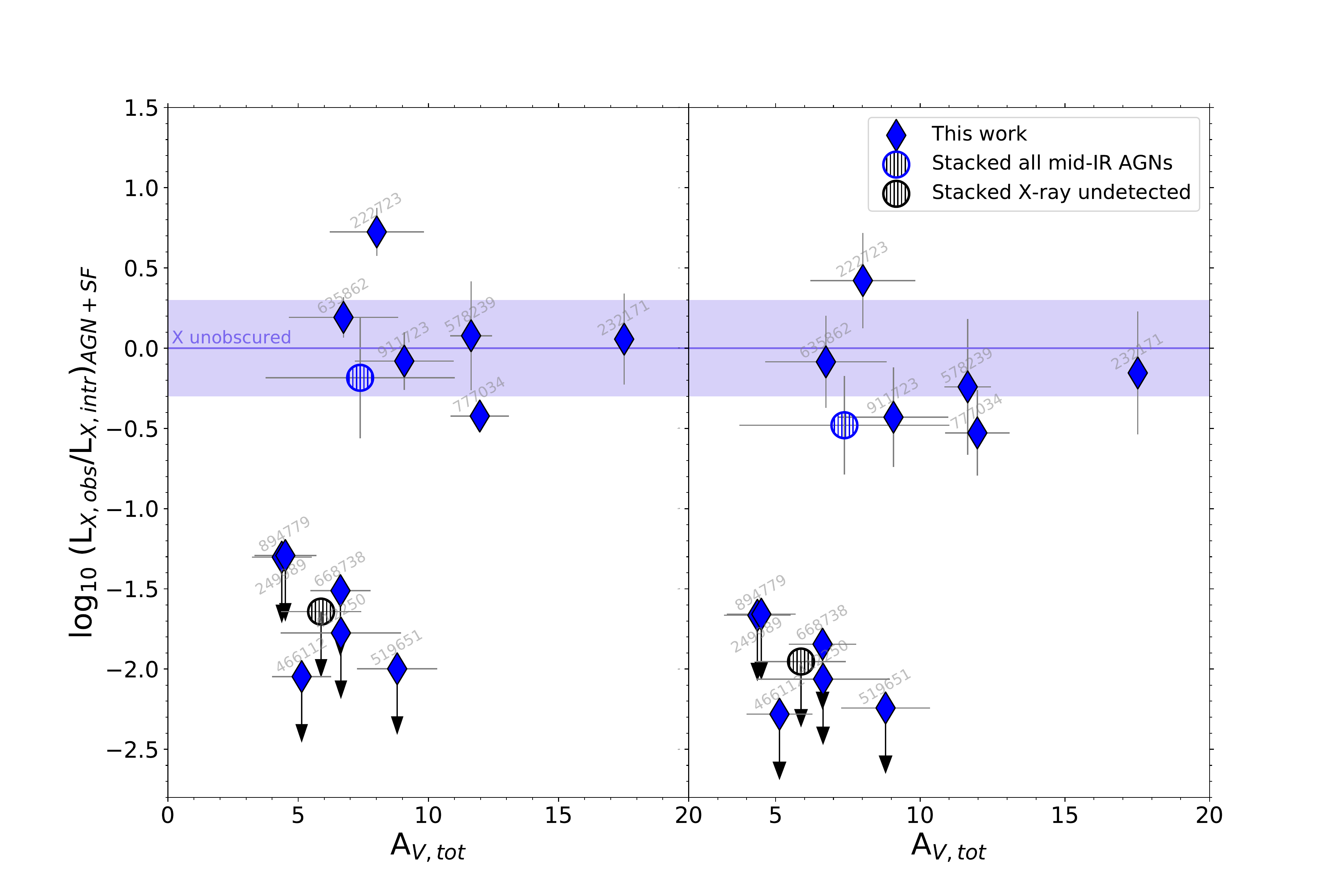}
  \caption{\small \textit{Left:} Comparison between the observed X-ray luminosity (L$_\text{X,obs}$) and the distance from the Mass-size relation of LTGs at $z\sim0.7$, for our $6$ starbursts detected in X-rays. Upper limit on L$_\text{X,obs}$ for $6$ mid-IR AGNs undetected in X is shown with a black circle, where the horizontal segment represents the range of dist$_\text{LTG}$ spanned by this subset. The intrinsic X-ray luminosity due to star-formation is highlighted with a gray line for the median SFR of the sample ($\pm 0.4$dex scatter from \citet{mineo14}), and may dominate the total X-ray observed emission for the X-undetected starbursts; \textit{Center:} X-ray attenuation L$_{X,obs}$/L$_{X,int}$ as a function of the infrared-based attenuation A$_\text{V,tot}$ (in a mixed model geometry and towards the center) for our sample of mid-IR detected AGNs. We assumed here a bolometric correction factor L$_{\text{X,intr,AGN}}=0.04\times$ L$_{\text{BOL,AGN}}$ \citep{vasudevan07}. Stacks on the whole sample and on the X-undetected subset are displayed with hatching circles, while the violet shaded regions indicate the area of no obscuration, which incorporates a factor of 2 uncertainty in the conversion between intrinsic X-ray and bolometric AGN luminosity; \textit{Right:} Same diagram as before, but assuming an L$_{\text{BOL,AGN}}$- dependent bolometric correction \citep{lusso12}, as explained in the text.  
  }\label{distltg}
\end{figure*}

\begin{figure}[h!]
  \centering
  \includegraphics[angle=0,width=\linewidth,trim={0.1cm 0cm 0cm 1.3cm},clip]{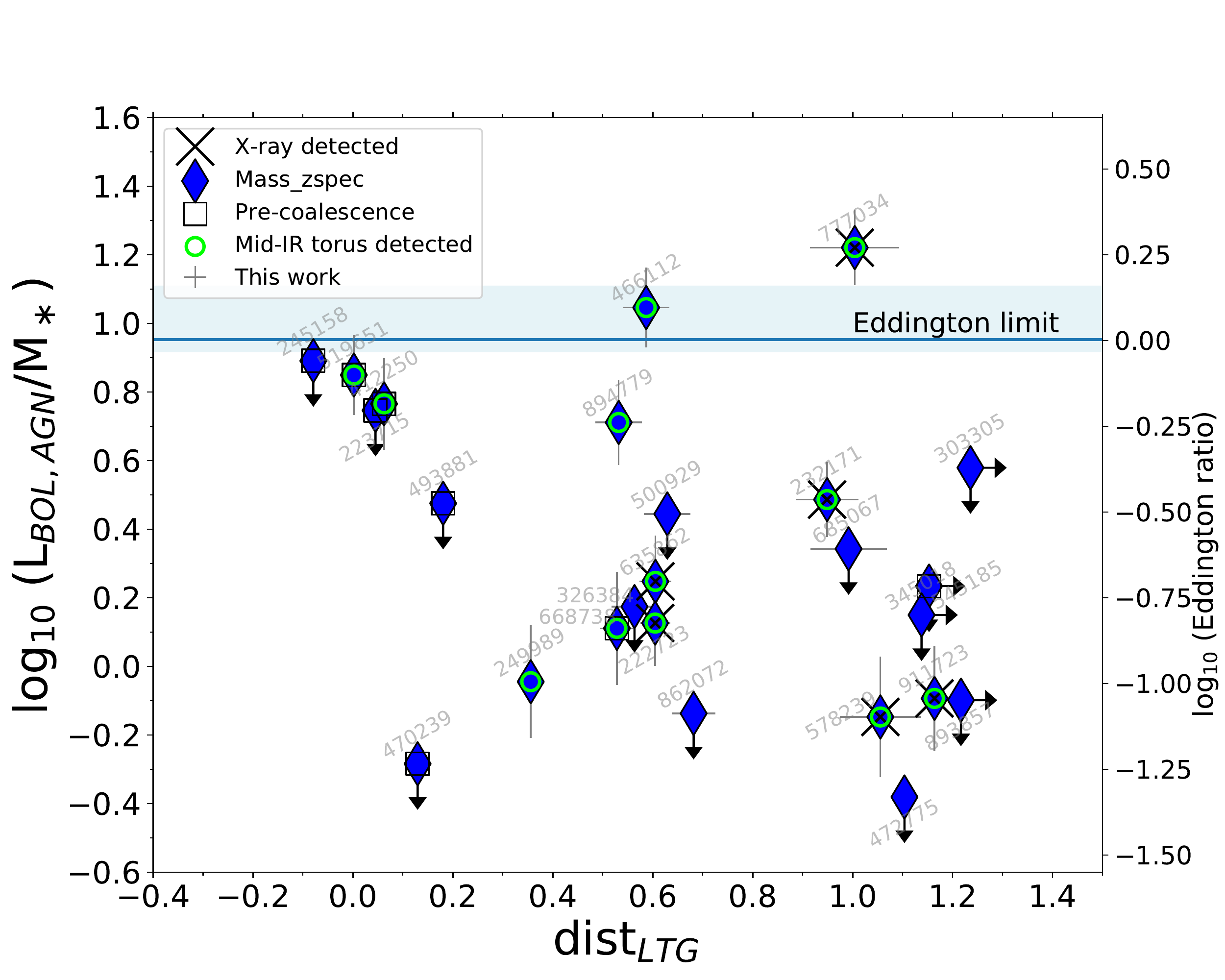}
  \caption{\small L$_{\text{BOL,AGN}}$/M$_\ast$ vs. distance from the Mass-size relation of LTGs (dist$_{LTG}$) for our SBs sample. The Eddington limit is shown with a blue horizontal line, while the shaded area takes into account the spread of M$_\ast$ among our sample and the uncertainty of the relation between stellar mass and BH mass by \citet{reines15}. The Eddington ratio is shown on the right y-axis.}\label{distltg2}
\end{figure}

\subsection{QSOs in formation at $z \sim 0.7$ ?}\label{QSO_in_formation}


In the starburst selection phase, we discarded several quasars because of the impossibility to study the properties of their host galaxies (dust attenuation, SFR, stellar mass), as discussed in Section \ref{cartoon_section}. In order to overcome these limitations, several authors have extensively studied also the transitional moments (just preceding the final blow-out) in the dress of type-I and warm ULIRGs \citep{kawakatu06,sanders88}. Similarly, we can have some clues of the inner black-hole activity just before the hypothetical QSO by looking at the AGN diagnostics in our starburst sample. 

As mentioned in Section \ref{AGN_identification}, we detected the mid-IR dusty torus AGN emission in $12$ galaxies, and simultaneous X-ray emission in $6$ of them. Notably, the latter are the only ones (among mid-IR AGNs) whose host galaxies lie below the ETG relation in the Mass-size diagram (see Fig. \ref{Mass_Size}), at systematically smaller sizes than ellipticals at z $\leq 0.7$. 
This suggests that during early merger stages the AGNs are predominantly obscured, while they start to appear in X-rays toward intermediate stages (i.e. when the host starbursts are more compact and obscured), possibly driven by rapid AGN feedback clearing the gas and dust around the black hole.

This can be seen better in Fig. \ref{distltg}-\textit{left}, where all the host galaxies of X-ray detected AGNs are located at larger distances from the Mass-size relation of LTGs compared to X-ray undetected AGNs. Moreover, they have X-ray luminosities at least 1 order of magnitude higher ($\sim +1.5$ on average) than what inferred from their SFR. For the X-undetected galaxies instead, the upper limit on L$_X$ $=$10$^{41.7}$ erg/s, determined by average-stacking their fluxes in the $2$-$10$ keV band at the median $z$ of the sample, is consistent with emission produced by star-formation only, suggesting that in this band the AGN is completely obscured.

In order to assess the level of obscuration, we computed the ratio between the observed and the intrinsic X-ray luminosity L$_{\text{X,obs}}$/L$_{\text{X,intr}}$, comparing this quantity to the total dust attenuation A$_{V,tot}$ inferred in a mixed model from Pa$\beta$ and the bolometric IR luminosity (Paper I). L$_{\text{X,intr}}$ comprises the contribution from both star-formation (as explained in Section \ref{AGN_identification} and from the AGN, assuming L$_{\text{X,intr,AGN}}=0.04\times$ L$_{\text{BOL,AGN}}$ \citep{vasudevan07} and a bolometric AGN luminosity L$_{\text{BOL,AGN}}$ $=1.5\times$ L$_{AGN,IR}$ \citep{elvis94} (Fig. \ref{distltg}-\textit{center}). 
Alternatively, we considered the bolometric correction of \citet{lusso12} for type-2 AGNs, which depends on L$_{\text{BOL,AGN}}$ itself through the following, nonlinear equation:
\begin{equation}
    \text{log}_\text{10}\left(\frac{\text{L}_\text{BOL}}{\text{L}_\text{2-10keV}}\right)=0.217x-0.022x^2-0.027x^3+1.289
\end{equation}
, where x $=$ log$_{10}$(L$_\text{BOL}$) $-12$ and the scatter of the relation is $0.26$ dex.
However, the results derived with this second assumption (Fig. \ref{distltg}-\textit{left}) do not change significantly compared to the first case.

The figures presented above confirm that the total attenuation inferred from the L$_\text{X}$ may be the discriminating parameter between X-ray detected and undetected mid-IR AGNs: while the former are relatively unobscured, the X-ray emission from the second is suppressed at least by a factor of $30$. 
Interestingly, this transition does not seem to be related to an increased bolometric luminosity, since no correlations are observed with this quantity.
We also remind that for this analysis we are following the standard procedure, which does not take into account shock contribution to the X-ray luminosity. A possible non negligible shock emission at these high energies (which in any case is difficult to model) would result in an underestimation of the true effective X-ray attenuation towards the AGN.



The previous results suggest that the X-ray attenuation decreases as the starburst becomes more dust-obscured (probed by A$_\text{V,tot}$) during the last merger phases. In a standard framework (i.e., if we exclude dominant contribution from shocks to the X-ray luminosity), this apparent contradiction can be reconciled by considering the different timescales of our diagnostics.
On the one hand, Pa$\beta$ (used to calculate A$_\text{V,tot}$), yields a luminosity (or, equivalently, a SFR, by applying the \citet{kennicutt94} conversion) that is averaged over a timescale of $20$-$30$ Myr. Conversely, the AGN luminosity that we measure in X-rays gives an istantaneous information of the AGN activity.
As a consequence, with the X-ray analysis we are able to probe the current dust attenuation level, while A$_\text{V,tot}$ traces the obscuration in the recent past ($\lesssim 30 Myr$). According to this speculation, the X-ray luminosities measured for a subset of $6$ late stage mergers indicate that the AGN-induced blow-out may have already started since a few Myr ago, clearing the surrounding gas and dust content, and that we might be very close to the final QSO phase.


Furthermore, in Fig.\ref{distltg2} we display the AGN accretion efficiencies of our galaxies as a function of their distance from the LTGs Mass-size relation.
The efficiencies were estimated by comparing the observed L$_{\text{BOL,AGN}}$/M$_\ast$ ratios to the maximum value allowed by Eddington (L$_{\text{BOL,AGN}}$/M$_\ast$|$_{EDD}\simeq$1.5), from which we derived the so-called Eddington ratio (L/L$_\text{EDD}$)|$_{AGN}$.
We assumed the typical correlation for AGNs between the stellar mass M$_\ast$ and black hole mass M$_{BH}$ of \citet{reines15} and a spherically symmetric accreting BH, yielding log(L$_{EDD}$/M$_\ast$) $\simeq$ 0.9685+0.05 log$_{10}$(M$_\ast$). The M$_\ast$ in the second term can be approximated with the median value of the sample M$_{\ast,\text{median}}$, leaving a small secondary dependence on stellar mass which, for our mass ranges (10$^{10}$-10$^{11}$ M$_\odot$) produces variations of $<5\%$. This variation, added to the uncertainty on the relation between M$_\ast$ and M$_\text{BH}$ reported by \citet{reines15} (cfr. their equations 4 and 5), is highlighted in Fig.\ref{distltg2} with a blue shaded area around the Eddington limit (blue line) calculated above. 

From this analysis, we found that $2$ AGNs have an Eddington ratio higher than 1 ($1.2<$ (L/L$_\text{EDD}$)|$_{AGN}$ $<1.85$), while additional $3$ AGNs are radiating between $57\%$ and $79\%$ of their maximum luminosity. However, all these $5$ AGNs are still consistent within $2\sigma$ errors with (L/L$_\text{EDD}$)|$_{AGN}=1$ if we also consider the uncertainty on the conversion factor from the ratio L$_{\text{BOL,AGN}}$/M$_\ast$, as discussed before. We remark that additional uncertainties on the M$_\ast$- M$_\text{BH}$ relation can come from the assumptions on the BH accretion geometry, which is not taken into account here.
The remaining 7 IR-detected AGNs have instead lower Eddington ratios between $0.35$ and $0.08$. 

The galaxies which are undetected in X, radio and mid-IR may contain low-active AGNs, even though current upper limits on the Eddington ratio are not so stringent and do not allow to discriminate them from the detected subset. 
The intrinsic variability of AGN accretion may thus explain why we are currently missing many of these sources in our sample, and that only deeper X-ray observations can potentially reveal. The duty cycle above $30\%$ and $1\%$ L$_{\text{EDD}}$ seems to be at least $\sim 25\%$ and $\sim 50\%$, respectively.

\section{Summary and conclusions}

Using our unique sample of 25 starburst galaxies (typically 7 times above the star-forming Main Sequence) at z=0.5-0.9 with near-IR rest frame spectroscopy of Paschen lines, we found in Paper I that they span a large range of attenuations toward the core centers from A$_V=2$ to A$_V=30$, forming a sequence which is consistent with a mixed model geometry of dust and stars. In this paper we have investigated the nature of this attenuation sequence, comparing A$_V$ with other physical properties, such as the radio size (which traces the extension of the starburst), the emission lines velocity widths and [\ion{N}{II}]/H$\alpha$ ratios (which reflect the increasing potential well depth and likely shock contribution towards the final merger stages), and finally the EW of hydrogen absorption lines, which is sensitive to the luminosity-weighted age of the stellar populations surrounding the optically obscured core.

We summarize the main results of this paper as follows:
\begin{itemize}
    \item We found that the physical quantities introduced above, namely the radio sizes (FWHM$_{\text{radio}}$), the line velocity widths (FWHM$_{\text{line}}$), the [\ion{N}{2}]/H$\alpha$ ratios (N2) and the equivalent widths of Paschen/Balmer lines (EW$_{Balmer,Paschen}$), all correlate each other (Fig.\ref{correlations}), defining a one-parameter sequence of z$\sim$0.7 starburst galaxies.
    \item These correlations can be interpreted as a time-evolutionary sequence of merger stages. As the merger evolves, the starburst becomes more compact and dust obscured, while the deep potential wells created by merging nuclei produce, according to the Virial Theorem, an increase of the kinetic energy and shocks in the system. At the same time, intermediate aged A-type stars in the outer starburst core regions are primarily responsible for the stronger optical+near-IR absorption lines in later phases. 
    \item 4 galaxies are outliers simultaneously in 3 of the 10 main correlations, which involve A$_\text{V,tot}$ and, respectively, N2, FWHM$_\text{line}$ and EW(Pa$\beta$). Having the largest dust attenuations and among the smallest radio sizes in our sample, these outliers may represent the very end phases of the merger evolution, where the above 3 relations may reach a saturation level. 
    \item Using sky-subtracted 2D spectra, we identified a subset of $7$ pre-coalescence mergers by the presence of spatially separated or kinematically detached (i.e., rotation-driven tilted lines with different inclination angles) H$\alpha$ components, representing earlier, less obscured phases of the interaction. The radio sizes measured for these systems are likely tracing the separation between the merging nuclei rather than the dimensions of single cores. However, our sample may contain additional double nuclei which we are not currently able to resolve.
    \item Half of our sample comprises extremely compact starbursts, with average half-light radii of $600$ pc ($6$ galaxies have only upper limits), similar to the sizes of starbursting cores observed in local ULIRGs. These sizes are also $\sim 0.5$ dex smaller than ETGs at redshift $\sim$0.7 and below, indicating that our merger-driven starbursts cannot be direct progenitors of the population of massive ellipticals formed in the last $\sim 7$ Gyr. On the contrary, they are more consistent with typical sizes and masses of bulge structures \citep{graham08}, suggesting a possible evolutionary connection between our starburst cores and bulges. 
    \item In our sample, we detected at $>3\sigma$ the mid-IR dusty torus AGN emission in $12$ starbursts, with Eddington ratios ranging from $1.9$ to less than $0.08$. Among them, only 6 galaxies are simultaneously detected (at $3\sigma$) in X-rays. 
    Intriguingly, the latter have the largest departures from the Mass-size relation of LTGs (at $z\leq$ 0.7), suggesting that AGNs start to appear in X-rays during the latest (compact) merger phases, as the blow-out of surrounding dust/gas may precede a possible final QSO.  
\end{itemize}
Overall, the relations among the above physical parameters converge toward a time-evolutionary sequence of merger stages, which represents an observational evidence (translated at higher redshift) of the theoretical merger-induced starbursts framework of \citet{hopkins08a,hopkins08c,dimatteo05}, and the evolutionary sequence postulated by \citet{toomre72}. The future advent of JWST will allow to test this scenario up to very high redshift, where the conditions of the Universe and gas content of galaxies were even different compared to the epochs studied here. 

\smallskip

\begin{acknowledgements}
We thank the anonymous referee for useful suggestions that improved the quality of this manuscript, G.Rudie for assistence with Magellan observations, Nicol{\'a}s Ignacio Godoy for data reduction, and Daniela Calzetti for discussions. M.O. acknowledges support from JSPS KAKENHI Grant Number JP17K14257. N.A. acknowledges support from the Brain Pool Program, funded by the Ministry of Science and ICT through the Korean National Research Foundation (2018H1D3A2000902). M.B. acknowledges FONDECYT regular grant 1170618. R.C. acknowledges financial support from CONICYT Doctorado Nacional No. 21161487 and CONICYT PIA. ACT172033. A.C. acknowledges RadioNet conference funding. This research has made use of the zCosmos database, operated at CeSAM/LAM, Marseille, France.
\end{acknowledgements}


\begin{sidewaystable*}
\centering
\Large{Main physical quantities derived for our Magellan sample of starbursts}
\rule{0cm}{1.cm}
\centering
\resizebox{\textwidth}{!}{%
\centering
\rule{+1cm}{0.cm}
\renewcommand{\arraystretch}{1.}
\begin{tabular}{|l|l|l|l|l|l|l|l|l|l|l|l|l|}
\hline
& & & & & & & & & & & & \\
\textbf{ID} & \textbf{\Large{z$_\text{spec}$}} & \textbf{A$_\text{V,tot}$} & \textbf{FWHM$_{\text{radio}}$} & \textbf{FWHM$_{\text{line}}$} & \textbf{EW(H$\alpha$)} & \textbf{EW(Pa$\beta$)} & \textbf{EW(H$\delta$)} & \textbf{[\ion{O}{III}]5007} & \textbf{H$_\beta$} & \textbf{log$_{10}$(L$_{\textbf{bol,AGN}}$)} & $\bm{\sigma}_\textbf{\tiny{AGN,IR}}$ & \textbf{M$_{\text{type}}$} \\
 & & mag & kpc & km/s & $\AA$ & $\AA$ & $\AA$ & erg cm$^{-2}$s$^{-1}$ & erg cm$^{-2}$s$^{-1}$ & L$_\odot$ & & \\
 \hline
(1) & (2) & (3) & (4) & (5) & (6) & (7) & (8) & (9) & (10) & (11) & (12) & (13) \\
\hline
\textbf{245158}	 & 0.5172  & 1.9 $\pm$ 0.1  &	14.43	$\pm$	0.24 $\ddagger$	&	378.1	$\pm$	20.4	&	91.1	$\pm$	11.3	&	33.1	$\pm$	3.6	&	-- 	&	--  	&	--   	&	<11.59	& <3	&	S, m$^\ast$ \\	
\textbf{470239}	 & 0.6609  & 6.2 $\pm$ 0.2  &	7.72	$\pm$	0.26	&	372.0	$\pm$	6.7	    &	81.1	$\pm$	8.6	    &	24.5	$\pm$	2.4	&	2.0	$\pm$	1.6	    &	17.7	$\pm$	5.7	&	22.7	$\pm$	2.3	&	<10.13	            & <3	&	MIII \\
\textbf{493881}	 & 0.6039  & 2.7 $\pm$ 0.2  &	8.81	$\pm$	0.26	&	347.4	$\pm$	3.5	    &	103.2	$\pm$	9.6	    &	44.4	$\pm$	9.6	&	0.3	$\pm$	1.3	    &	23.1	$\pm$	3.7	&	59.6	$\pm$	2.6	&	<11.38	            & <3	&	S, m \\
\textbf{578239}	 & 0.5578  & 11.7 $\pm$ 0.3  &	1.29	$\pm$	0.24 $\dagger$	&	514.0	$\pm$	21.5	&	68.1	$\pm$	6.3	    &	14.8	$\pm$	0.9	&	-2.6	$\pm$	1.7	&	50.1	$\pm$	5.9	&	13.5	$\pm$	2.9	&	10.94 $\pm$ 0.14	        & 3.0    	&	no-HST \\
\textbf{635862}	 & 0.5508  & 6.8 $\pm$ 0.6  &	3.43	$\pm$	0.25	&	971.4	$\pm$	32.6	&	31.2	$\pm$	5.0   	&	8.0	$\pm$	1.9	    &	-2.5	$\pm$	2.3	&	15.0	$\pm$	4.0	&	8.4	$\pm$	1.8	    &	11.22	$\pm$	0.09	& 4.9 &	m      \\
\textbf{685067}	 & 0.3735  & 10.9 $\pm$ 0.3  &	1.06	$\pm$	0.19 $\dagger$	&	537.2	$\pm$	35.7	&	12.1	$\pm$	1.8	    &	4.2	$\pm$	0.2	    &	-5.9	$\pm$	0.5	&	8.8	$\pm$	4.1	    &	--               	&	<10.76	          & <3  	&	MV     \\
\textbf{777034}	 & 0.6889  & 12 $\pm$ 0.4  &	1.25	$\pm$	0.26	&	641.7	$\pm$	25.2	&	83.8	$\pm$	10.0	&	20.3	$\pm$	1.8	&	-4.2	$\pm$	2.3	&	43.6	$\pm$	4.9	&	21.7	$\pm$	1.6	&	12.02	$\pm$	0.04 & 9.7	&	MIV$^\ast$ \\ 
\textbf{862072}	 & 0.6811  & 7.8 $\pm$ 0.4  &	2.98	$\pm$	0.30	&	561.2	$\pm$	15.2	&	55.8	$\pm$	6.0	    &	15.6	$\pm$	2.1	&	-7.5	$\pm$	1.2	&	--               	&	3.7	$\pm$	0.8   	&	<10.91	            & <3	&	m, S$^\ast$ \\ 
\textbf{545185}	 & 0.5337  & 22.5 $\pm$ 1.0  &	<0.70		            &	345.6	$\pm$	11.7	&	33.8	$\pm$	4.8	    &	20.5	$\pm$	2.6	&	--                	&	--               	&	--                	&	<10.56	            & <3	&	MIII$^\ast$ \\
\textbf{222723}	 & 0.5254  & 8.1 $\pm$ 0.5  &	3.38	$\pm$	0.23	&	168.3	$\pm$	3.9	    &	52.4	$\pm$	4.0	    &	13.5	$\pm$	2.6	&	--                	&	--               	&	--                	&	11.07	$\pm$	0.07 & 5.8	&	MV$^\ast$ \\ 
\textbf{223715}	 & 0.5174  & 3.6 $\pm$ 0.2  &	10.95	$\pm$	0.28	&	237.0	$\pm$	6.3	    &	54.9	$\pm$	6.0	    &	16.3	$\pm$	4.3	&	--                	&	--               	&	--                	&	<11.47	             & <3	&	m, S$^\ast$ \\ 
\textbf{249989}	 & 0.6656  & 4.4 $\pm$ 0.2  &	5.32	$\pm$	0.30	&	261.2	$\pm$	10.1	&	206.3	$\pm$	26.0	&	38.5	$\pm$	6.8	&	--                	&	--               	&	--                	&	10.66	$\pm$	0.13  & 3.3	&	MIII$^\ast$	\\
\textbf{326384}	 & 0.8042  & 9.8 $\pm$ 1.1  &	2.78	$\pm$	0.29	&	321.7	$\pm$	7.8	    &	174.2	$\pm$	23.7	&	36.8	$\pm$ 11.4  &	--                	&	--               	&	--                	&	<10.55	            & <3	&	S$^\ast$ \\
\textbf{345018}	 & 0.7521  & 23.7 $\pm$ 3.4  &	<0.86		            &	286.4	$\pm$	11.9	&	82.3	$\pm$	8.2	    &	18.9	$\pm$	5.5	&	-7.2	$\pm$	1.1	&	16.7	$\pm$	4.4	&	12.0	$\pm$	1.8	&	<10.81	            & <3	&	m, S$^\ast$ \\
\textbf{412250}	 & 0.8397  & 6.7 $\pm$ 0.6  &	8.84	$\pm$	0.31 $\ddagger$	&	299.1	$\pm$	6.5	    &	244.7	$\pm$	21.9	&	18.2	$\pm$	5.4	&	--                	&	--               	&	--                	&	11.14	$\pm$	0.09 & 4.9	&	MIII$^\ast$	\\
\textbf{466112}	 & 0.7607  & 5.2 $\pm$ 0.3  &	2.64	$\pm$	0.28	&	298.1	$\pm$	3.7	    &	229.1	$\pm$	17.6	&	57.8	$\pm$	9.6	&	-3.9	$\pm$	1.1	&	34.0	$\pm$	4.0	&	9.0	$\pm$	1.6	    &	11.42	$\pm$	0.06 & 7.3	&	MIII$^\ast$	\\
\textbf{472775}	 & 0.6604  & 18.0 $\pm$ 1.3  &	<1.05	  &	498.3	$\pm$	27.9	&	65.7	$\pm$	6.3	    &	17.3	$\pm$	2.7	&	-- 	&	--     	&	--    	&	<10.52	 & <3	&	MIII  \\
\textbf{519651}	 & 0.6709  & 8.8 $\pm$ 0.5  &	10.94	$\pm$	0.26 $\ddagger$	&	191.1	$\pm$	3.6	    &	86.2	$\pm$	7.4	    &	32.0	$\pm$	5.0	&	-2.9	$\pm$	1.3	&	38.9	$\pm$	7.1	&	27.3	$\pm$	1.8	&	11.37	$\pm$	0.06 & 7.3	&	MIV	  \\
\textbf{668738}	 & 0.7481  & 6.7 $\pm$ 0.2  &	3.68	$\pm$	0.28	&	484.2	$\pm$	22.3	&	92.0	$\pm$	13.3	&	24.0	$\pm$	3.4	&	--   	&	28.0	$\pm$	4.4	&	11.3	$\pm$	3.0	&	10.87	$\pm$	0.13 & 3.3	&	MIII$^\ast$ \\	
\textbf{894779}	 & 0.5506  & 4.5 $\pm$ 0.2  &	2.41	$\pm$	0.26	&	132.0	$\pm$	10.5	&	140.8	$\pm$	21.4	&	35.4	$\pm$	6.2	&	--    	&	--   	&	--     	&	10.66	$\pm$	0.07 & 5.8	&	MIV$^\ast$ \\
\textbf{911723}	 & 0.6606  & 9.1 $\pm$ 0.6  &	<0.88 	&	440.8	$\pm$	10.3	&	60.5	$\pm$	6.8	    &	23.9	$\pm$	4.5	&	--  	&	--  	&	-- 	&	10.73	$\pm$	0.12 & 3.7	&	E$^\ast$  \\
\textbf{303305}	 & 0.5306  &  >13.9   &	<0.71	&	877.7	$\pm$	85.1	&	11.0	$\pm$	2.4	    &	--  &	-9.3	$\pm$	2.6	&	11.6	$\pm$	1.3	&	--  	&	<11.30	 & <3	&	E  \\
\textbf{500929}	 & 0.9498  & >6.2   &	2.99	$\pm$	0.32	&	547.6	$\pm$	48.0	&	42.3	$\pm$	10.8	&	--	                &	--                	&	--               	&	--               	&	<11.26	            & <3	&	MIII$^\ast$	\\
\textbf{893857}	 & 0.8512  &  >11   &	<0.93	            	&	655.1	$\pm$	29.6	&	39.6	$\pm$	2.5	    &	--	                &	--                	&	--               	&	-               	&	<11.08	            & <3	&	E$^\ast$  \\  
\textbf{232171}	 & 0.5251  &  >17.6   &	1.62	$\pm$	0.23	&	374.4	$\pm$	33.8	&	11.3	$\pm$	3.3	    &	--	                &	--                	&	--               	&	--                	&	11.54	$\pm$	0.04 & 9.7	&	MII$^\ast$ \\
\hline
\end{tabular} 

}
\hfill
\caption{\normalsize Table columns: (1) Identification number from \citet{laigle16}, as in \citet{calabro18}; (2) Spectroscopic redshift inferred from Magellan spectra; (3) Total attenuation A$_{\text{V,tot}}$ towards the center in a mixed model geometry, calculated in \citet{calabro18} and explained in Section \ref{methodology} ; (4) FWHM size in radio $3$ GHz band (from VLA-COSMOS); $\dagger$: fitted with a single 2D Gaussian but with fixed axis ratio and position angle (1 and 0); $\ddagger$: fitted with a Double 2D Gaussian, we give here the total FWHM (average of single component sizes + separation between the two); we put a $3\sigma$ upper limit for unresolved sources, while the remaining starbursts are fitted with a single 2D Gaussian and free parameters ; (5) line velocity width of emission lines, derived from fitting the Magellan spectra with MPFIT (for double Gaussians, this quantity is the sum of the single Gaussian FWHM and the separation in velocity between the two peaks); (6,7,8) Observed equivalent width of H$\alpha$,Pa$\beta$ and H$\delta$, the latter coming from zCOSMOS or SDSS optical spectra; (9,10) Fluxes of [\ion{O}{III}]$\lambda$5007$\AA$ and H$\beta$ (in units of $10^{-17}$) inferred from optical spectra. H$\beta$ fluxes have been corrected for underlying absorption assuming EW$_{abs}=$5$\AA$ as determined from \citet{bruzual03} synthetic stellar spectra; (11) Bolometric AGN luminosity derived as $1.4\times$L$_{AGN,IR}$, the latter being the AGN luminosity in the infrared inferred from SED fitting (see \citet{liu18} and \citet{jin18} for the methodology); (12) Significance of mid-IR dusty-torus detection from SED fitting ($<3$ means that it is not detected); (13) Morphological type following the visual criteria of \citet{kartaltepe10} (objects with a $\ast$ have been already classified in the same paper). We remind that the coordinates of our targets, their stellar masses, SFRs and H$\alpha$ (Pa$\beta$) flux measurements can be retrieved from \citet{calabro18}.}\label{table2}
\end{sidewaystable*}

\clearpage

\appendix

\section{Additional plots}\label{additional_plots}

\begin{figure}[h!]
    \centering
    \includegraphics[angle=0,width=\linewidth,trim={0.cm 0.4cm 0cm 1.9cm},clip]{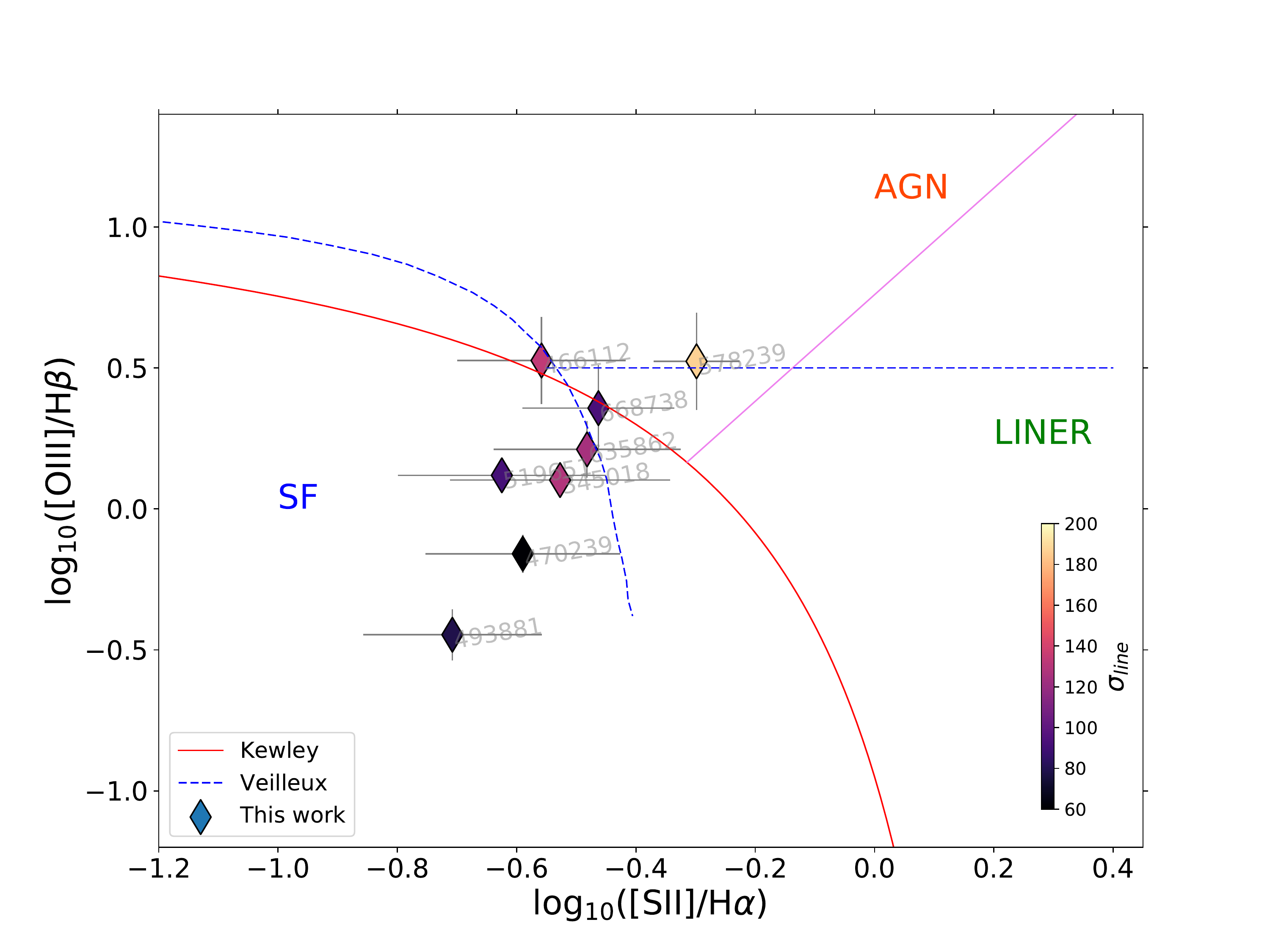}
    \includegraphics[angle=0,width=\linewidth,trim={0.cm 0.cm 0cm 1.9cm},clip]{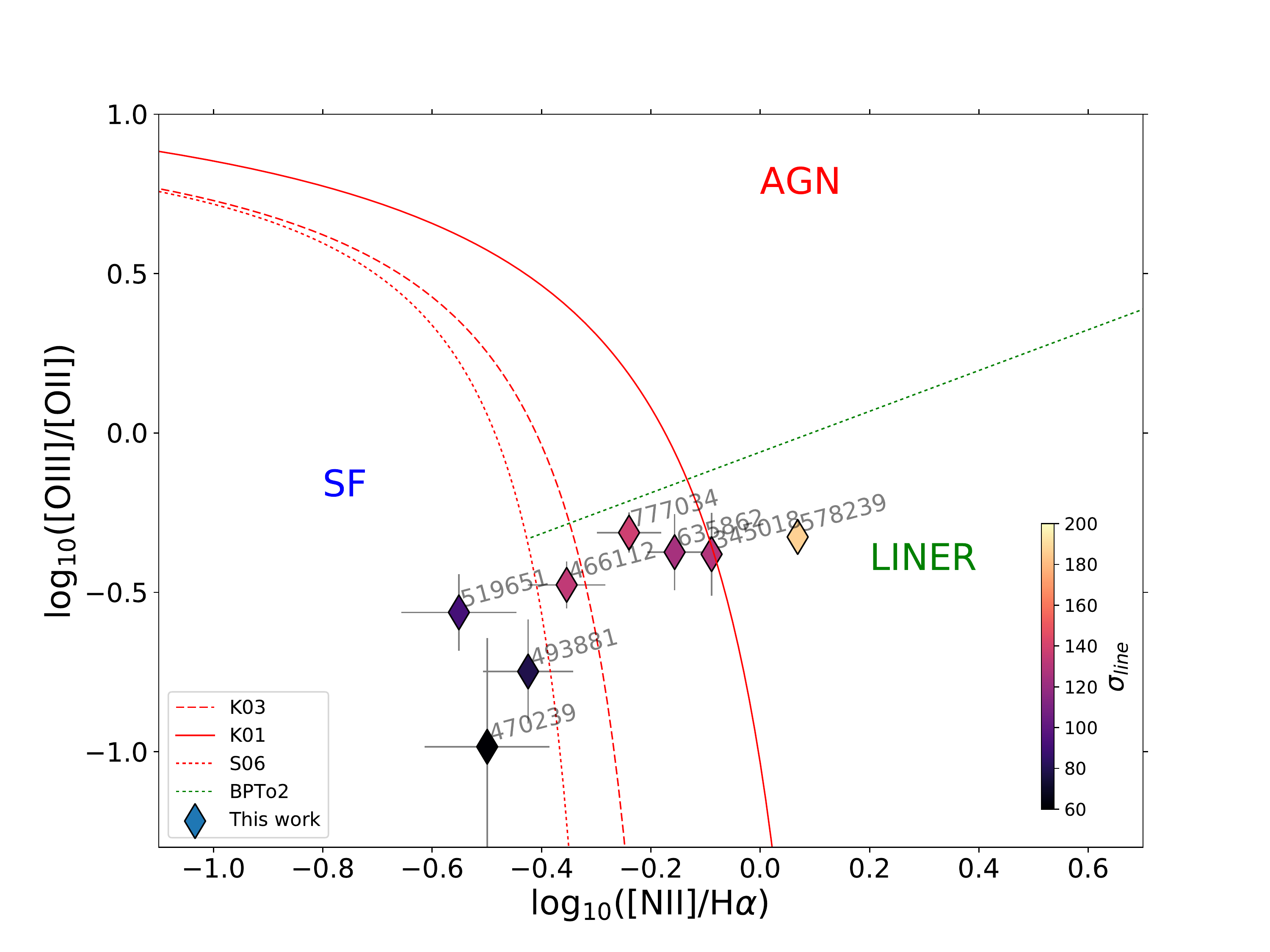}
    \caption{\small Diagram of [\ion{O}{III}]/H$\beta$ compared to [\ion{S}{II}]6717+6731/H$\alpha$ (\textit{Top}) and [\ion{O}{III}]5007/[\ion{O}{II}]3727+3729 vs. [\ion{N}{II}]6583/H$\alpha$ (\textit{bottom}) for $8$ galaxies of our sample with optical spectra available and contemporary detection of [\ion{O}{III}] and one between [\ion{S}{II}]+H$\beta$ or [\ion{O}{II}].  The two diagrams are also called the S2 and the O2 BPT diagrams, respectively. As in the classical BPT diagram in Fig. \ref{BPT1}, they show that our starbursts have different line excitation properties, and those with higher line velocity widths are generally shifted towards the AGN/LINER regions, according to empirical separation lines derived in the local ($z\lesssim0.3$) Universe. }\label{BPTdiagram}
\end{figure}

We display in Fig. \ref{BPTdiagram} two different versions of the BPT diagram, involving the [\ion{S}{II}]6717+6731/H$\alpha$ line ratio (top panel), and the [\ion{O}{III}]5007/[\ion{O}{II}]3727+3729 ratio (bottom panel), color coded according to the total velocity width of the lines ($\sigma_{line}$).  

In Fig.\ref{SEDfitting} we show the SED fitting of all the $25$ starbursts in our sample, performed as explained in Section \ref{sample_selection}. The majority of our galaxies are fitted with a GN20 template from \citet{magdis12}. However, for six objects (ID 245158, 223715, 470239, 493881, 635862 and 862072), a MS template gives a better $\chi^2$ than the GN20 SED. This is due to the fact that this particular template is not universal for all starbursts, and the true SB SED can have multiple dust temperatures. 

Finally, we display in Fig.\ref{EW_EW} the correlations among H$\delta$, H$\beta$, H$\alpha$ and Pa$\beta$ equivalent widths.

\begin{figure*}[ht!]
    \centering
    \includegraphics[angle=0,width=0.98\linewidth,trim={0cm 9.6cm 10cm 0.cm},clip]{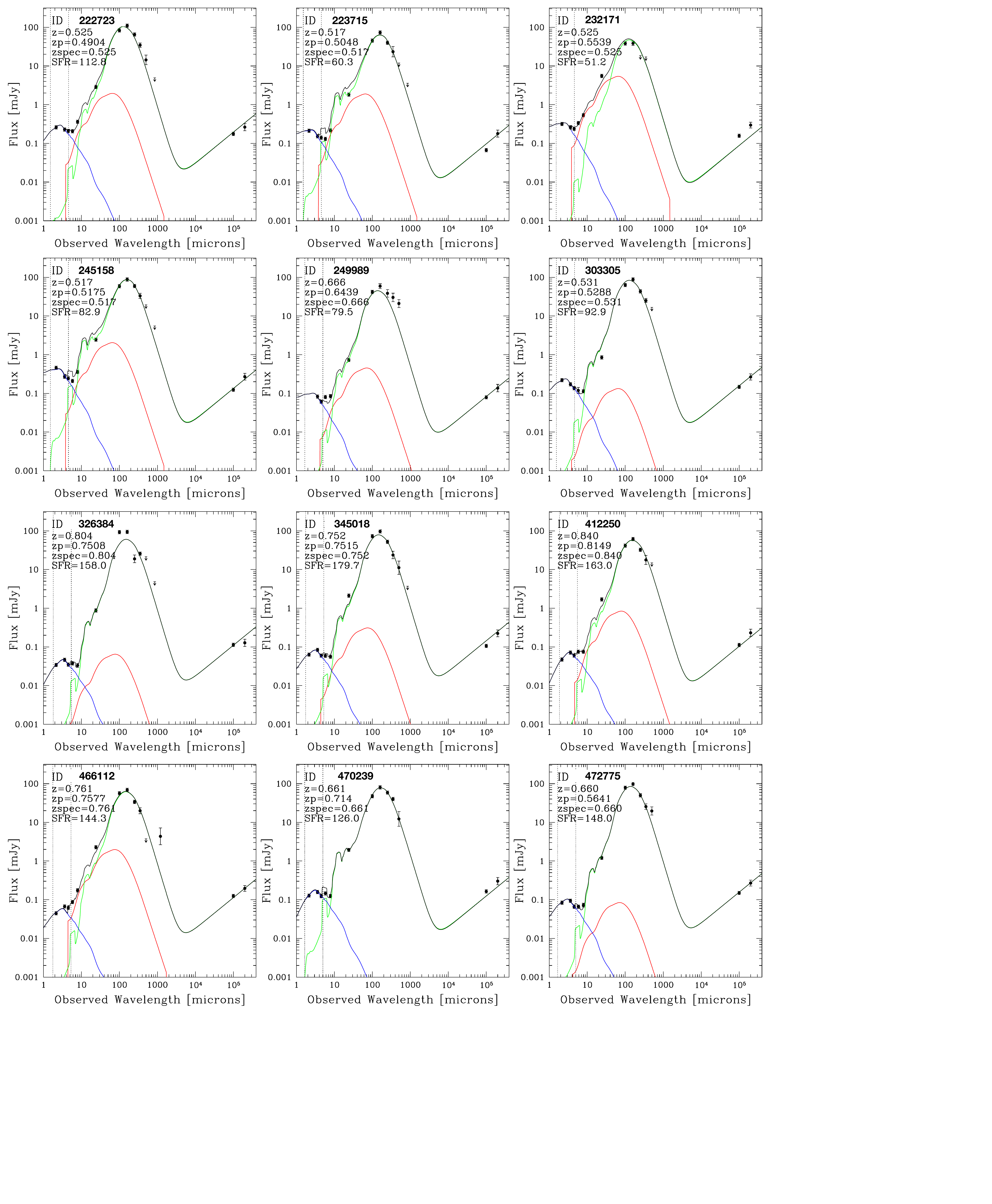}
    \caption{}
\end{figure*}    

\begin{figure*}[ht!]
    \centering
    \includegraphics[angle=0,width=0.98\linewidth,trim={0cm 9.6cm 10cm 0.cm},clip]{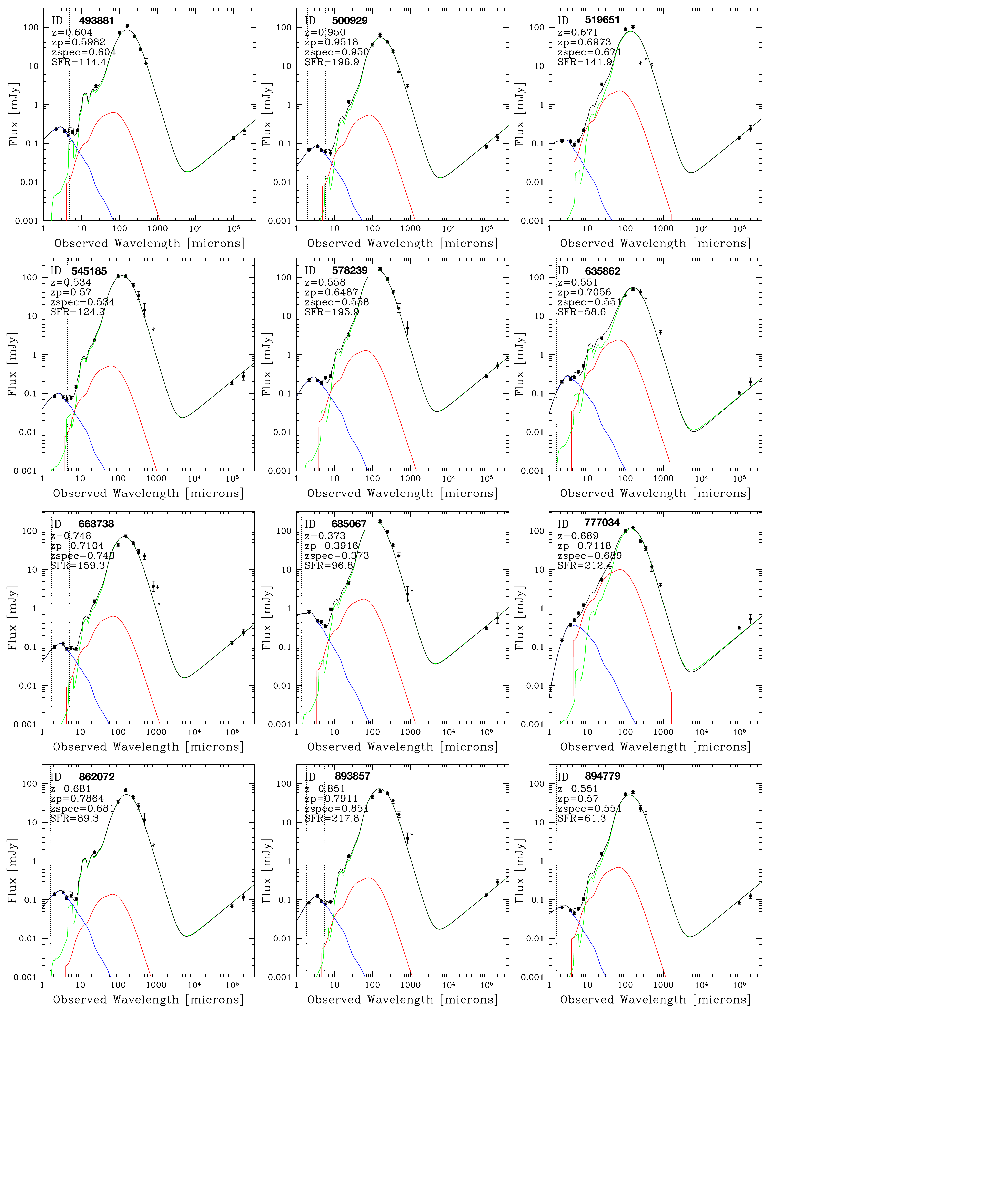}
    \caption{}
\end{figure*}   

\begin{figure*}[ht!]
    \centering
    \includegraphics[angle=0,width=0.98\linewidth,trim={0cm 47cm 10cm 0.cm},clip]{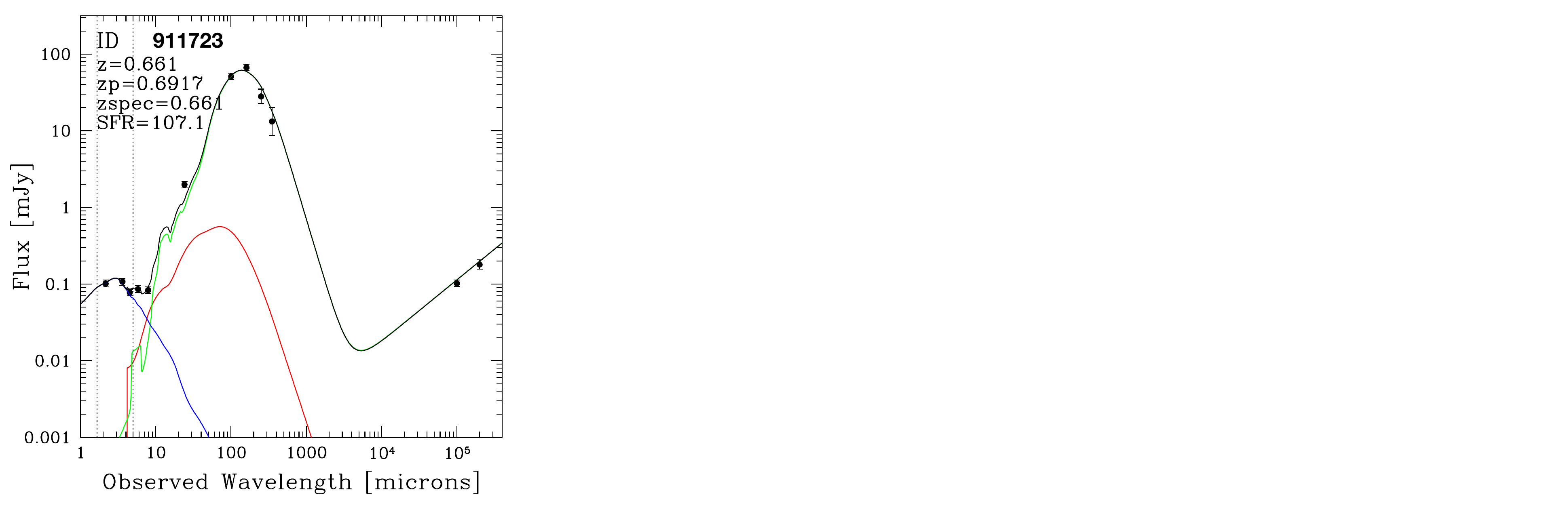} 
    \vspace{-0.2cm}
    \caption{\small SED fitting of the $25$ starburst galaxies studied in this paper, presented in numerical order. The SED fitting procedure is described in Section~2.1. Blue and red curves show the stellar component \citep{bruzual03} and the AGN torus emission \citep{mullaney11}, while the dust continuum emission is shown in green \citep{magdis12}. Downward arrows show the 2$\sigma$ upper limit photometry at given wavelength. 
    In each panel, $z_\text{spec}$ is the spectroscopic redshift from our Magellan \textbf{spectra}, $z_{phot}$ is the photometric redshift from \citet{laigle16}, $z$ is the output redshift of the best-fit SED (which in our cases was fixed to the spectroscopic value), and SFR is the best-fit infrared star-formation rate. The vertical dotted lines indicates the optimized range for fitting the stellar SED only (in order to avoid contamination from AGN torus and dust emission). The rightmost line also indicates the starting wavelength for fitting the AGN + dust SED in our paper. We remind the reader that the radio photometry (at $1.4$ and $3$ GHz) was not used in the fit, as explained in Section \ref{sample_selection}.  
    }\label{SEDfitting}
\end{figure*}

\begin{figure*}[ht!]
    \bigskip
    \bigskip
    \centering
    \includegraphics[angle=0,width=6.5cm,trim={0.45cm 0.cm 0.cm 0.4cm},clip]{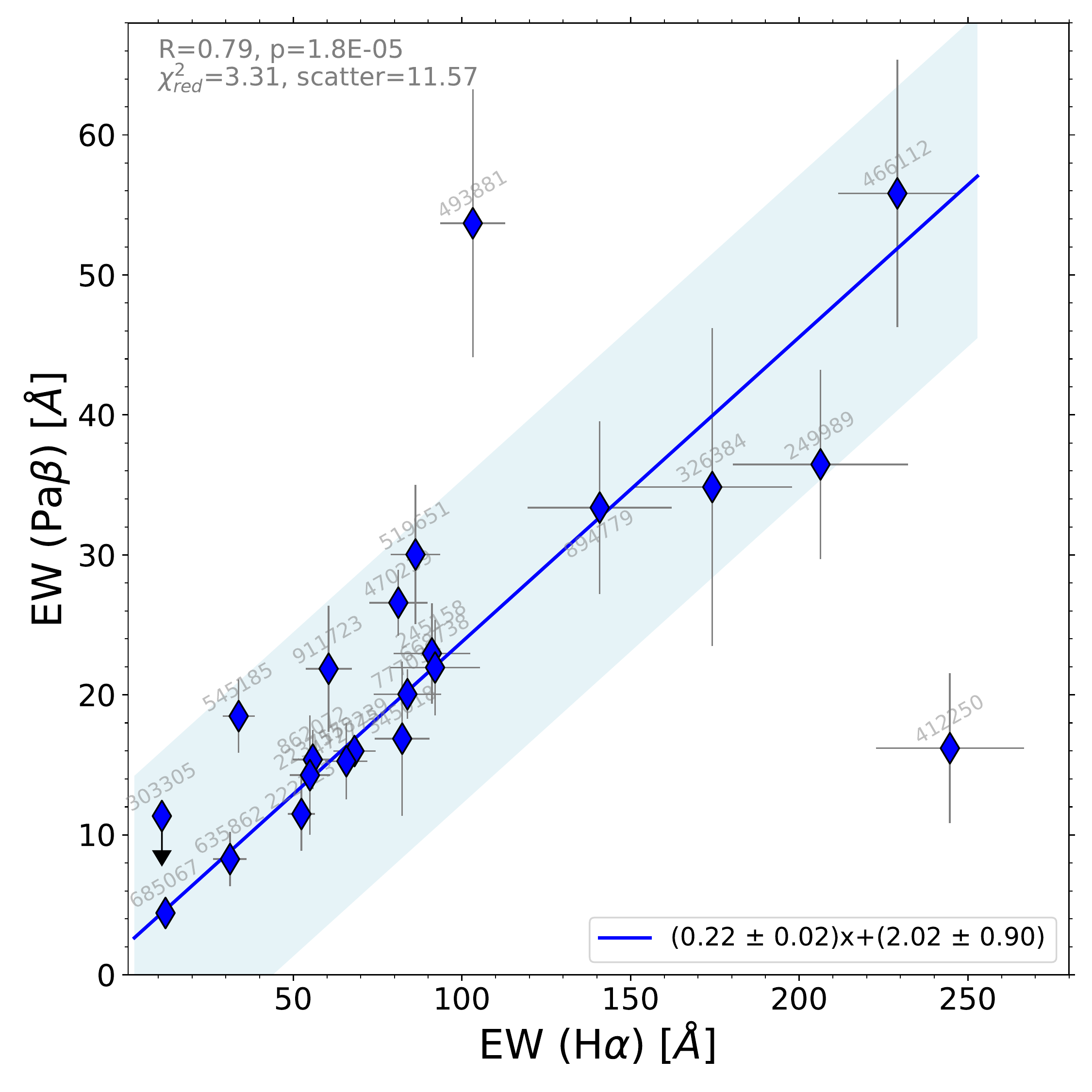}
    \includegraphics[angle=0,width=6.5cm,trim={0.15cm 0.cm 0.4cm 0.4cm},clip]{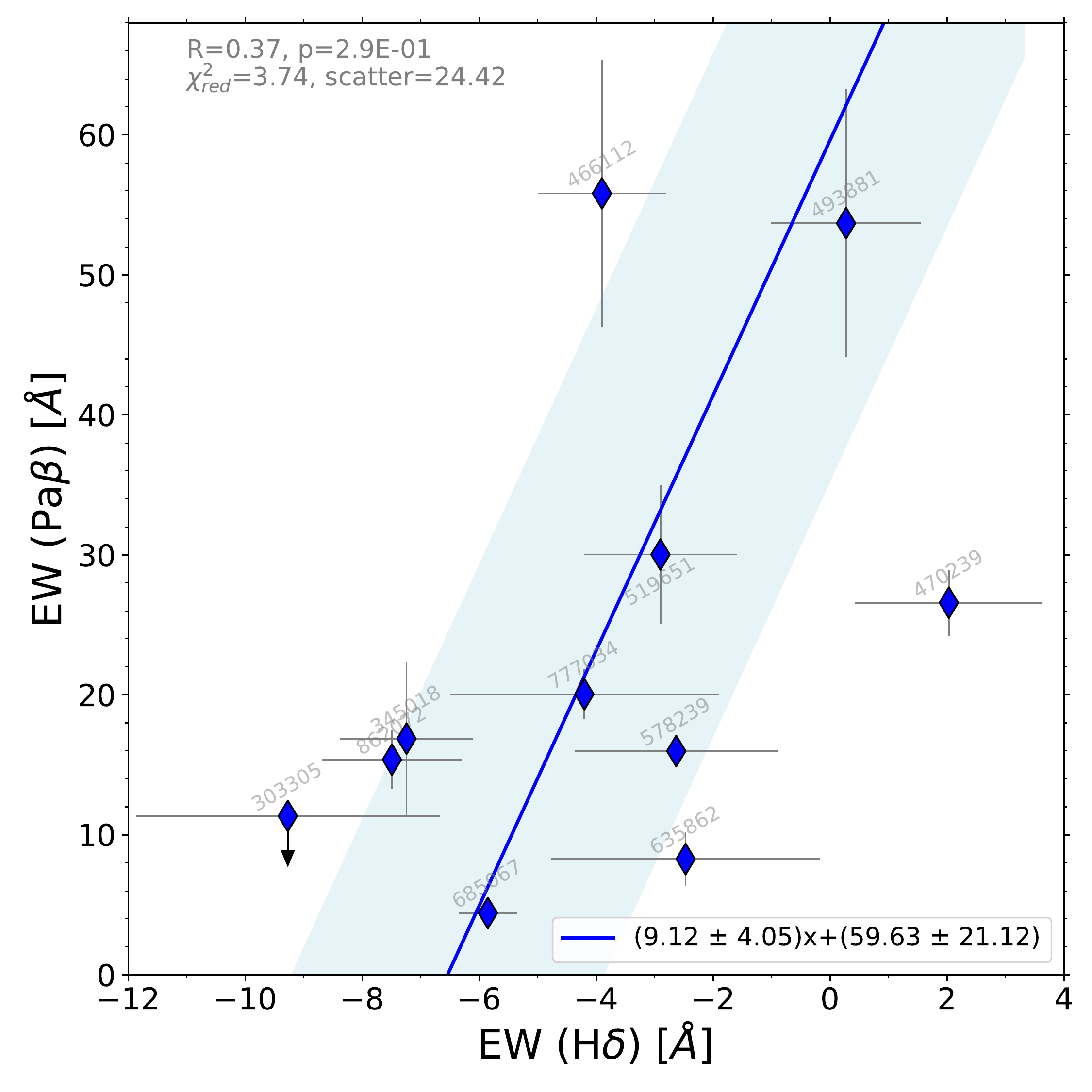}
    \includegraphics[angle=0,width=6.5cm,trim={0.45cm 0.cm 0.cm 0.3cm},clip]{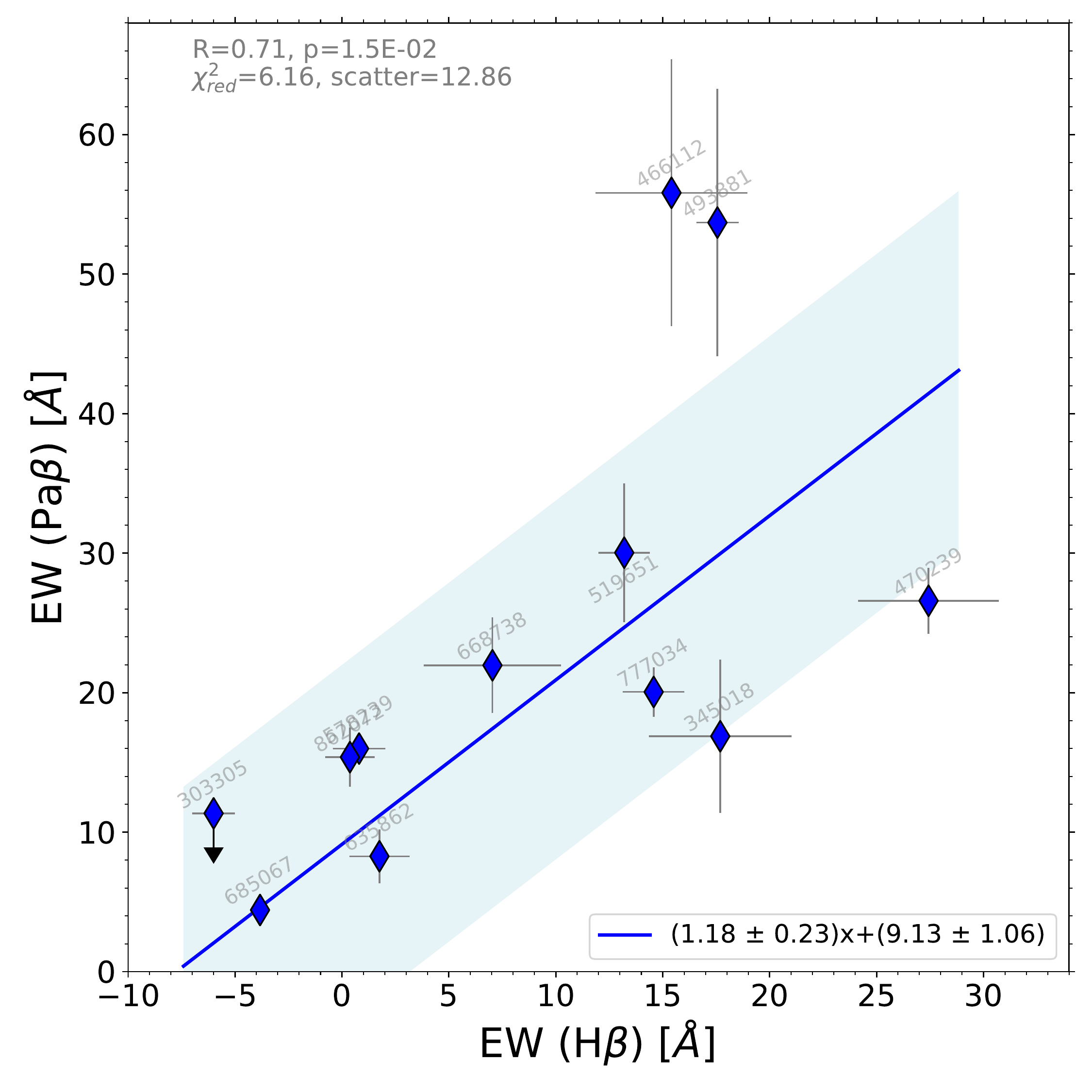}
    \includegraphics[angle=0,width=6.5cm,trim={0.15cm 0.cm 0.4cm 0.3cm},clip]{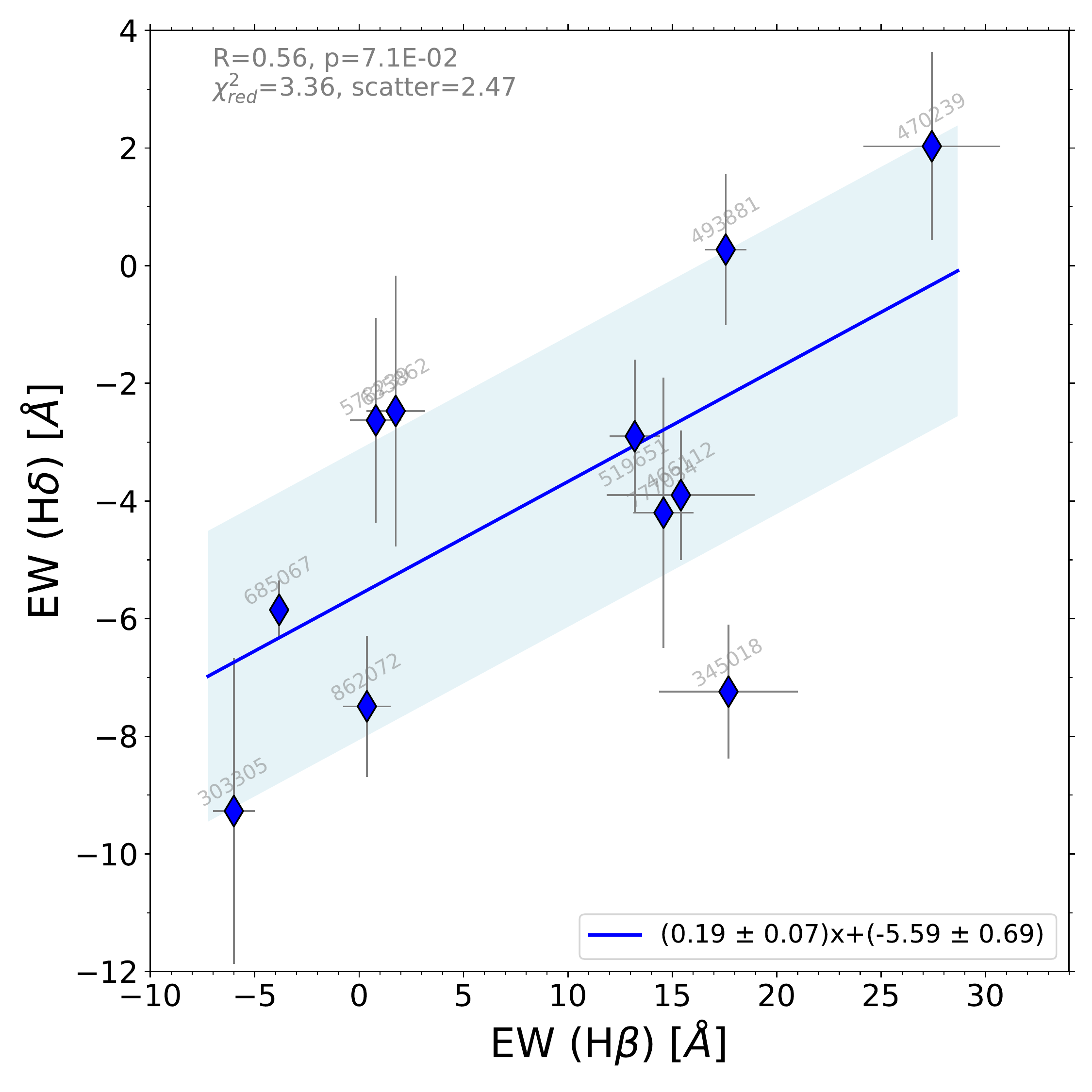}
    \vspace{-0.2cm}
    \caption{\small Comparison between the EWs of H$\delta$, H$\beta$, H$\alpha$ and Pa$\beta$ lines. The blue lines represent the best-fit linear relation, while the blue shaded areas indicate the $\pm1\sigma$ dispersion of the galaxies around the best-fit line. We display in top-left corner of each panel the Spearman correlation coefficient `R' and corresponding `p'-value, the residual $\chi^2$ of the correlation (`$\chi_\text{red}^2$') and the $1\sigma$ scatter around the best-fit line, whose equation (comprising $1\sigma$ errors on the best-fit coefficients) is shown in the bottom-right corner. In the four diagrams, all the points with available measurements (excluding upper limits) were used in the fit.}\label{EW_EW}
\end{figure*}

\section{H$\alpha$ and Pa$\beta$ emission line fits}\label{emissionlinefits}

\begin{figure*}[ht]
  \centering
  \includegraphics[angle=0,width=0.9\linewidth,trim={0cm 12cm 0cm 0cm},clip]{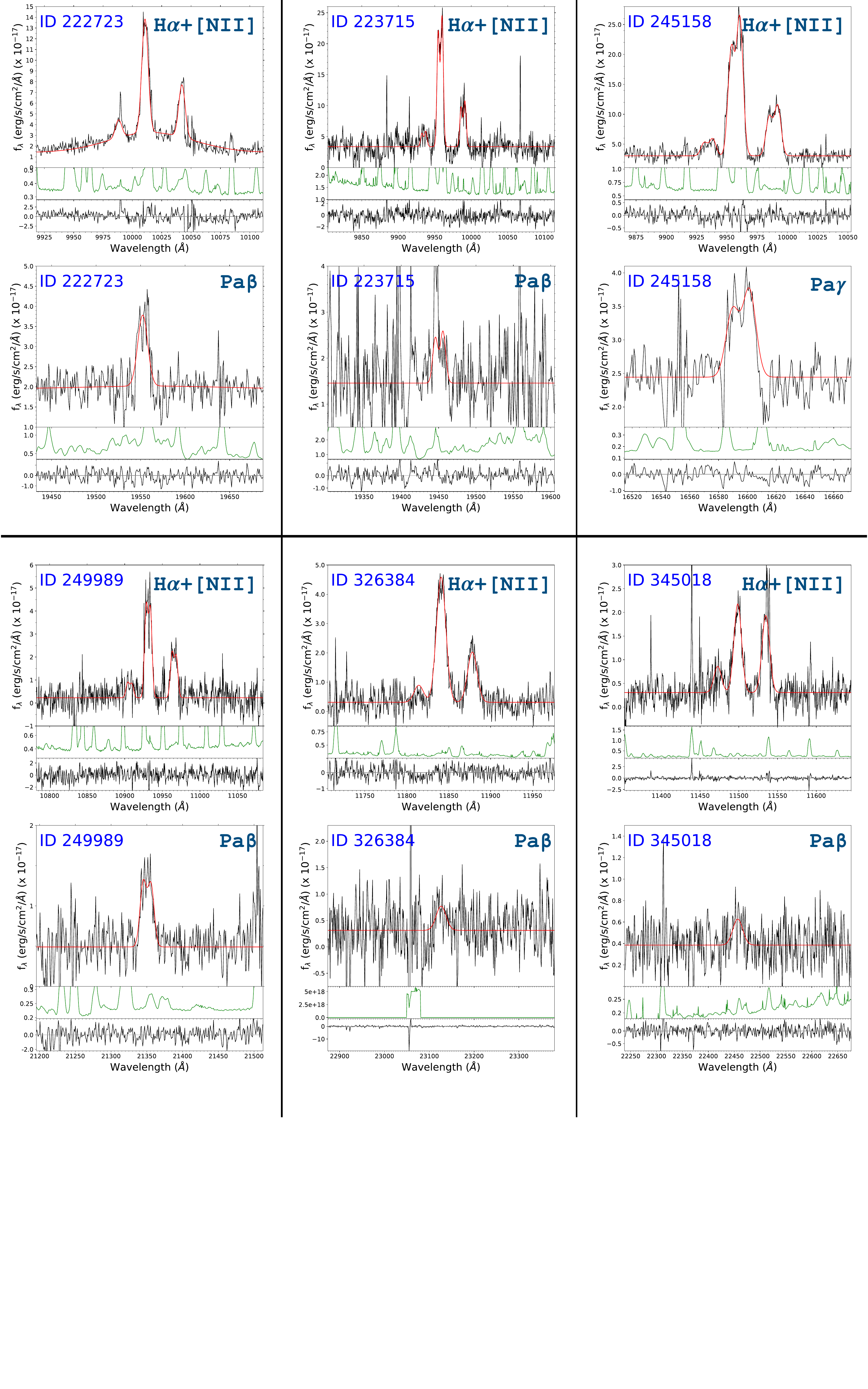} 
  \caption{}\label{spectra1}
\end{figure*}

\begin{figure*}[ht]
  \centering
  \includegraphics[angle=0,width=0.9\linewidth,trim={0cm 12cm 0cm 0cm},clip]{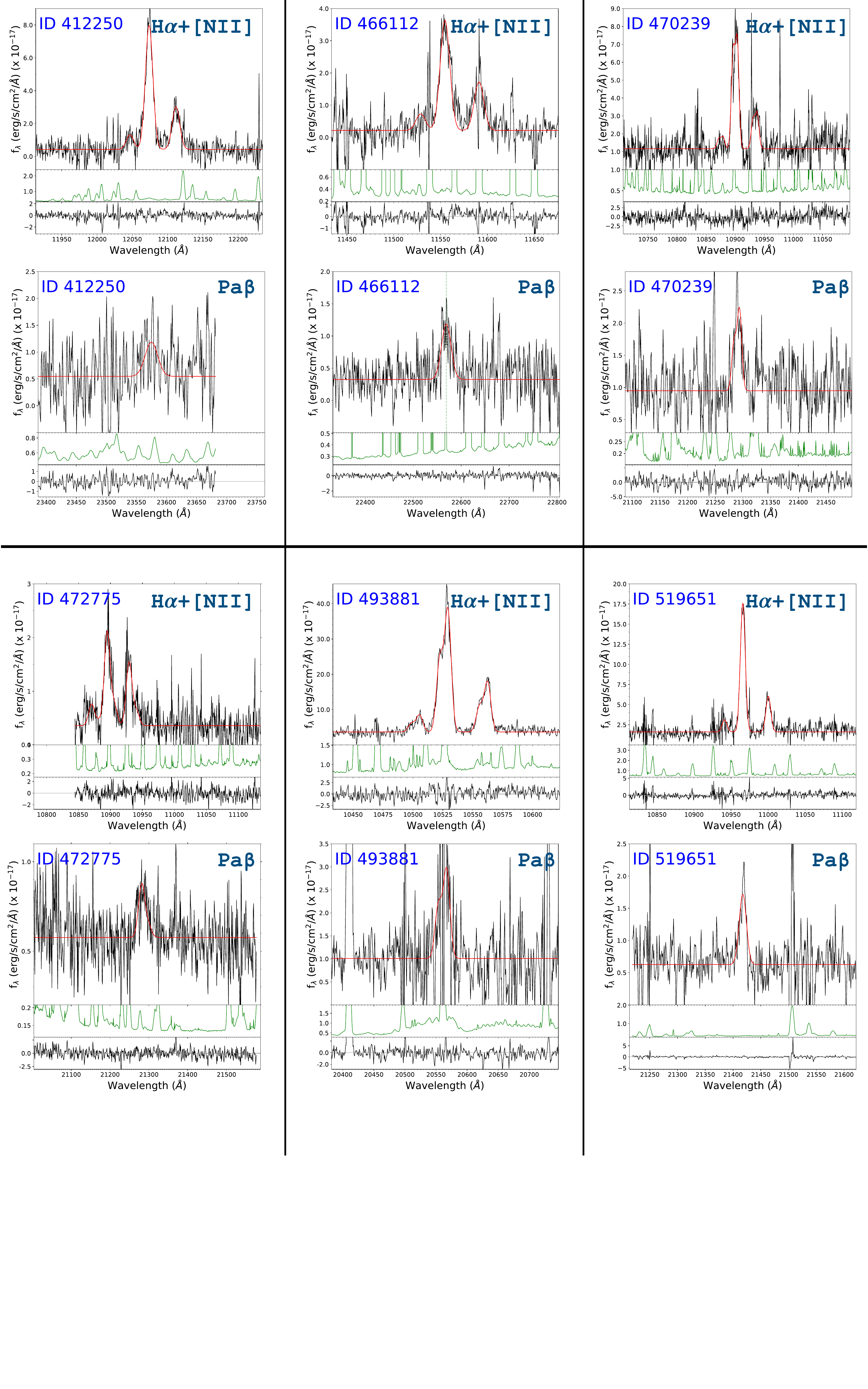}
  \caption{}\label{spectra2} 
\end{figure*}

\begin{figure*}[ht]
  \centering
  \includegraphics[angle=0,width=0.9\linewidth,trim={0cm 12cm 0cm 0cm},clip]{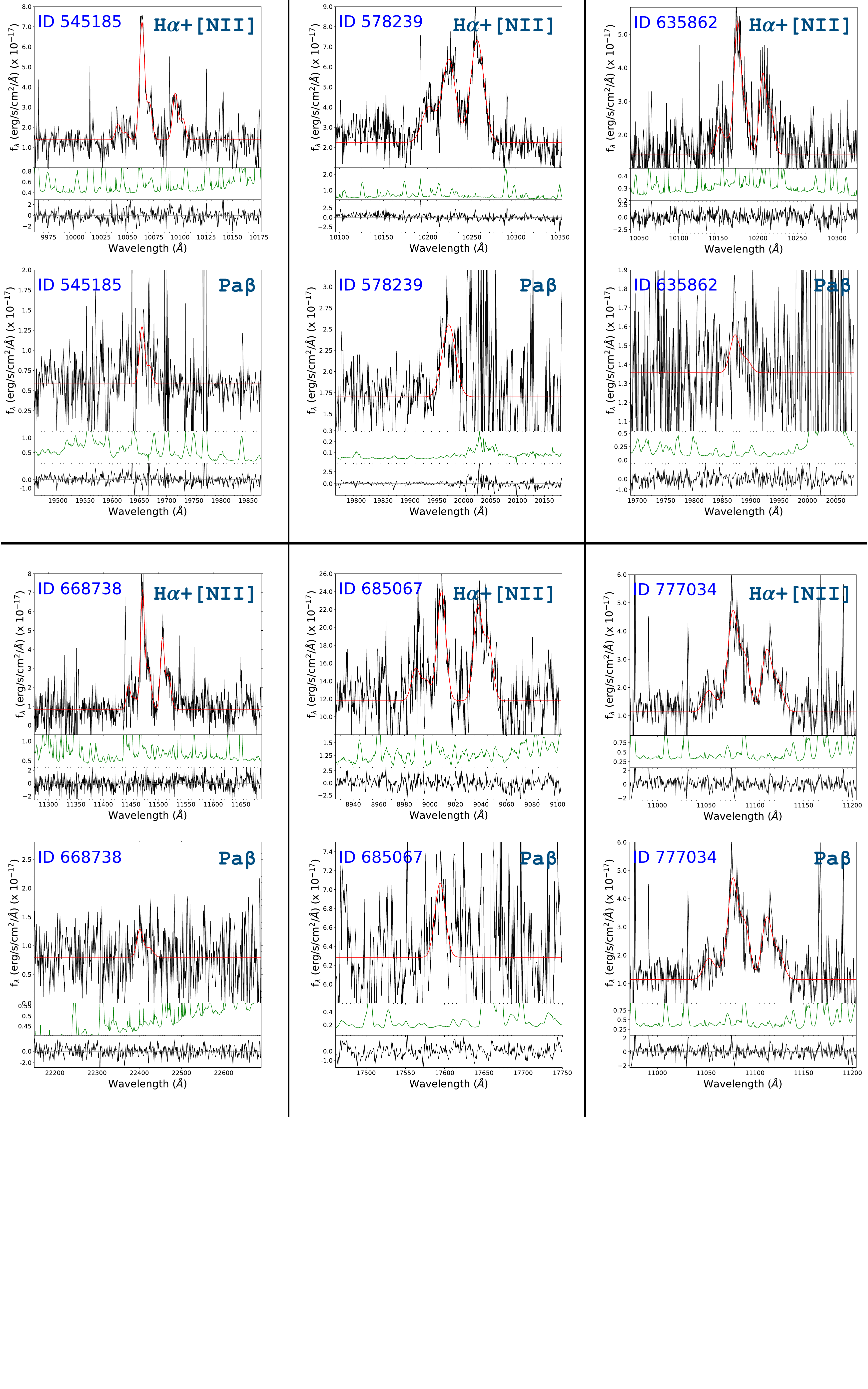}
  \caption{}\label{spectra3} 
\end{figure*}

\begin{figure*}[ht]
  \centering
  \includegraphics[angle=0,width=0.9\linewidth,trim={0cm 10cm 0cm 0cm},clip]{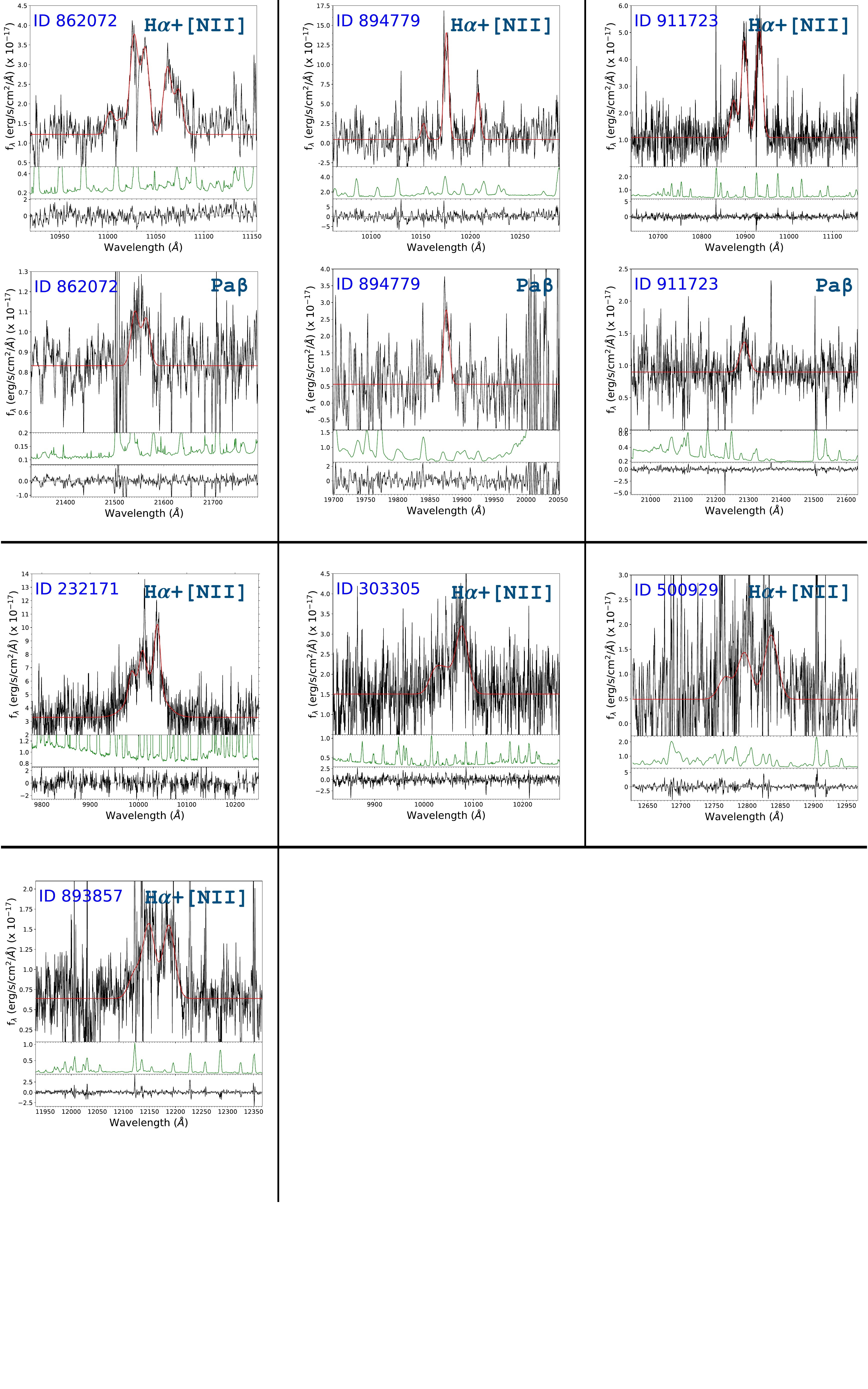}
  \caption{\small For each galaxy, we show in two contiguous panels (\textit{Top} and \textit{Bottom}) the H$\alpha$+[\ion{N}{II}]$\lambda 6583$\AA\ and Pa$\beta$ emission lines from the Magellan-FIRE spectrum (Pa$\gamma$ is shown for the galaxy ID 245158, since this line is used to infer the attenuation). Each panel is further divided into three sections: in the upper part is shown with a black line the 1-D portion of the spectrum (fully calibrated) close to H$\alpha$ or Pa$\beta$ line, along with the best-fit Gaussian superimposed (red line). In the second part below we display the noise spectrum for the same spectral range of the object (green line), while the fit residuals (noise normalized, as described in Calabr\`o et al. 2018) are shown in the bottom section with a black line. The galaxies are presented in numerical order for double detections, while those with only H$\alpha$+[\ion{N}{II}] detected are shown at the end.}\label{spectra4}
\end{figure*}

In this section, we show in Fig.\ref{spectra1}, \ref{spectra2}, \ref{spectra3}, \ref{spectra4}, the H$\alpha$+[\ion{N}{II}]$\lambda 6583$\AA\ and Pa$\beta$ emission lines (if detected) coming from the Magellan-FIRE spectra of our starbursts sample, fitted with single or double Gaussian profiles by using the tool MPFIT (see the Section \ref{line_measurements} for the procedure).
We show the best-fit Gaussians with red lines, superimposed on the fully reduced and calibrated spectra (black line). The noise spectrum in the same spectral range of the object is drawn in the lower panel for each cutout (green line), while the fit residuals (noise normalized, as described in Calabr\`o et al. 2018) are displayed in the bottom panel with a black line.  

For $12$ galaxies in our sample, we fitted their emission lines with a double Gaussian function (see Section \ref{line_measurements}), among which:
\begin{itemize}
    \item 6 of them are also pre-coalescence SBs (ID 245158, 493881, 470239, 223715, 545185, 668738) (see Section \ref{pre-coalescence}), and we interpret the two 1D Gaussians as coming from different merger components 
    \item for the starbursts ID 862072 and 249989, the shape of the H$\alpha$ line in the 2D spectra is consistent with global rotation, as it shows a spatially extended H$\alpha$ line with a single, uniform inclination. Additionally, their 1D line profiles are nearly symmetric, i.e., the two Gaussian components have the same flux within the uncertainties. Even so, we cannot definitely exclude that these signatures are also produced by identical merging pairs rotating around the common barycenter with opposite relative velocities. 
    \item in $4$ cases (ID 635862, 777034, 472775, 685067) we were not able to apply the visual criteria 1) and 2) presented in Section \ref{pre-coalescence}, because of the poor S/N of their sky-subtracted 2D spectra. However, we noticed that $3$ of them have very asymmetric 1D emission line profiles, with fluxes of the two Gaussian components differing by $>50\%$. Since they are less likely to be produced by a single rotating disk, we suggest they might be due to different merger components.
\end{itemize}

\end{document}